\documentclass[twocolumn]{aastex62}

\usepackage{amsmath}
\usepackage{amssymb}
\usepackage{amsfonts}
\usepackage{float}
\usepackage{tensor}
\usepackage{bm}
\usepackage{physics}

\newcommand{\hi}{H{\sc i} }

\newcommand{\ddfrac}[2]{\dfrac{\displaystyle #1}{\displaystyle #2}}

\DeclareMathOperator{\Arg}{Arg}

\begin{document}

\title{Ionospheric Attenuation of Polarized Foregrounds in 21 cm Epoch of Reionization Measurements: A Demonstration for the HERA Experiment}

\author{Zachary E. Martinot}
\affiliation{Dept. of Physics and Astronomy, University of Pennsylvania, Philadelphia, PA, USA}

\author{James E. Aguirre}
\affiliation{Dept. of Physics and Astronomy, University of Pennsylvania, Philadelphia, PA, USA}

\author{Saul A. Kohn}
\affiliation{Dept. of Physics and Astronomy, University of Pennsylvania, Philadelphia, PA, USA}

\author{Immanuel Q. Washington}
\affiliation{Dept. of Physics and Astronomy, University of Pennsylvania, Philadelphia, PA, USA}

\correspondingauthor{Zachary E. Martinot}
\email{zmarti@sas.upenn.edu}

\begin{abstract}

Foregrounds with polarization states 
that are not smooth functions of frequency present a challenge to \hi Epoch of Reionization (EoR) power spectrum measurements if they are not cleanly separated from the desired Stokes I signal.
The intrinsic polarization impurity of an antenna's electromagnetic response limits the degree to which components of the polarization state on the sky can be separated from one another, leading to the possibility that this frequency structure could be confused for \hi emission.
We investigate the potential of Faraday rotation by the Earth's ionosphere to provide a mechanism for both mitigation of, and systematic tests for, this contamination.
Specifically, we consider the delay power spectrum estimator, which relies on the expectation that foregrounds will be separated from the cosmological signal by a clearly demarcated boundary in Fourier space, and is being used by the Hydrogen Epoch of Reionization Array (HERA) experiment. 
Through simulations of visibility measurements which include the ionospheric Faraday rotation calculated from real historical ionospheric plasma density data, we find that the
incoherent averaging of the polarization state over repeated observations of  the sky may attenuate polarization leakage in the power spectrum by a factor of $10$ or more. Additionally, this effect provides a way to test for the presence of polarized foreground contamination in the EoR power spectrum estimate.

\end{abstract}


\section{Introduction}
\label{sec:intro}

Experiments seeking to observe the redshifted \hi signal from the Epoch of Reionization (EoR) must contend with foregrounds that are $\sim10^4$ times brighter than the cosmological signal by employing foreground removal or avoidance strategies \citep[e.g.][]{Santos05, Bernardi.09, Bernardi.10, Pober13, Dillon14}. These techniques rely on the smooth frequency structure of the foreground emission, in contrast to the spectrally structured cosmological signal \citep[e.g.][]{Datta.10, Morales.12,Trott.12, Pober.14, Liu.14a, Liu.14b, Nithya.15a, Nithya.15b}. While the total intensity (Stokes I) of foreground radiation is spectrally smooth, Faraday rotation during propagation through our galaxy produces frequency structure in the linear polarization state (Stokes $Q$ and $U$) at low frequencies \citep{jelic2010}. Although extra-galactic point-sources appear largely depolarized, the large scale synchrotron emission within the Milky Way appears to retain a significant level of polarization by the time it reaches an observer on Earth \citep{bernardi13, lenc16}.

The cosmological signal is expected to be effectively unpolarized given current experimental sensitivities \citep{BabichLoeb2005, Hirata.17} and thus will be detected by measurements of Stokes I on the sky. On its own, frequency structure in the polarization state would not seem to be a concern when the objective is a measurement of the total intensity. However, the dipole antenna elements used in low radio frequency interferometers generally have significant sensitivity over the full sky when compared to the faintness of the EoR emission - bright foreground emission off of boresight in the instruments beam may still be relatively bright compared to the cosmological emission along the antenna's boresight. Additionally, these dipole antennae are necessarily imperfect polarimeters over the full sky, and do not naturally produce measurements of the incident radiation field in an orthogonal basis - a necessary condition to properly measure Stokes I. This imperfection in the measurements, commonly referred to as "polarization leakage", means that even though we would like to make obtain a pure measurement of the the Stokes I intensity field on the sky, the measured visibilities will always involve a coupling to the polarization state of incident radiation. Although the sky is thought to be largely depolarized in the low frequency radio spectrum \citep{Farnes.14}, even a polarization fraction of $p\approx 10^{-2}$, which implies a polarized brightness that is "small" compared to total intensity, is not necessarily negligible compared to the cosmological signal, and therefore has the potential to produce contamination that is comparable to the EoR signal. This coupling must be understood and appropriately addressed to ensure that frequency structure in the polarization state of astrophysical foregrounds will not be mistaken for the cosmological power spectrum.

As a successor to the PAPER experiment \citep{Parsons.10} the HERA experiment \citep{deBoer17} plans to use a delay spectrum based estimator \citep{parsons2012} to make measurements of the EoR power spectrum. In contrast to other efforts to observe the EoR that pursue imaging-based methods, the delay spectrum analysis approach does not involve precision imaging and thus has not included detailed modeling and subtraction of polarized foregrounds. This makes potential contamination due to polarization leakage particularly concerning for the HERA experiment. However, in \citet{Moore17} it was proposed that the natural density fluctuations of the plasma in the Earth's ionosphere will produce a kind of polarization filter that can attenuate the coupling of visibility measurements to the polarization state of the sky. 

In this paper we seek to understand the magnitude of this ionospheric attenuation effect in visibility measurements and the derived power spectra. We simulate interferometric visibilities based on models that include the wide-field effect of the ionosphere on the polarization state of diffuse foregrounds, and the full-polarizaion instrumental response of an early HERA antenna design. This paper is organized as follows: in Section~\ref{sec:theory}, we review the relevant mathematical description of polarization in interferometric measurements including ionospheric Faraday rotation, and present a pedagogical picture of its attenuating effect on the measured polarized power. In Section~\ref{sec:numerical}, we discuss our implementation of this formalism which involves modeling of the instrumental response, the diffuse polarized foreground emission on the sky, and calculations using archival data of Faraday rotations based on real ionospheric behavior. Section~\ref{sec:results} presents the results of our simulations and analysis of the effect of ionospheric behavior on HERA observations. We conclude in Section~\ref{sec:conc}.

\section{Preliminary Formalism}
\label{sec:theory}

\subsection{Ionospheric Variation and Polarization Attenuation}

The ionosphere is a turbulent upper region of the Earth's atmosphere that is ionized by solar radiation \citep[e.g.][]{kintner85,loi15}. The permeation of this ionized medium by the persistent magnetic field of the Earth then produces a magnetized plasma which will induce a rotation in the linear polarization state of electromagnetic plane waves propagating through it - the effect known as Faraday rotation. The rotation angle of the electric vector is $\varphi \lambda^2$ where $\lambda = \frac{c}{\nu}$ is the wavelength and $\varphi$ is the rotation measure (RM) which is given - in SI units - by \citep{TMS}
\begin{equation}
\varphi(\vu{s}) = \frac{e^3}{8\pi^2 \epsilon_0 m_e^2 c^3} \int \rho_e(s,\vu{s}) \ \va{B}(s,\vu{s}) \mkern2mu{\vdot}  \vu{s}  \dd{s} 
\label{eqn:RM}
\end{equation}
where we have written the position vector $\va{s} = s \vu{s}$ and $\varphi$ has units of rad/m$^2$. Here the integral is taken along the line-of-sight $\vu{s}$, the function $\rho_e(\vu{s}, s)$ is the free electron density at a radial distance $s$ through the ionosphere, and $\va{B}(\vu{s},s)$ is the geomagnetic field.

The primary effect of the ionosphere that has concerned EoR power spectrum measurements so far is the refractive effect of the ionosphere \citep{vedantham15a,vedantham15b}. Here we are instead concerned with the effect of the ionosphere on the polarization state of the sky and with the short baselines ($\sim$ 10's of wavelengths) of the compact HERA array which are most sensitive to the large scale cosmological signal. The polarized emission at low radio frequencies appears to be dominated by large-scale diffuse Galactic emission rather than many unresolved point sources and the effect of small refractive shifts on such spatially smooth emission are thus expected to be negligible. We focus here on the changing Faraday rotation due to variations over long time scales in the ionospheric RM.

Driven by the heating from the sun, the free electron density $\rho_e$ varies quasi-cyclically with the rotation of the Earth at any fixed geographic location - the plasma density increases when the Sun is up and decreases at night but the ionosphere will not return to exactly the same state. This means observations of a polarized source on the sky made on different days will always be made through an ionospheric screen that is at least slightly different than the previous day.

\citet{Moore17} proposed that the effect of the ionospheric Faraday rotation on polarization leakage in a visibility could be estimated by approximating the RM over the sky as a constant $\varphi(\vu{s}) \approx \bar{\varphi}$ and additionally that the level of polarization leakage attenuation could be estimated without regard for the details of the instrumental response. While this simple approximation turns out to be an inadequate description of real visibilities it is equivalent to considering the effect of the ionosphere for a single source on the sky and is a good way to build some intuition. This will be useful for interpreting the results of the detailed simulations in Section \ref{sec:results}.

Suppose you used a good polarimeter to repeatedly observe a polarized source with a linear polarization state $(Q,U)$ on each of $N$ different days. Propagating through the ionosphere on the $n$-th day will rotate the polarization state by an angle $2 \varphi_n \lambda^2$ so that the observed polarization state is
\begin{equation}
Q_n + i U_n = e^{2 i \varphi_n \lambda^2} (Q + iU)
\label{eqn:LinPolRotation}
\end{equation}
where $\varphi_n \in \{\varphi_1, \ldots, \varphi_N \}$ is a sequence of different ionospheric rotation measures towards the source on the $n$-th day. If one then averages over all these observations the resulting quantity would be
\begin{align}
\overline{Q} + i \overline{U} & = \frac{1}{N}\sum_{n=1}^{N}(Q_n + i U_n) \\
& = (Q + i U)\frac{1}{N}\sum_{n=1}^{N} e^{2 i \varphi_n \lambda^2}.
\end{align}
While this would not be a sensible thing to do if one were actually interested in measurements of the polarization state, this averaging process is realized in the standard processing of HERA visibility data for power spectrum estimation. It is straight-forward to see that this decreases the magnitude of the polarization $\overline{L} = \abs{\overline{Q} + i \overline{U}}$ since the magnitude of the sum in the second line is always $\leq$ 1. The ratio of the intrinsic polarized power $L^2 = Q^2 + U^2$ to the power of the incoherently averaged polarization state $\overline{L}^2 = \overline{Q}^2 + \overline{U}^2$ is then
\begin{align}
A^2(N, \lambda, \varphi_1, \ldots , \varphi_N) & = \frac{\overline{L}^2}{L^2}\\
& =\frac{1}{N^2} \abs{\sum_{n=1}^{N} e^{2 i \varphi_n \lambda^2}}^2\\
& = \frac{1}{N} + \frac{2}{N^2} \sum_{k=1}^N \sum_{l = k+1}^N \cos(2\lambda^2 (\varphi_k - \varphi_l)) \label{eqn:AttenuationAmplitude}
\end{align}
We can think of the varying RM as defining a set of steps in a 2D plane with unit-length steps where the position after the $N$-th step is
\begin{equation}
Z(N) = \sum_{n=1}^{N} e^{2 i \varphi_n \lambda^2}.
\label{eqn:ComplexWalk}
\end{equation}
Then the attenuation is the (squared) ratio of the actual distance traveled from the origin $\abs{Z(N)}$ to the maximum distance $N$ that could have been traveled
\begin{equation}
A^2(N) = \abs{\frac{Z(N)}{N}}^2
\end{equation}
\begin{figure*}[p]
	\centering
	\begin{tabular}{c}
		\includegraphics[width=\textwidth]{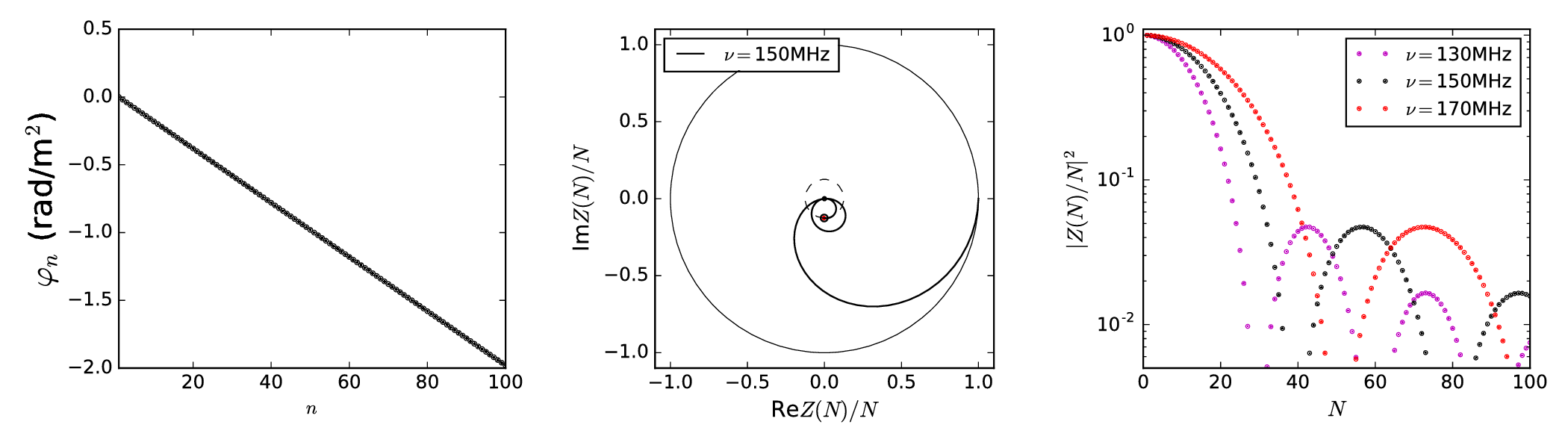}\\
		\includegraphics[width=\textwidth]{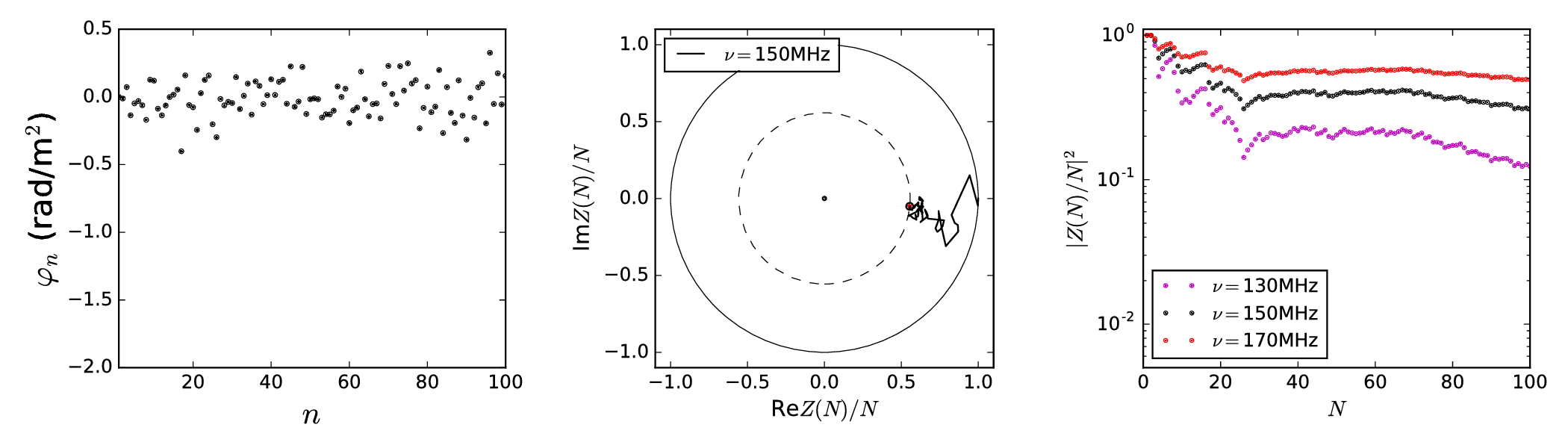}\\
		\includegraphics[width=\textwidth]{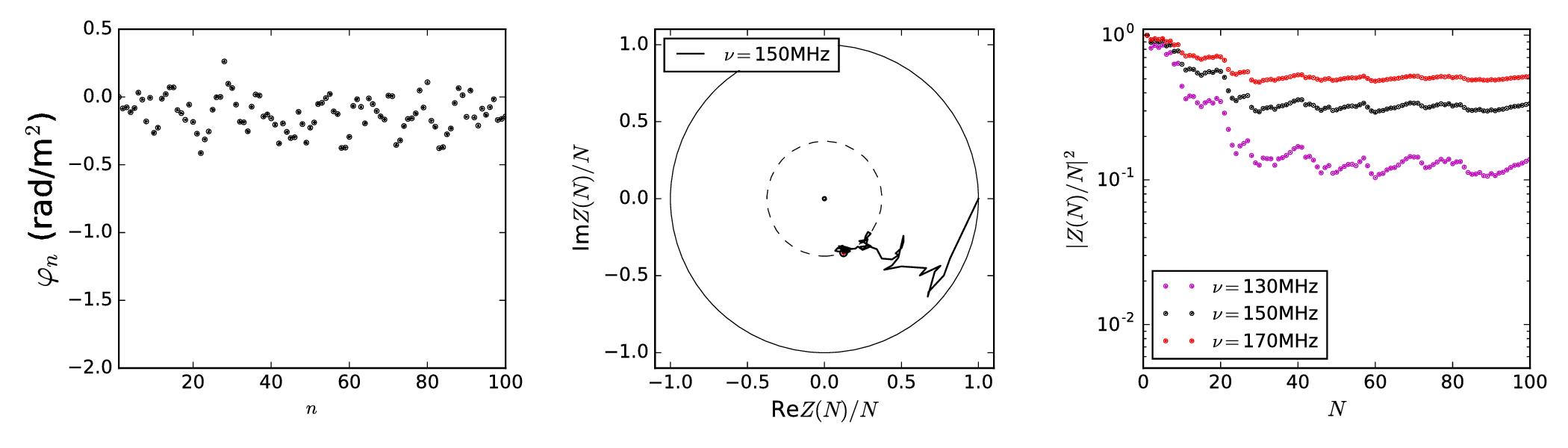}\\
		\includegraphics[width=\textwidth]{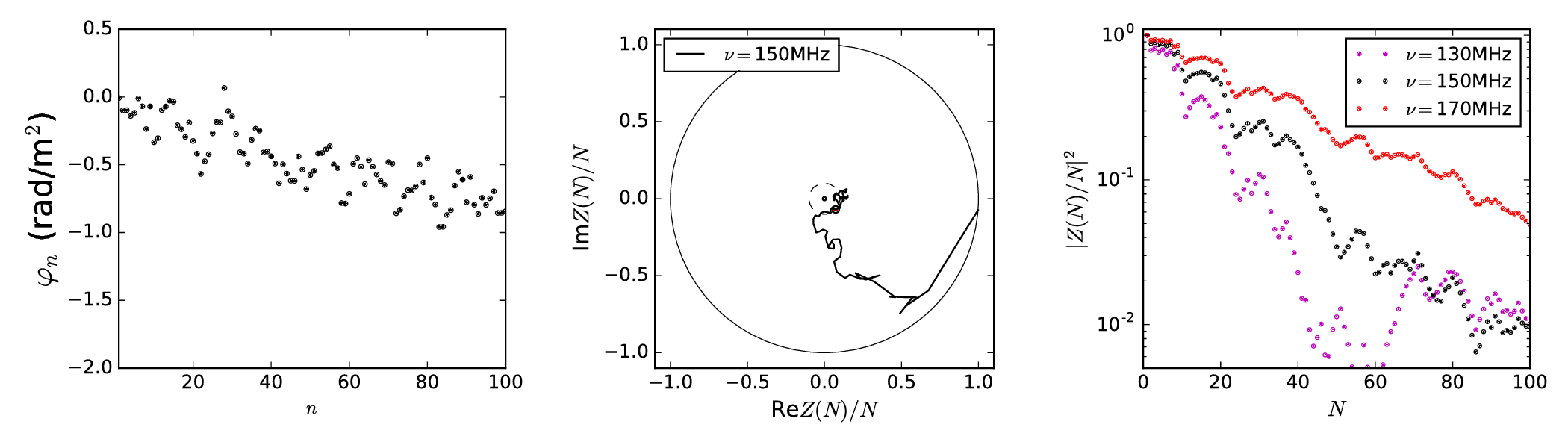}\\
	\end{tabular}
	\caption{Examples of four different types of normalized walks $Z(N)/N$ in the unit disk described in the text. The left-hand plots show the path of the walk over 100 steps, while the right-hand plots show the (squared) distance from the origin at each step - i.e. the attenuation of the polarization amplitude for a source observed through through an ionospheric variation characterized by the walk in left-hand plot. From top to bottom the walks are generated by a linear phase angle, an uncorrelated Gaussian distributed sequence of angles, a correlated Gaussian distributed sequence, and finally the same correlated Gaussian sequence with an additional linear trend added. In the center column the solid circle is the circle of radius $1$ while the dashed circle denotes the final radius of the walk at the position $Z(100)/100$, which is marked by a red dot. }
	\label{fig:walks}
\end{figure*}
Examples of walks for several distributions of the $\varphi_n$'s are shown in Figure \ref{fig:walks} along with the resulting attenuation curves as a function of $N$. From the top, the first panel shows the walk when $\varphi_n$ is simply a linear function
\begin{equation}
\varphi_n = \alpha n
\end{equation}
for some slope $\alpha$. In this case the expression for the attenuation can be simplified by summing the geometric series
\begin{align}
Z(N) & = \sum_{n=1}^N \qty(e^{2 i \alpha \lambda^2})^n \\
& = e^{2 i \alpha \lambda^2} \frac{1 - e^{2 i \alpha N \lambda^2}}{1 - e^{2 i \alpha \lambda^2}}.
\end{align}
The attenuation factor is then 
\begin{align}
A^2(N, \lambda, \alpha)  = \ddfrac{\sin[2](2 \lambda^2 \alpha N)}{\sin[2](2 \lambda^2 \alpha) N^2}
\end{align}
and the top right-hand panel plots $A^2$ with $\alpha = 0.02$ rad m$^{-2}$.

The second panel from the top shows the random walk generated by a realization of a sequence of $N$ uncorrelated Gaussian random variables $\varphi_n \sim \mathcal{N}(0, \sigma^2)$ with $\sigma = 0.2$ rad m$^{-2}$. In this case it is straightforward to compute the expectation value of the attenuation factor in Equation \ref{eqn:AttenuationAmplitude} which is
\begin{equation}
\expval{A^2} = \frac{1}{N} + e^{-4\sigma^2 \lambda^4} \qty(1 - \frac{1}{N}).
\end{equation}
While the real ionospheric RM sequences of interest to us are not necessarily well described as a Gaussian random variable or a purely linear trend, the features of these simple models are worth noting. In the case of the linear trend, the attenuation oscillates as the walk passes near the origin, but is bounded by a $\sim \frac{1}{N^2}$ envelope. For the Gaussian distribution with finite variance $\sigma^2$, the attenuation eventually approaches an asymptote $A^2 \rightarrow  e^{-4\sigma^2 \lambda^4}$ as $N \rightarrow \infty$. On the other hand taking $\sigma^2 \rightarrow \infty$ produces the limit $\expval{A} \rightarrow \frac{1}{N}$, corresponding to a uniform distribution over the angle $2 \varphi_n \lambda^2$. 

The third panel from the top then shows the walk generated by a sequence of correlated Gaussian random variables $(\varphi_1, \ldots, \varphi_N) \in \mathcal{N}(0, \bm{\Sigma})$ where the covariance matrix $\bm{\Sigma}$ is given by
\begin{align}
\Sigma_{kl} = \sigma_c^2 e^{-\frac{(k - l)^2}{\ell_1^2}} + \sigma_c^2 e^{-\frac{(k-l)^2}{\ell_2^2}},\\
\text{ } \ell_1 = 1, \text{ } \ell_2 = \sqrt{5}, \text{ } \sigma_c = 0.1\text{ rad m}^{-2}.
\end{align}
Then the panel at the bottom shows the result of adding a linear trend to the exact same $\varphi_n$ sequence as in the third panel, $\varphi_n \rightarrow \varphi_n + \alpha n$ with $\alpha = 0.007$ rad m$^{-2}$.

These walks and the resulting attenuation curves provide some intuition for the features of the attenuation curves obtained from the more complex visibility simulations - varying degrees of smoothness, discontinuities, oscillation, and lack of consistent monotonicity in frequency. The attenuation of polarization in a visibility can then be seen as a function of each of the random walks taken by the polarization in each direction on the sky - and this function is explicitly the visibility measurement equation.

\subsection{Polarization in Interferometric Visibilities and the Delay Spectrum}
The fundamental measurement made by interferometric arrays is the correlation function of the electric field. The van Cittert-Zernike theorem \citep{born1999principles}, suitably generalized to include the four possible 2-point correlation functions for pairs of isolated and identical dual-feed dipole antennae \citep{carozzi2009, smirnov2011a, smirnov2011d} relates the polarized intensity distribution on the sky to a measured visibility matrix by:
\begin{align}\label{eqn:VisMat}
\bm{\mathcal{V}}(t,\nu,\va{b}) & = \int_{\mathbb{S}^2} \bm{\mathcal{J}}(t,\nu,\vu{s}) \bm{\mathcal{C}}(t,\nu,\vu{s}) \bm{\mathcal{J}}^\dagger(t,\nu,\vu{s}) e^{- 2 \pi i \frac{\nu}{c} \va{b} \vdot \vu{s}} \\
& = \mqty*(\mathcal{V}_{aa}(t,\nu,\va{b}) & \mathcal{V}_{ab}(t,\nu,\va{b}) \\ \mathcal{V}_{ba}(t,\nu,\va{b}) & \mathcal{V}_{bb}(t,\nu,\va{b}) \\)
\end{align}
where the integral is taken over the unit sphere  $\mathbb{S}^2 = \{\vu{s} : \vu{s} \in \mathbb{R}^3,  \norm{\vu{s}} = 1\}$ (a.k.a "the sky"). Here the vector $\va{b}$ denotes a baseline between the two antennas, $\nu$ is the sampled frequency, $\bm{\mathcal{J}}$ is a direction-dependent Jones matrix, $a$ and $b$ label  two different antenna feed orientations, and $\bm{\mathcal{C}}$ is the polarized brightness field on the sky expressed as the rank-2 coherency tensor field
\begin{align}
\bm{\mathcal{C}} & =
\expval*{\mathcal{E}_\delta \mathcal{E}_\delta^*} \vu{e}_\delta \otimes \vu{e}_\delta + 
\expval*{\mathcal{E}_\delta \mathcal{E}_\alpha^*} \vu{e}_\delta \otimes \vu{e}_\alpha + \ldots \nonumber \\ 
& \qquad{} \qquad{} \ldots
\expval*{\mathcal{E}_\alpha \mathcal{E}_\delta^*} \vu{e}_\alpha \otimes \vu{e}_\delta +
\expval*{\mathcal{E}_\alpha \mathcal{E}_\alpha^*} \vu{e}_\alpha \otimes \vu{e}_\alpha\\
& = \mqty*(\expval*{\mathcal{E}_\delta \mathcal{E}_\delta^*} & \expval*{\mathcal{E}_\delta \mathcal{E}_\alpha^*} \\ \expval*{\mathcal{E}_\alpha \mathcal{E}_\delta^*} & \expval*{\mathcal{E}_\alpha \mathcal{E}_\alpha^*}).
\end{align}
The symbol $\otimes$ denotes a tensor product of unit vectors and the $\expval{.}$ denotes an ensemble average of the incoherent celestial radiation fields. The functions $\mathcal{E}_\delta, \mathcal{E}_\alpha$ are the projections of the complex-valued electric-vector amplitude
\begin{equation}
\va{\mathcal{E}}(\vu{s}) = \mathcal{E}_\delta(\vu{s}) \vu{e}_\delta + \mathcal{E}_\alpha(\vu{s}) \vu{e}_\alpha
\end{equation} 
of a plane wave with wave-vector $\propto \vu{s}$ and components specified in the normalized tangent basis $\{\vu{e}_\delta, \vu{e}_\alpha\}$ induced by the equatorial coordinates Right Ascension $\alpha$ and Declination $\delta$.  As a Hermitian matrix the coherency matrix is by definition specified by the frequency and direction dependent Stokes parameters $I,Q,U,V$ so that
\begin{align}
\bm{\mathcal{C}} & = \frac{1}{2} \mqty*(I + Q & U - iV \\ U + iV & I - Q\\)\\
& = \frac{1}{2}\qty(I \bm{\sigma}_I + Q \bm{\sigma}_Q + U \bm{\sigma}_U + V \bm{\sigma}_V)
\end{align}
where the $\bm{\sigma}_\mathcal{S}$ matrices are the Pauli matrices
\begin{alignat}{2}
\bm{\sigma}_I & = \mqty*(1 & 0 \\ 0 & 1\\), & \quad \bm{\sigma}_Q & = \mqty*(1 & 0 \\ 0 & -1), \\
\quad \bm{\sigma}_U & = \mqty*(0 & 1 \\ 1 & 0), &\quad \bm{\sigma}_V & = \mqty*(0 & -i \\ i & 0).
\end{alignat}
For our purposes here it is useful to adopt the point of view of a fixed observer under a rotating sky. Therefore we will think of $\bm{\mathcal{C}}$ as a time $t$ dependent function while the instrumental response of a drift-scanning antenna is fixed with respect to $t$. The coherency tensor is then a periodic function of the time $t$ of the observation with a period $T$ which is the rotational period of the Earth
\begin{equation}
\bm{\mathcal{C}}(t) = \bm{\mathcal{C}}(t + T).
\end{equation}
On the other hand the ionospheric RM is only quasi-cyclic and thus not periodic in $t$. Since visibility data is averaged over multiple days of observation at the same LST it is useful to break the time variable into the LST $t \in [0, T)$ and an integer $n$ that indexes sidereal days. So from here on we will use the $t$ variable to refer only to the LST of an observation. Then we write $\bm{\mathcal{V}}(n, t, \nu, \va{b})$ as the visibility matrix observed on the $n$'th sidereal day at the LST $t$.

The effect of the ionospheric Faraday rotation is described by a Jones matrix 
\begin{equation}
\bm{R}_n(\nu, t, \vu{s}) = \mqty*(\cos(\varphi(n,t,\vu{s}) \frac{c^2}{\nu^2}) & \sin(\varphi(n,t,\vu{s}) \frac{c^2}{\nu^2}) \\ -\sin(\varphi(n,t,\vu{s}) \frac{c^2}{\nu^2}) & \cos(\varphi(n,t,\vu{s}) \frac{c^2}{\nu^2}))
\end{equation}
which describes the rotation of a plane wave vector amplitude $\va{\mathcal{E}} \rightarrow \bm{R}_n \va{\mathcal{E}}$ upon propagation through the ionosphere. In this work we compute the visibilities resulting from 
\begin{equation}
\bm{\mathcal{J}}(n,t, \nu, \vu{s}) = \bm{J}(\nu, \vu{s}) \bm{R}_n(t, \nu, \vu{s})
\end{equation}
where $\bm{J}$ is the instrumental Jones matrix that describes the response of the instrument to polarized plane wave excitations. We do not include further direction-independent Jones matrices so that our analysis concerns an idealized limit of data that has been calibrated for the direction independent receiver-chain effects, nor do we include a thermal noise term in order to isolate the effect of the ionospheric Faraday rotation.

\begin{figure*}[p]
\centering
\includegraphics[scale=1.0]{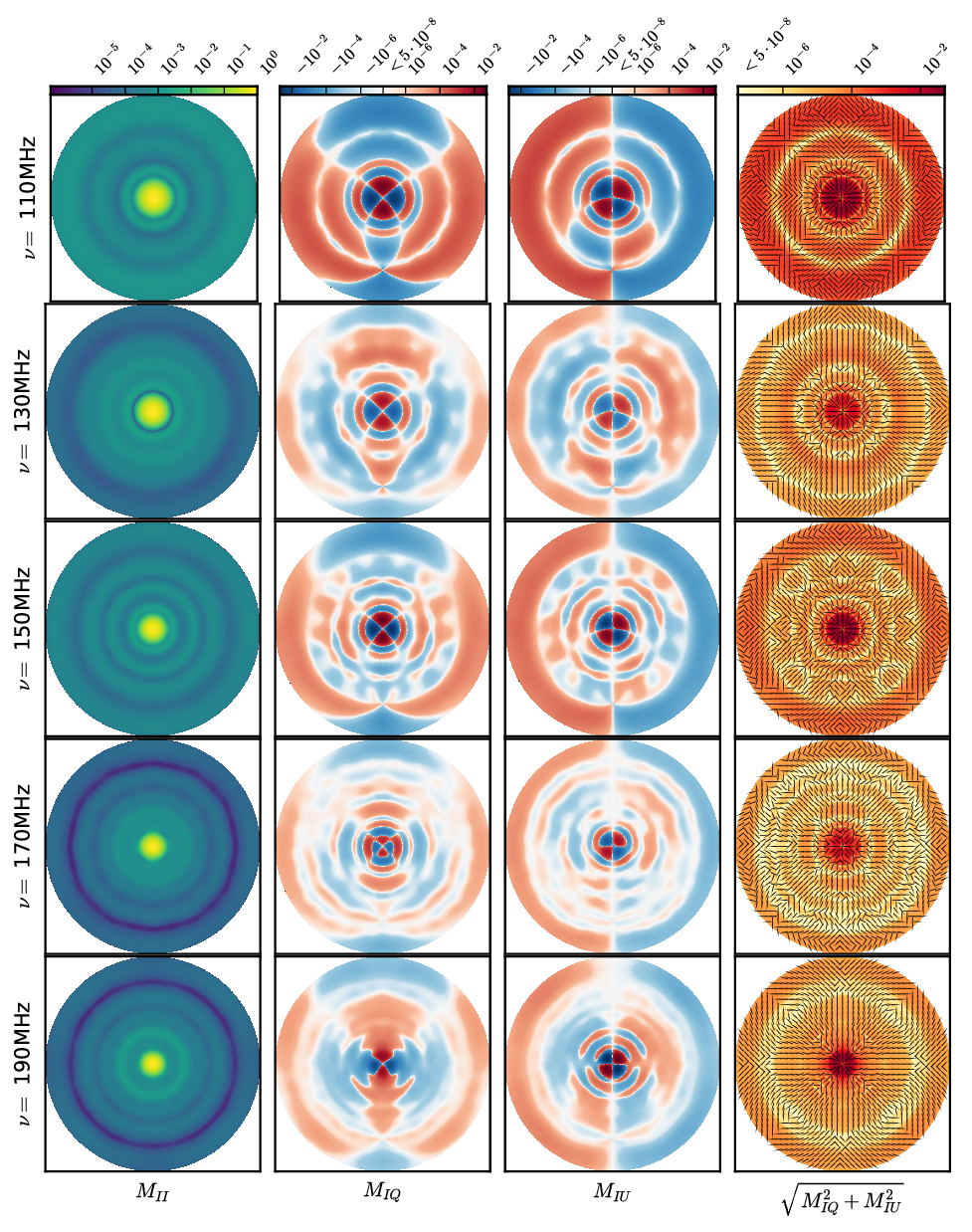}

\caption{The Mueller matrix elements for the $\{\vu{e}_\delta, \vu{e}_\alpha\}$ basis (see Appendix \ref{sec:InstrumentAppendix}) in Equation \ref{eqn:MuellerIntegral} derived from a simulation of a transmitting HERA antenna's far-field electric vector fields. Each image is a Lambert equal-area projection of the function on the hemisphere centered on the antenna's bore-sight - the inscribed circle is the antenna's local horizon. The rows are different frequencies 110,130,150,170, and 190 MHz, from top to bottom. Since the $M_{IQ}, M_{IU}$ elements take values in a range that is symmetric about 0, the color scale is a symmetric $\log_{10}$ that spans six orders of magnitude on each side, with linearized values in $10^{-6} - 10^{-8}$. All values are \textless 1 in absolute value so the sign is unambiguous. Overlaid each image in the last column on the right is a unit tensor field that shows the geometric orientation of the polarization leakage. The plotted vector is headless and thus symmetric under a rotation by an angle $\pi$. This reflects the symmetry of the polarization state $Q+iU$ of the \textit{source} of incident radiation under a rotation by $\pi$.}
\centering
\label{fig:HERAmueller}
\end{figure*}

\begin{figure*}[p]
\centering
\begin{tabular}{c}
\includegraphics[height=0.48\textheight]{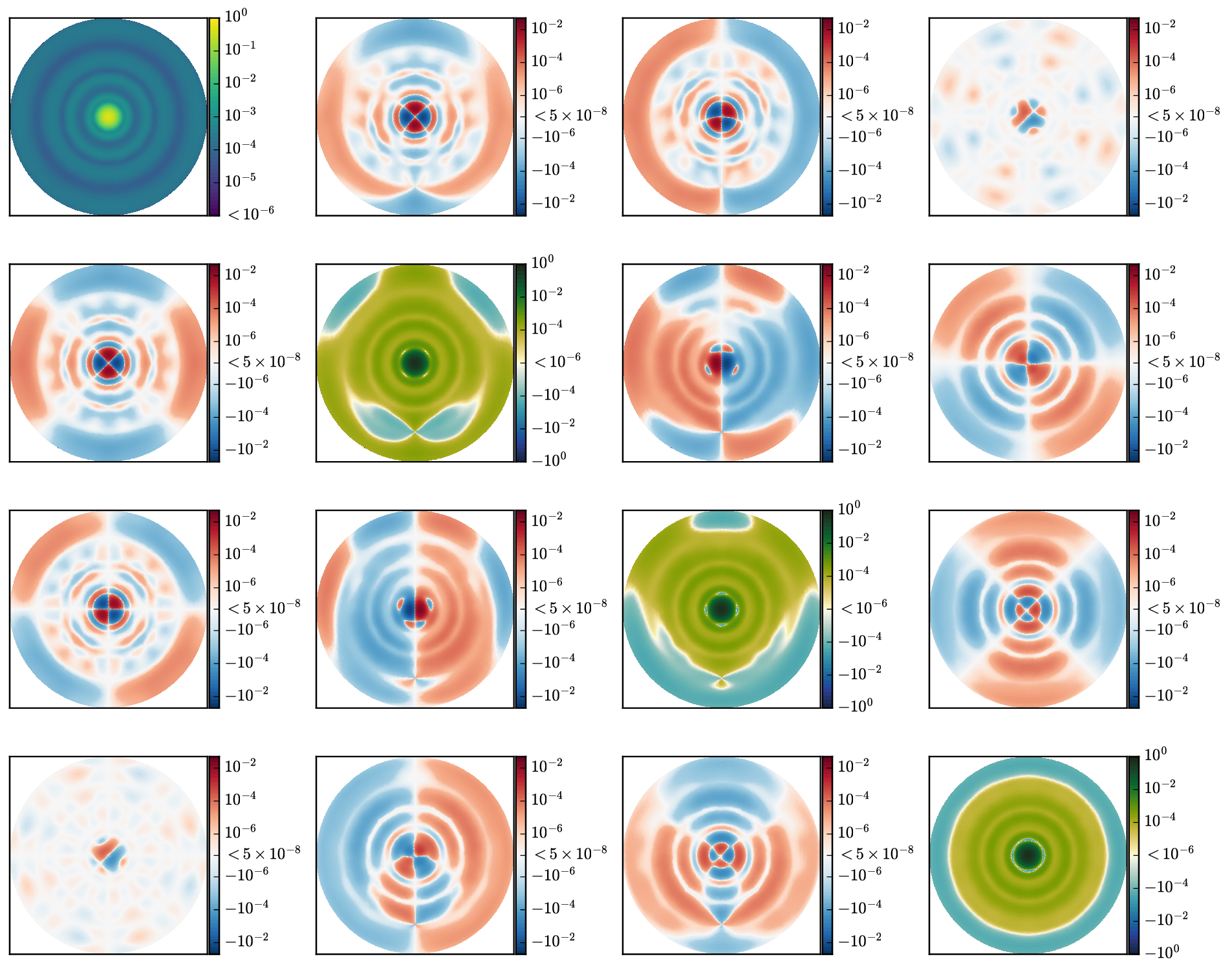}\\
\hline \\
\includegraphics[height=0.48\textheight]{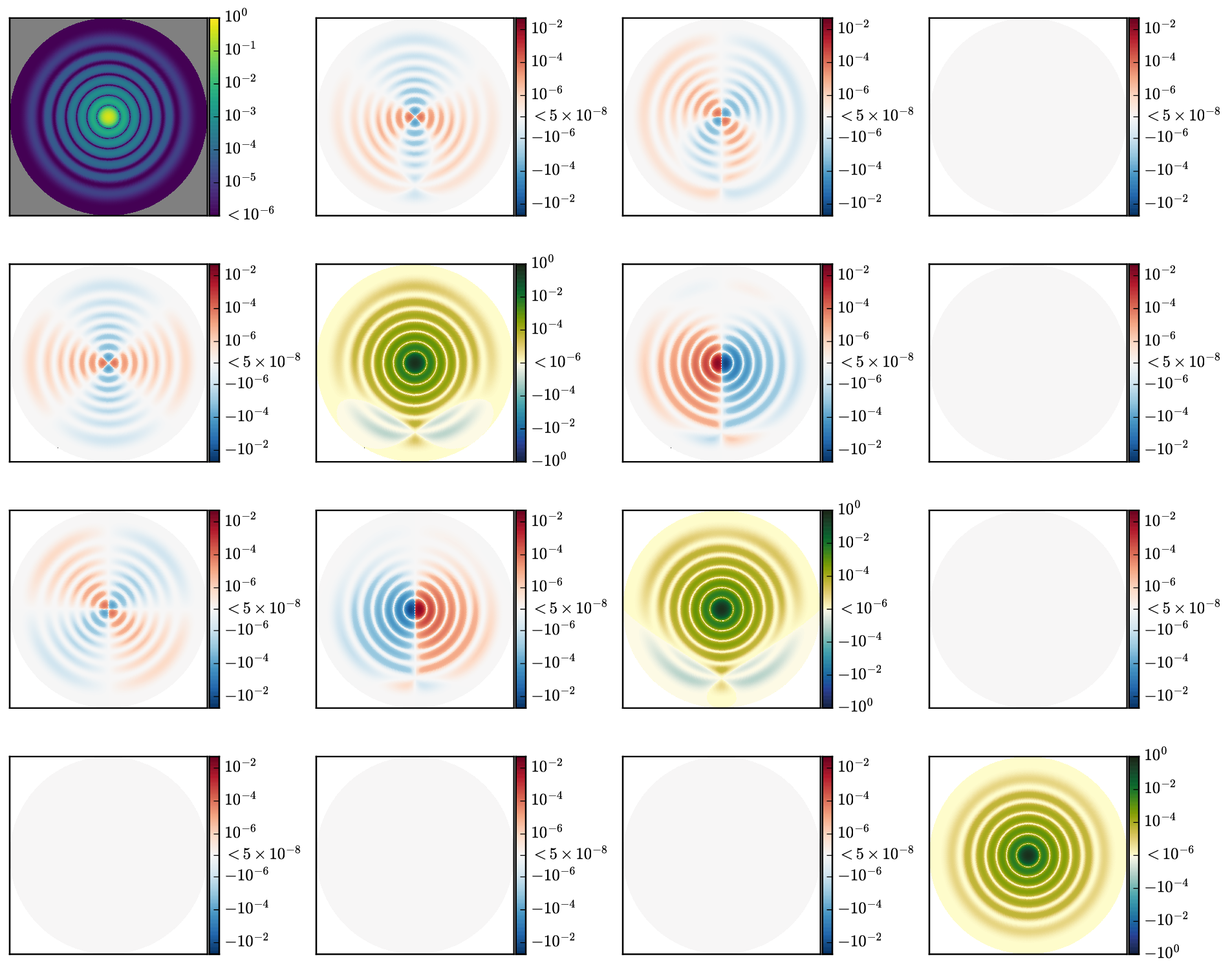}\\
\end{tabular}

\caption{Mueller matrices at $150$MHz as defined in Equation \ref{eqn:MuellerComponents} for the HERA model (top) and the analytically defined Airy beam dipole (bottom). As in Figure \ref{fig:HERAmueller} the matrix elements are shown in the $\{\vu{e}_\delta, \vu{e}_\alpha\}$ basis (see Appendix \ref{sec:InstrumentAppendix}). The rows are the kernels for each of the Vokes parameters i.e the first row is $M_{II},M_{IQ},M_{IU}, M_{IV}$ from left to right, etc. The color scales are as described in Figure \ref{fig:HERAmueller}, but note that the diagonal elements have a different range than the off-diagonal components.}
\label{fig:fullHERAmueller}
\end{figure*}

From the visibilities we may form linear combinations analogous to the Stokes parameters, which we will refer to as "Vokes" parameters  e.g. "Vokes-I parameter". The Vokes parameters are
\begin{alignat}{2}
\mathcal{V}_I &\equiv \Tr(\bm{\sigma}_I \bm{\mathcal{V}}), &\quad  \mathcal{V}_Q &\equiv \Tr(\bm{\sigma}_Q \bm{\mathcal{V}}), \\
\quad \mathcal{V}_U &\equiv \Tr(\bm{\sigma}_U \bm{\mathcal{V}}), &\quad \mathcal{V}_V &\equiv \Tr(\bm{\sigma}_V \bm{\mathcal{V}}).
\end{alignat}
As noted in Section \ref{sec:intro} the cosmological signal we are interested in detecting is thought to be effectively unpolarized so Vokes-I provides the highest sensitivity to the cosmological Stokes-I signal, even though it is not generally a pure measurement of Stokes-I. Thus, the quantity used for estimation of the power spectrum in the delay spectrum estimator is,
\begin{align}
\mathcal{V}_I & = \Tr (\bm{\sigma}_I \bm{\mathcal{V}}) \\
& = \int_{\mathbb{S}^2} \Tr (\bm{\sigma}_I \bm{J} \bm{R}_n \bm{\mathcal{C}} \bm{R}_n^\dagger \bm{J}^\dagger) e^{- 2 \pi i \frac{\nu}{c} \va{b} \vdot \vu{s}} \\
& = \int_{\mathbb{S}^2} (M_{II}I + M_{IQ} Q_n + M_{IU} U_n + M_{IV}V) e^{- 2 \pi i \frac{\nu}{c} \va{b} \vdot \vu{s}} \label{eqn:MuellerIntegral}
\end{align}
where 
\begin{equation}
M_{ij} = \frac{1}{2}\Tr (\bm{\sigma}_i \bm{J} \bm{\sigma}_j \bm{J}^\dagger )
\label{eqn:MuellerComponents}
\end{equation}
are the instrumental Mueller matrix elements as shown in Figures \ref{fig:HERAmueller} and \ref{fig:fullHERAmueller}, and $Q_n, U_n$ are the linear polarization components after undergoing Faraday rotation in the ionosphere (Equation \ref{eqn:LinPolRotation}).

As an aside, it may also be useful to note that in the same way that the linear polarization on the sky may be described as a spin-2 field (in either Cartesian or polar form)
\begin{equation}
Q+iU  =  L e^{i 2\chi},
\end{equation}
the associated Mueller matrix elements describing the polarization impurity are also the components of a spin-2 field:
\begin{equation}
M_{IQ} + i M_{IU} =  M_{IL} e^{i 2 \psi}.
\end{equation} 
The Vokes-I polarization leakage terms can then be thought of as an inner-product between the polarization impurity of the instrument and the polarization state $Q$,$U$ of the incident radiation which is
\begin{equation}
M_{IQ}Q + M_{IU}U = M_{IL}L \cos(2\chi - 2\psi)
\end{equation}
The far-right column of Figure \ref{fig:HERAmueller} shows the scalar function $M_{IL}$ overlaid with a unit tensor field which shows the orientation on the sky of the instrumental impurity - geometrically, when the unit tensors of the instrumental impurity and the polarization state of the sky are at 45 degrees (or 135 degrees measured the other way), the inner product is 0. When the angle is 90 degrees, the inner product is negative and minimized i.e. $M_{IQ}Q + M_{IU}U = -M_{IL}L$. The visualization in Figure \ref{fig:HERAmueller} permitted by this representation may be useful for understanding the effect of the ionospheric Faraday rotation, which will be examined in more detail in Section \ref{sec:results}.

The power spectrum can then be estimated from $\mathcal{V}_I$ for each baseline  through the delay transform \citep{parsons2012}
\begin{align}\label{eqn:DelayTransform}
\widetilde{\mathcal{V}}_I(n, t, \tau, \va{b}, \mathcal{B}) = \int_{-\infty}^\infty \dd{\nu} W(\nu, \mathcal{B}) \mathcal{V}_I(n, t,\nu,\va{b}) e^{2 \pi i \tau \nu}
\end{align}
where $\mathcal{B}$ is the frequency band selected to estimate the power spectrum at a given redshift and $W(\nu, \mathcal{B})$ is a windowing function that accounts for the finite bandwidth and any further choice of tapering function. Then an estimator $\widehat{P}(k)$ for the spherically-averaged power spectrum $P(k)$ is obtained by averaging over time samples of the delay spectra (see e.g \citet{ali2015} for the additional analysis complexities required with real measurements, and \citet{liu2016} for more in-depth theoretical considerations)
\begin{equation}
\widehat{P}(k(\tau)) \propto \expval{\abs{\widetilde{\mathcal{V}}_I(\tau)}^2}_{t,\va{b}}.
\end{equation}
This method of estimating the power spectrum motivates the analysis in this paper, but since we will only be considering ratios of different power spectra the precise proportionality is unimportant here. 

\section{Visibility Simulation Components}
\label{sec:numerical}

We compute the visibility matrix in Equation \ref{eqn:VisMat} by a quadrature on a \textsc{HEALPix}\footnote{\url{http://healpix.sourceforge.net}} pixelization of the sky \citep{Gorski05}. The functions $\bm{J}\qty(\nu, \vu{s})$, $\bm{\mathcal{C}}(t,\nu, \vu{s})$, and $\bm{R}_n(t,\nu, \vu{s})$ are evaluated for each $\vu{s} = \vu{s}_p$ in the set of \textsc{HEALPix} pixels $\{\vu{s}_p\}_{p=1}^{N_p}$. Explicitly, Equation \ref{eqn:VisMat} is estimated as
\begin{widetext}
\begin{equation}
\bm{\mathcal{V}}(n, t, \nu, \va{b}) = \sum_{p=1}^{N_p} \bm{J}(\nu, \vu{s}_p) \bm{R}_n(t, \nu, \vu{s}_p) \bm{\mathcal{C}}(t, \nu, \vu{s}_p) \bm{R}_n^\dagger(t, \nu, \vu{s}_p)\bm{J}^\dagger(\nu, \vu{s}_p) e^{-2 \pi i \frac{\nu}{c} \va{b} \vdot \vu{s}_p} \Delta\Omega,
\end{equation}
\end{widetext}
where $\Delta \Omega = \frac{4 \pi}{N}$.
In this section we discuss our definition and evaluation of these three functions, as well as particular parameters of our simulations.

\subsection{Calculation of the Ionospheric RM from Archival TEC}
\label{sec:radionopy}

Over the past three decades methods to measure the total electron content (TEC) using global positioning system (GPS) dual-frequency receivers have been developed and improved \citep[e.g.,][]{royden84, lanyi88, mannucci98, schaer1999, iijima99, komjathy05, erdogan16}. These methods utilize the TEC-induced time delay between the arrival of radio waves of two closely-spaced frequencies to estimate the TEC value of the ionosphere above a GPS station. Repeating this method for stations around the world and interpolating spatially provides an estimate of the TEC above any location on Earth.  


Meanwhile, many generations of the International Geomagnetic Reference Field \citep[IGRF; e.g.,][]{Finlay10} have continually improved the model of the Earth's magnetic field. This model is composed by spatial interpolation of magnetic field measurements (in up to 13th-order spherical harmonic coefficients) reported by institutions around the world. 

\begin{figure*}
	\centering
	\includegraphics[width=\textwidth]{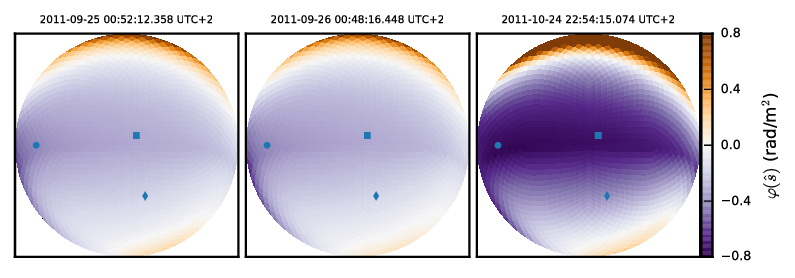}
	\caption{Images of $\varphi(\vu{s})$ at a fixed LST of 2.5 hours. The first two maps are consecutive sidereal days, the third is 30 days later. At the top of each panel is the civil time and date of the RM snapshot. The images are horizon-to-horizon in a Lambert equal-area projection centered on longitude $=+21.4283^\circ$, latitude$=-30.7215^\circ$. The square, diamond, and circle indicate the points for which RM sequences over the day-index $n$ are shown in Figure \ref{fig:RMskewers}. The images are oriented so that North is up and East is to the right. The places where $\varphi \rightarrow 0$ correspond to points where $\va{B}\vdot\vu{s} = 0$; the null to the North is near the equator.}
	\label{fig:RMmaps}
\end{figure*}

Based on the {\tt IonFR}\footnote{\url{sourceforge.net/projects/ionfarrot/}} package of \citet{sotomayor-beltran13}, we have developed {\tt radionopy}\footnote{\url{github.com/UPennEoR/radionopy}}, a \texttt{python} package to calculate ionospheric RM values (the function $\varphi$ in Section \ref{sec:theory}). Like {\tt IonFR}, {\tt radionopy} uses GPS-derived TEC maps (in the IONosphere Map EXchange format; {\sc ionex}) and the IGRF to estimate the value of $\varphi$ at a given latitude, longitude and date. Unlike its predecessor, {\tt radionopy} is written to calculate $\varphi$($\hat{s}$) over an arbitrary point-set of directions on the sky, allowing images of the full sky. Additionally \texttt{radionopy} implements the temporal interpolation scheme recommended in \citet{schaer1998} to obtain full-sky maps for arbitrary times between the 2-hour time resolution of the provided {\sc IONEX} data, such that the resulting $\varphi(n, t, \vu{s})$ is a fairly smooth function of $t$. The interpolation scheme is as follows. Let $\eta$ denote universal time (UT) and $i$ index the times at which the TEC maps
\begin{equation}
\overline{\rho}_e(\eta_i, \theta, \phi) = \int \rho_e(\eta_i, \theta, \phi, s) \dd{s}
\end{equation} 
are available as a function of the geocentric latitude $\theta$ and longitude $\phi$. Then the interpolated TEC map at an arbitrary time $\eta$ such that $\eta_i \leq \eta \leq \eta_{i+1}$ is a linear interpolation of the forward and backward rotated preceding and succeeding maps given by
\begin{multline}
\overline{\rho}_e(\eta, \theta, \phi) = \frac{\eta_{i+1} - \eta}{\eta_{i+1} - \eta_i} \overline{\rho}_e(\eta_i, \theta, \phi_i) + \ldots \\ 
\ldots \frac{\eta - \eta_i}{\eta_{i+1} - \eta_i} \overline{\rho}_e(\eta_{i+1}, \theta, \phi_{i+1})
\end{multline} 
where $\phi_k = \phi + \omega_\oplus(\eta - \eta_k)$ with $\omega_\oplus$ the angular speed of the Earth. The RM function $\varphi(n,t(\eta),\vu{s}(\theta, \phi))$ is then computed by the approximation of Equation \ref{eqn:RM} described in \citet{sotomayor-beltran13}.  Figure~\ref{fig:RMmaps} shows an example of \texttt{radionopy} output for the RM function $\varphi(n,t,\vu{s})$ evaluated in altitude / azimuth coordinates at the location of the HERA array, and Figure~\ref{fig:compareTEC} shows example output TEC maps over the Earth. 

While the intrinsic time and spatial resolution of the resulting RM maps is relatively low, the RM computed in this way has recently been validated as being reasonably close to more precise measurements using pulsar timing dispersion \citep{malins18}.

\begin{figure*}[h]
	\centering
	\begin{tabular}{ccc}
		\includegraphics[width=0.33\textwidth]{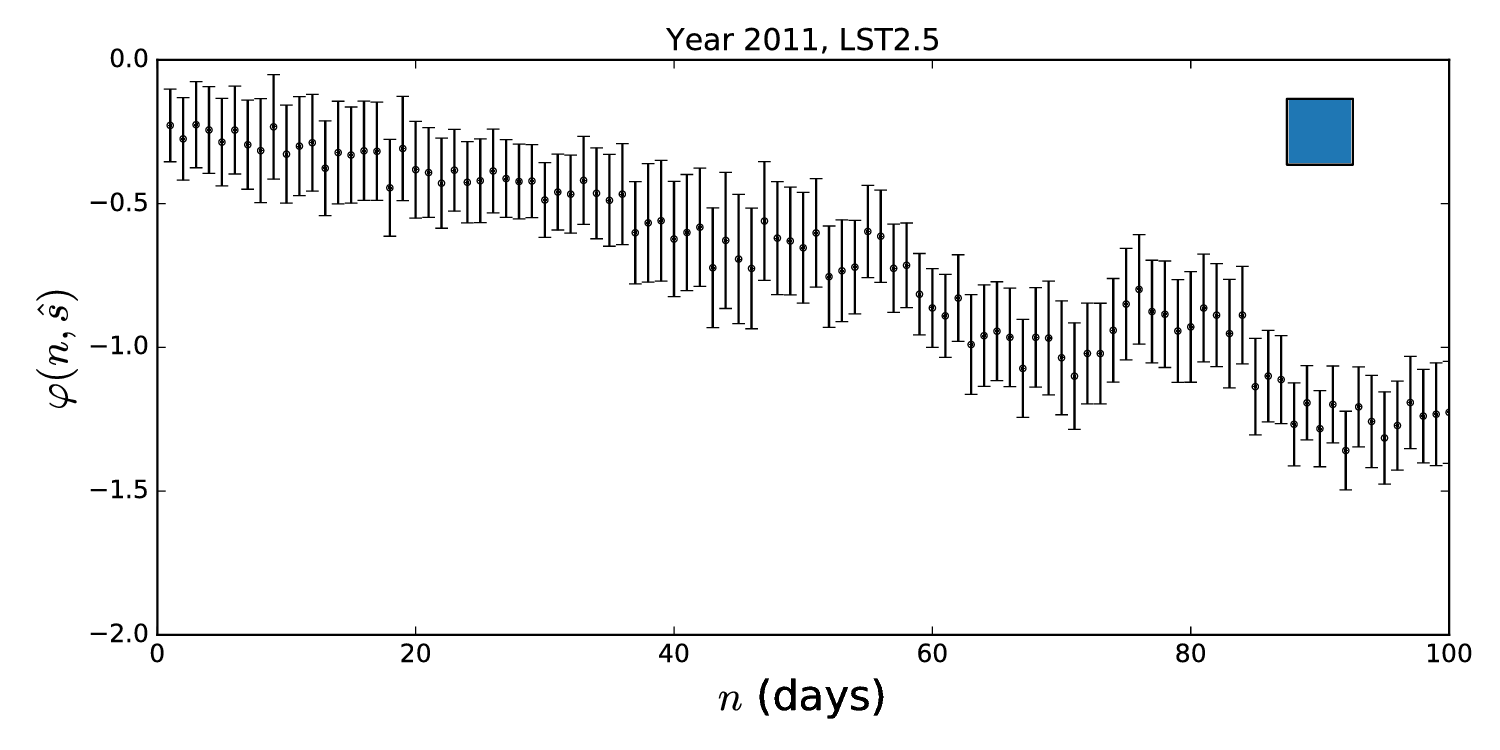}&
		\includegraphics[width=0.33\textwidth]{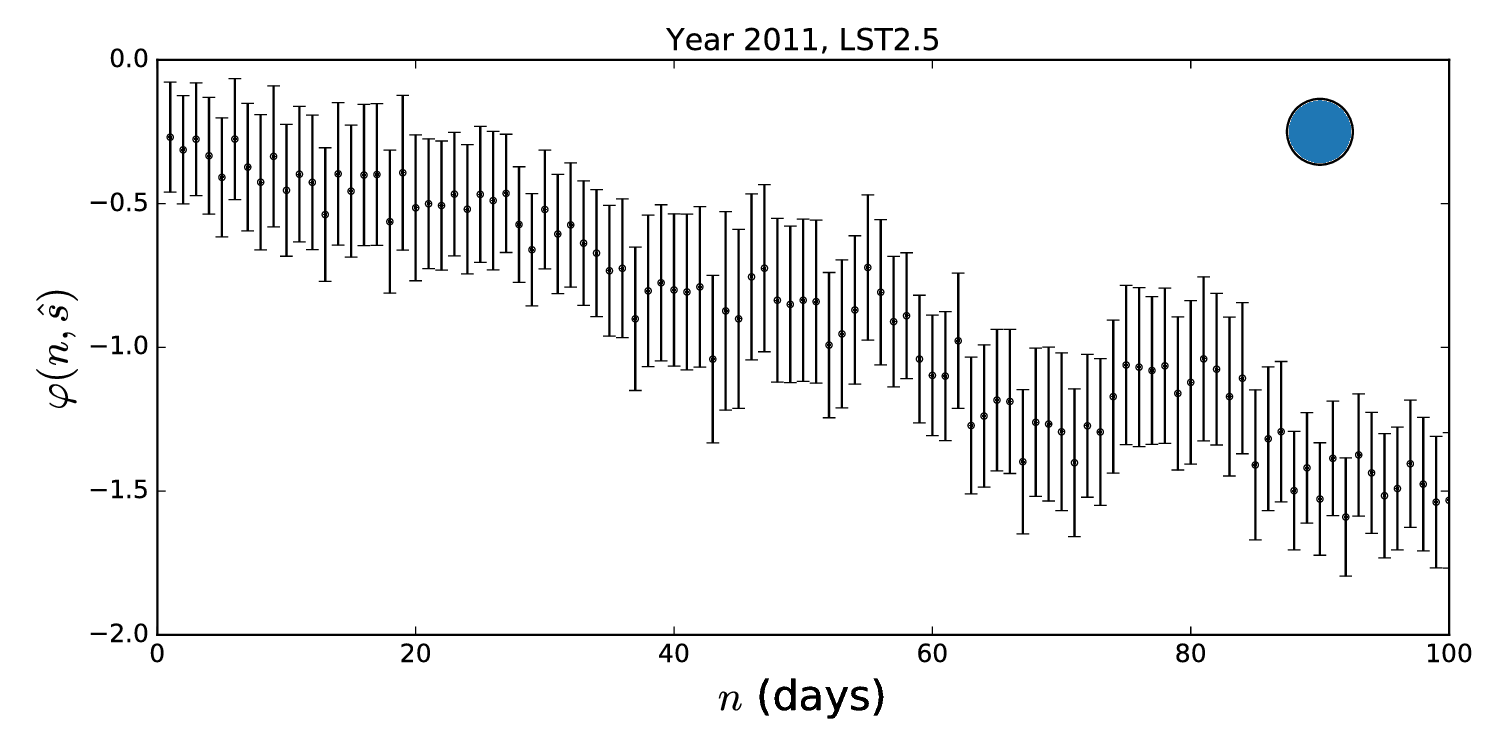}&
		\includegraphics[width=0.33\textwidth]{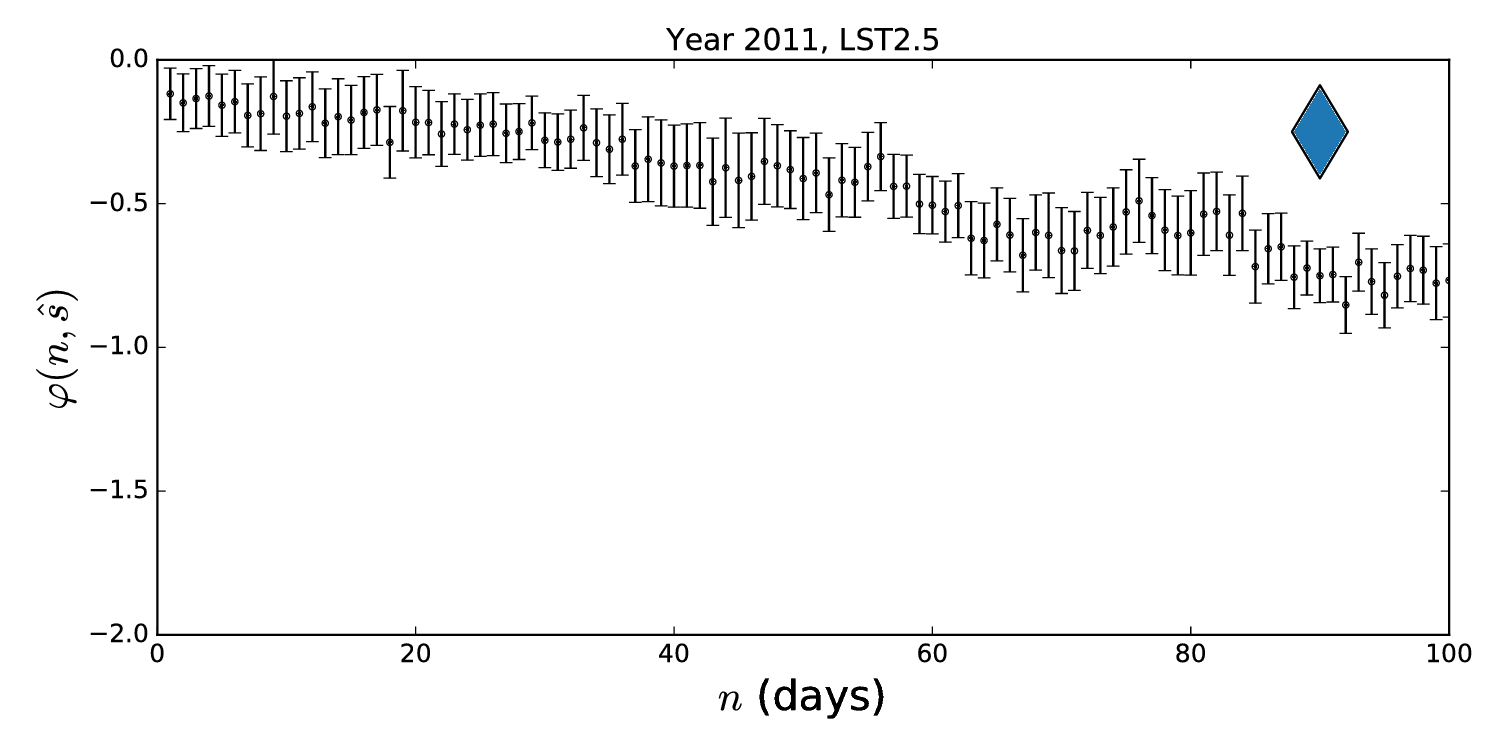}\\
		\includegraphics[width=0.33\textwidth]{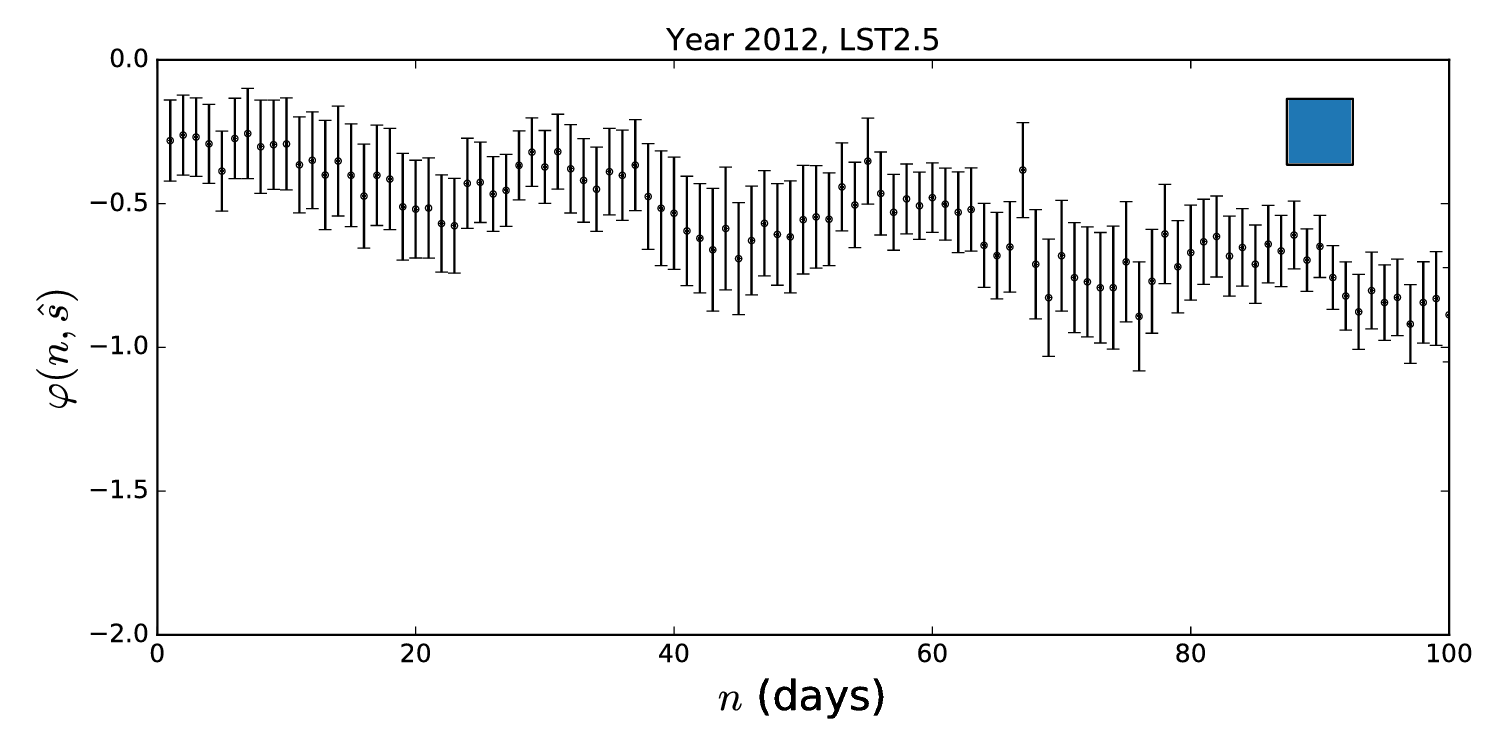}&
		\includegraphics[width=0.33\textwidth]{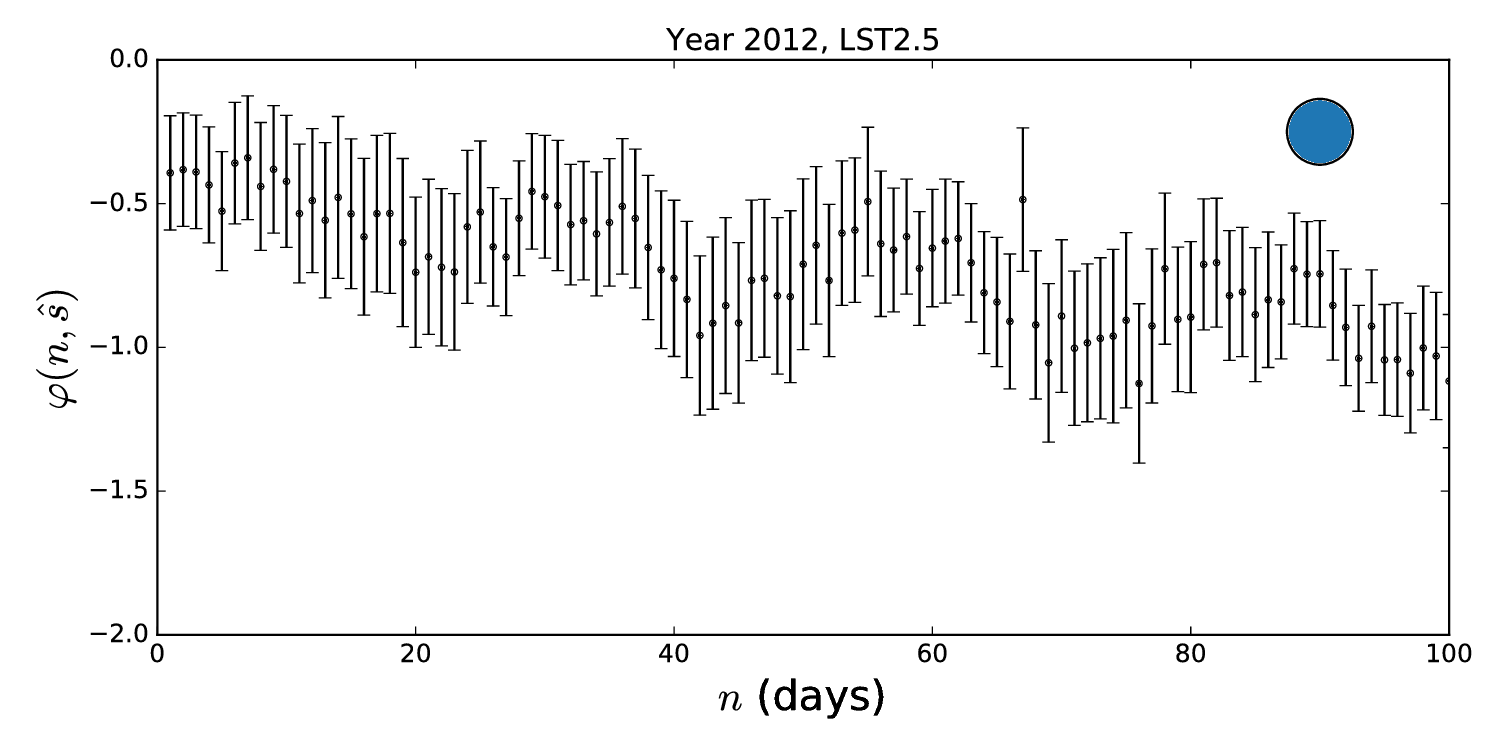}&
		\includegraphics[width=0.33\textwidth]{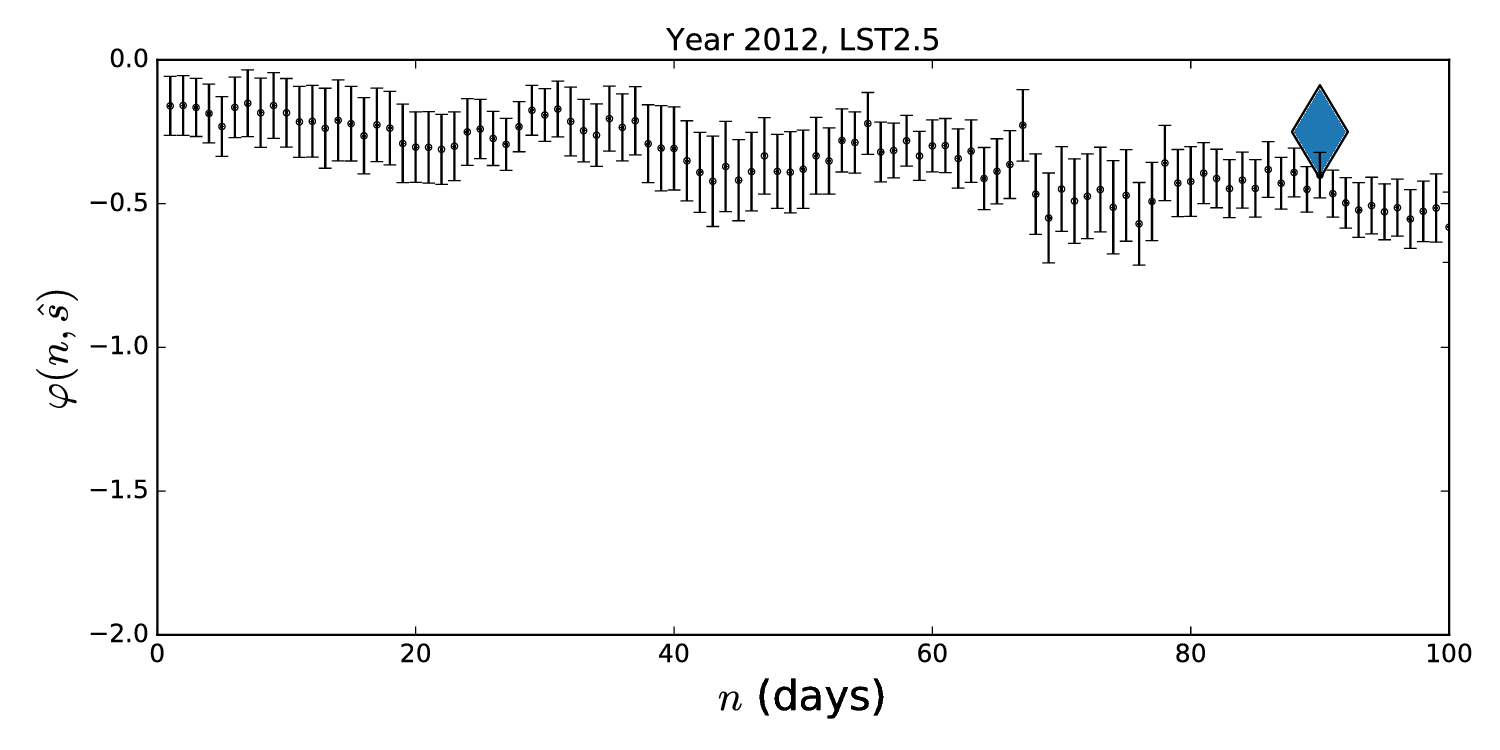}\\
		\includegraphics[width=0.33\textwidth]{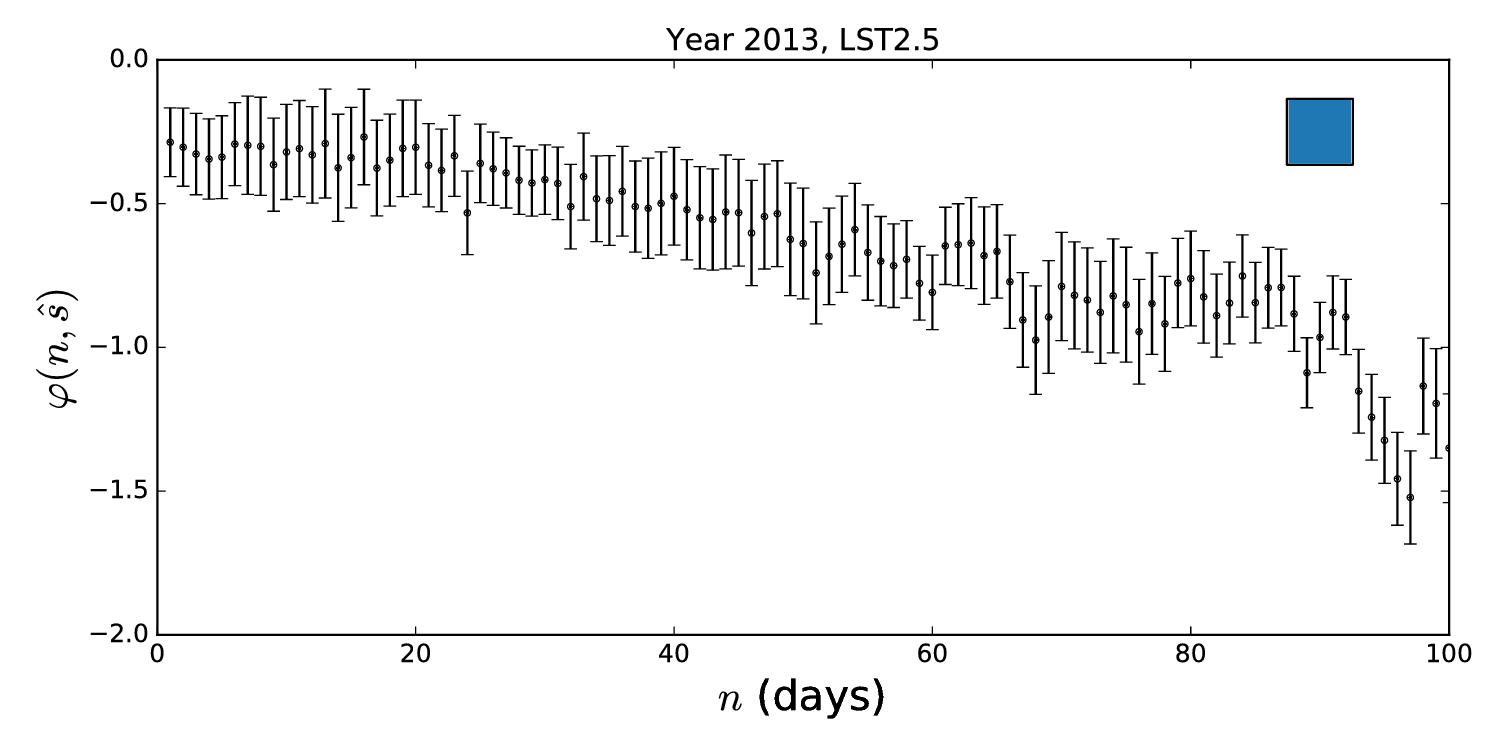}&
		\includegraphics[width=0.33\textwidth]{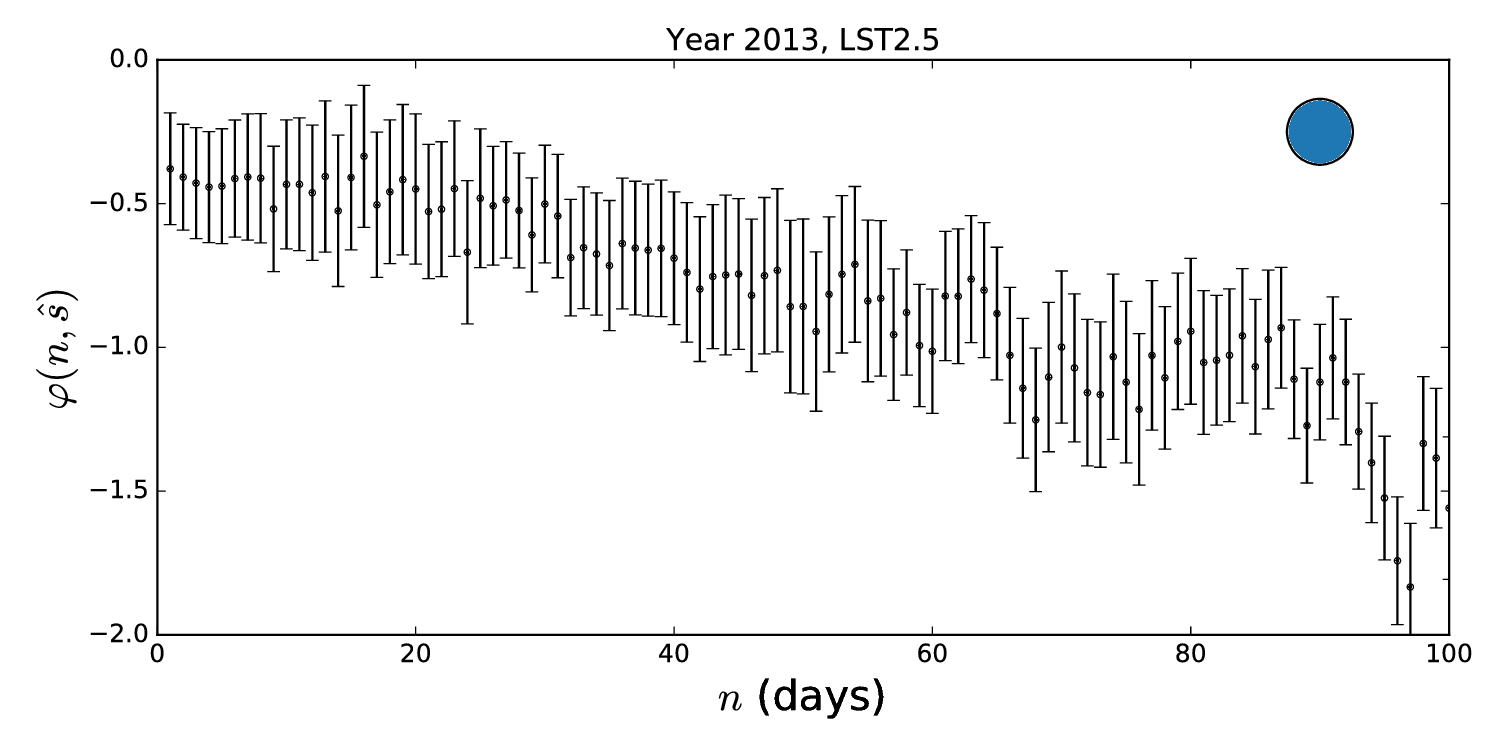}&
		\includegraphics[width=0.33\textwidth]{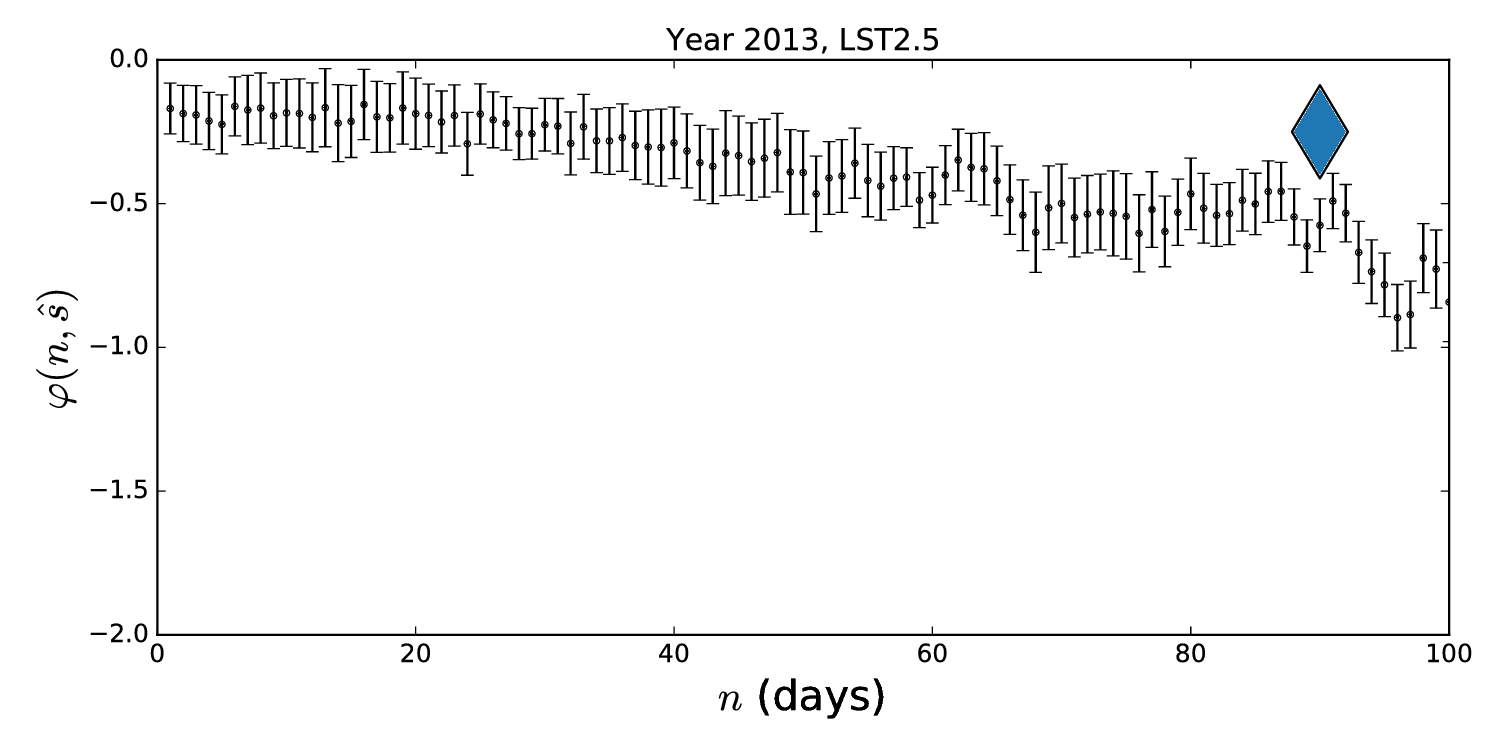}\\
		\includegraphics[width=0.33\textwidth]{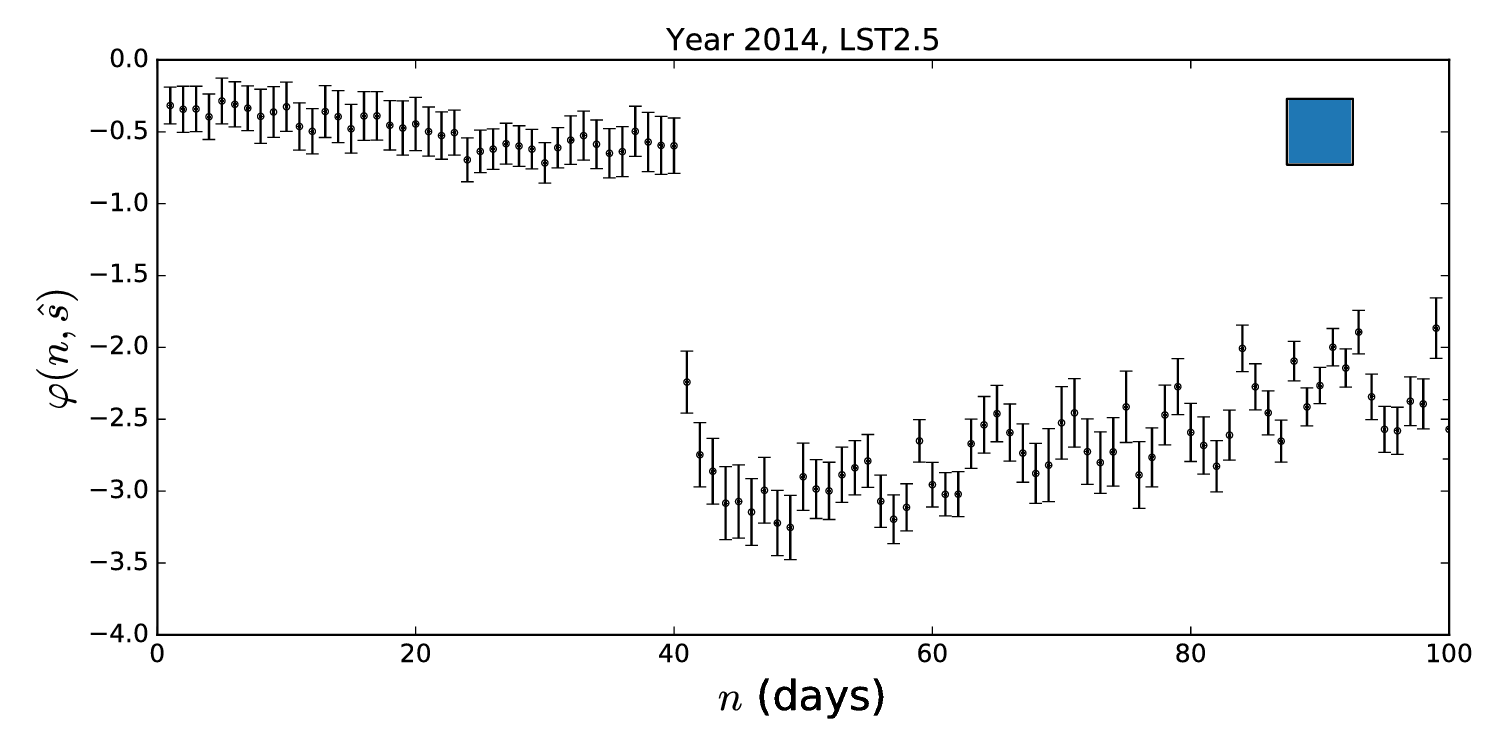}&
		\includegraphics[width=0.33\textwidth]{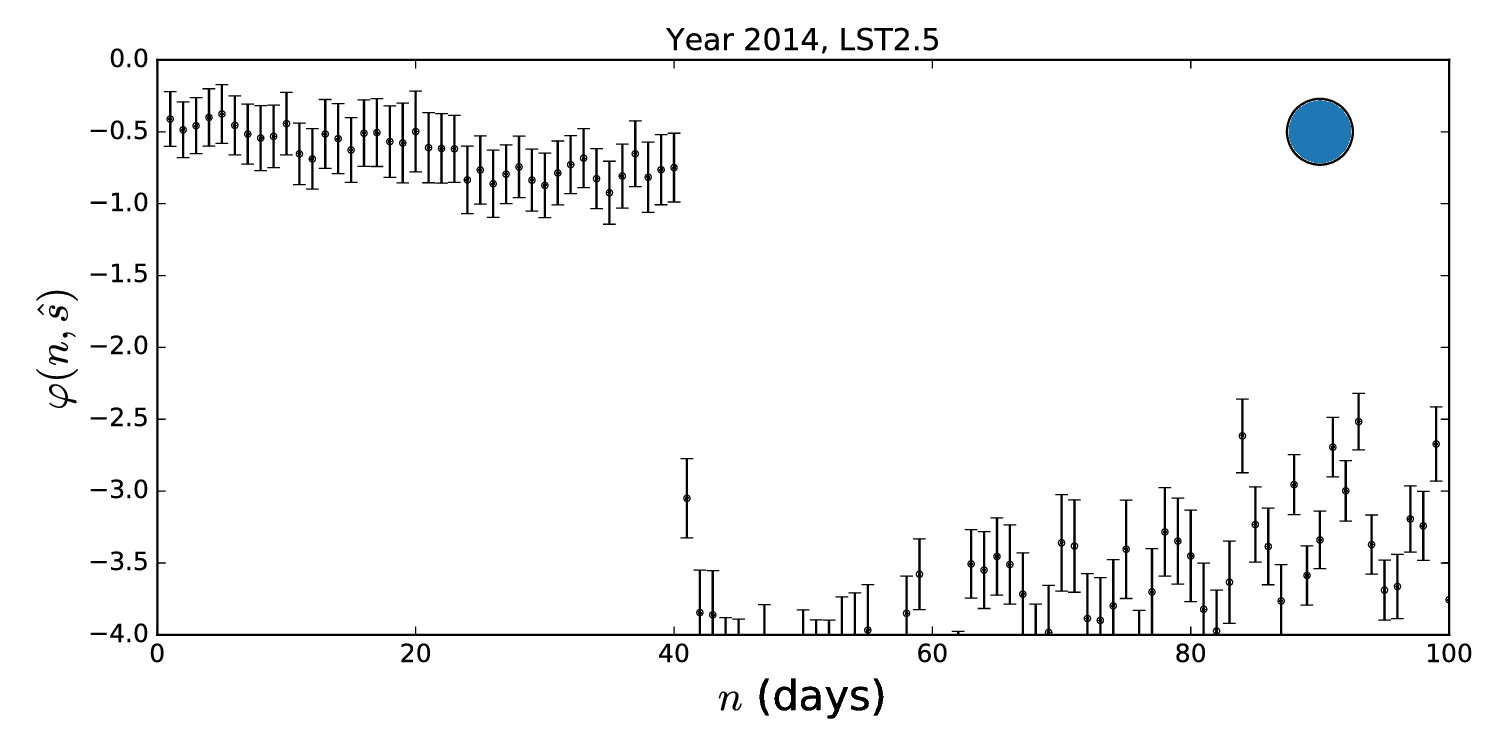}&
		\includegraphics[width=0.33\textwidth]{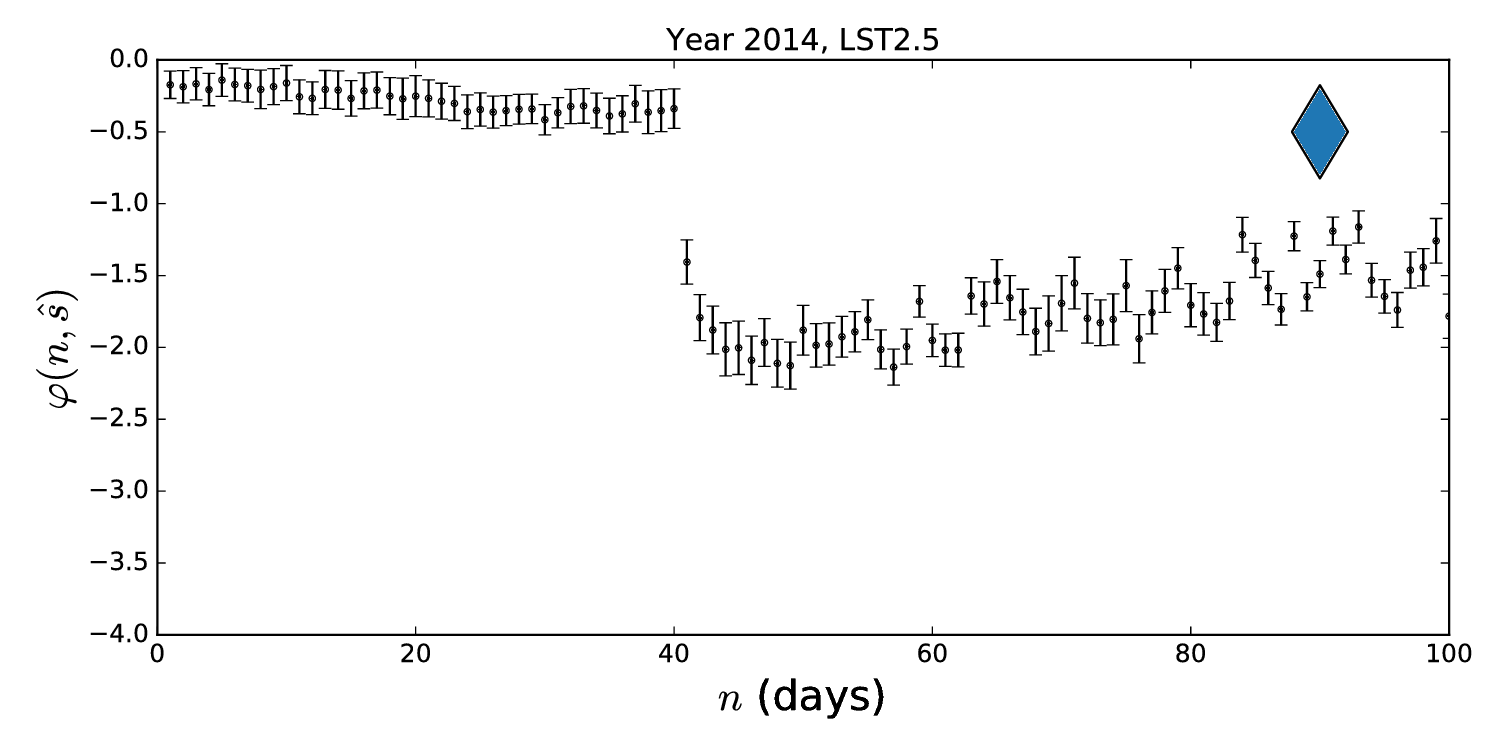}\\
	\end{tabular}
	
	\caption{Rotation measure sequences over 100 days at 3 different points on the sky at fixed LST for each year 2011-2014. The error bars are the 1-$\sigma$ error bars propagated from TEC ($\overline{\rho}_e$) uncertainties provided with the IONEX data. There is a clear trend along with the random variation. A significant solar event is observable as the large jump in 2014 (note the scale on the vertical axis of the bottom row of panels differs from the top three rows). This appears to correspond to a relatively large solar flare that was observed on Oct. 19, 2014 by NASA's Solar Dynamics Observatory which was followed by several weeks of abnormally high solar activity.}
	\label{fig:RMskewers}
\end{figure*}

Figure \ref{fig:RMskewers} shows a selection of RM sequences over 100 sidereal days at a fixed LST-hour of $2.5$ for several different years. The rotation measure due to the ionosphere is generally a random function over time with the underlying random variable being the TEC whose variation is driven by solar radiation. The RM  varies randomly from day to day, but follows a clear trend over the course of 100 days.

The cause of this trend can be understood broadly by noting that the magnitude of the RM goes inversely as the time since the sun went down. As the season progresses a given LST transit occurs progressively closer to the previous sunset. Since the sun is the driver of ionization in the atmosphere, as this proximity increases, the ionosphere has had less time for recombination to occur since it was last heated resulting in a higher free-electron density, and thus a higher magnitude of RM.

\begin{figure*}[h]
	\centering
	\begin{tabular}{cc}
		\includegraphics[width=0.4\textwidth]{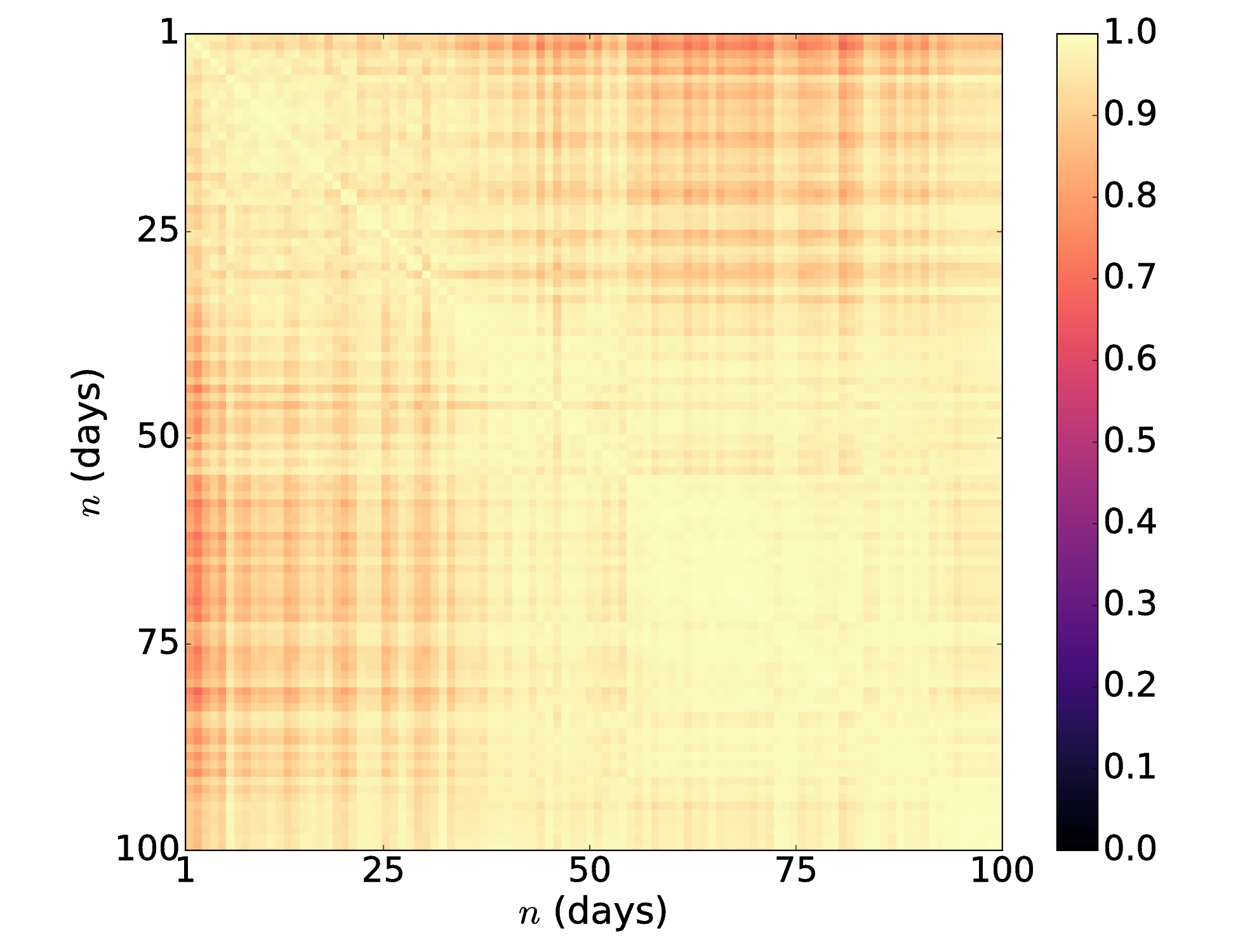} &
		\includegraphics[width=0.4\textwidth]{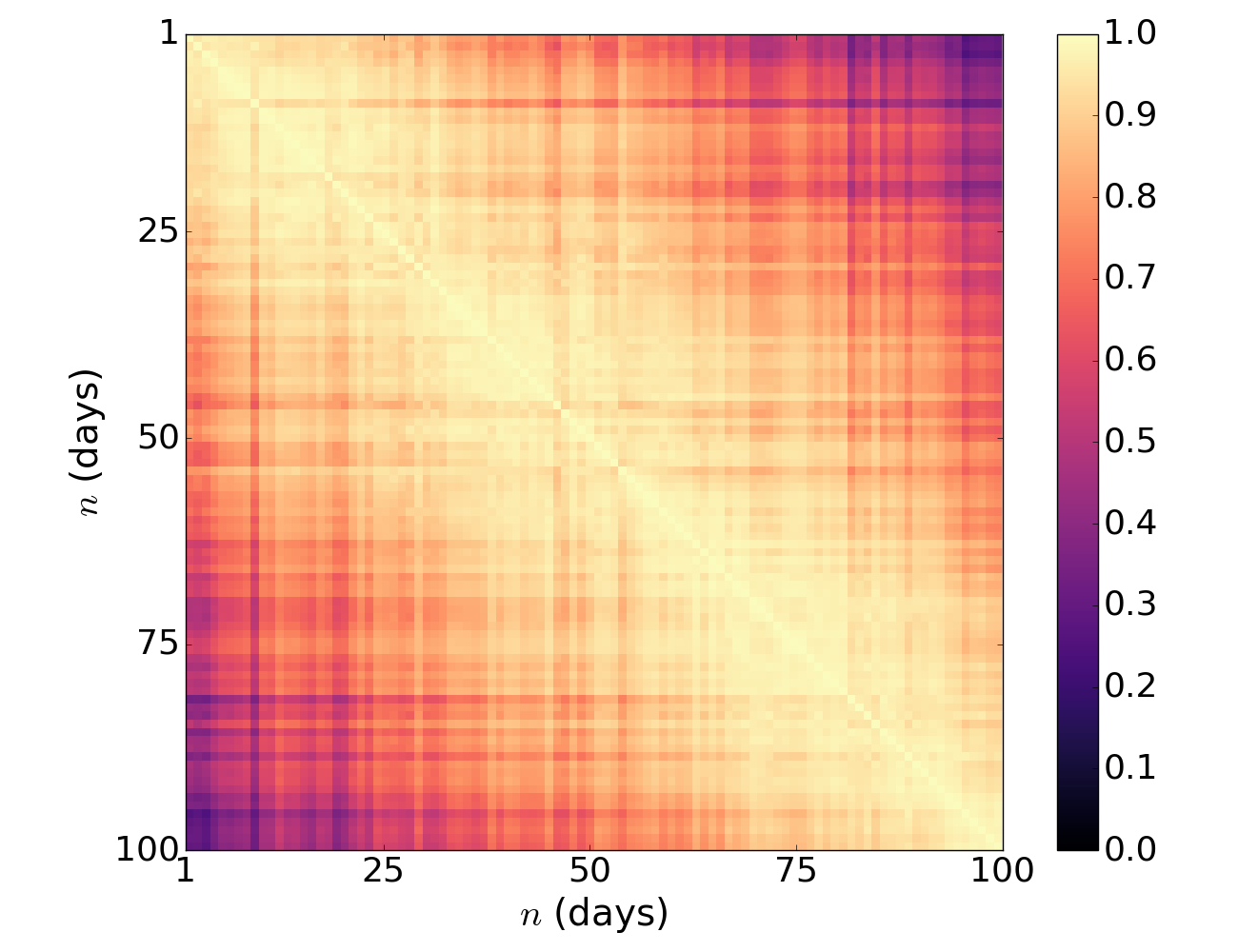} \\
	\end{tabular}
	\caption{Left: Spatial correlation matrix of the RM as defined in Equation \ref{eqn:CorrRM} at $t=$ LST 2.5 hours for 100 days starting on September 10, 2011. Right: The same spatial correlation function applied to the TEC column density $\overline{\rho}_e$.}
	\label{fig:CorrRM}
\end{figure*}

Careful inspection of Figure \ref{fig:RMskewers} would reveal that each of the three different points in the RM maps are not exactly rescalings of a common function of $n$ i.e. $\varphi$ is not a separable function of $n$ and $\vu{s}$. However, it is clear in Figure \ref{fig:RMmaps} that there is distinct average shape to the function $\varphi(\vu{s})$ which is largely due to the increasing path-length through the ionosphere with increasing zenith angle, and the projection of the geomagnetic field, along different lines of sight. These observations are quantified somewhat by considering the spatial correlation matrix 
\begin{equation}
c_{kl} = \frac{C_{kl}}{\sqrt{C_{kk}} \sqrt{C_{ll}}}, \label{eqn:CorrRM}
\end{equation}
where
\begin{align}
C_{kl}(t) & = \frac{1}{2 \pi} \int_{\mathbb{S}^2_+} \big(\abs{\varphi(k,t,\vu{s})} - \overline{\varphi}(k,t) \big) \big(\abs{ \varphi(l,t,\vu{s})} - \overline{\varphi}(l,t) \big), \\
\overline{\varphi}(n,t) & = \frac{1}{2 \pi} \int_{\mathbb{S}^2} \abs{\varphi(n,t,\vu{s})}.
\end{align}
The integral is taken over the observed hemisphere and the absolute value of $\varphi$ is taken because the sign does not vary between days. We can also compute this correlation for the TEC by replacing $\varphi \rightarrow \overline{\rho}_e$ in Equation \ref{eqn:CorrRM}. A representative example of the correlation matrices $c_{kl}$ for both the RM and TEC are shown in Figure \ref{fig:CorrRM} where we see that the shape of the RM field is not as variable between days as the underlying TEC field.

\subsection{Antenna Response Model}

The instrumental Jones matrix $\bm{J}(\nu, \vu{s})$ is derived from electromagnetic simulations of a HERA antenna element using the commercial software CST which solves for the far-field electric field radiated by an array antenna element when operated in transmission \citep{fagnoni2016}. By Lorentz reciprocity these electric field functions define the polarized response of the instrument to the incident plane waves produced by celestial sources \citep{born1999principles, deHoop1968, Potton2004, balanis2005}, and thus define the instrumental Jones matrix as described in Appendix \ref{sec:InstrumentAppendix}. 

We take the data for the fields output by CST and interpolate to a {\sc HEALPix} map. Only the simulation for a single feed is available, so we use the assumption that the antenna structure is symmetric under 90 degree rotations about the antenna bore-sight to derive the response of the second feed. Since the simulation was computed at $1$ MHz resolution in frequency (which was deemed sufficient to capture the frequency structure), an interpolation in frequency is performed by cubic spline fit to the components of the spherical harmonic transforms of the electric field components, and then synthesizing the fields at the desired frequency and spatial resolution. The CST simulation is done in free-space so the fields are defined over the full sphere. We apply a hard cut to the fields at the local horizon which is defined as the zenith angle of $\pi/2$ and set the instrumental response $\bm{J}$ to zero below this horizon.

Additionally, for comparison with a simplified model we use a Hertzian dipole with a Airy disk directivity taper. In terms of a set of Cartesian coordinates $(x,y,z)$ such that $\vu{e}_z$ is along the antenna's bore-sight and $\theta(\vu{s}) = \cos[-1](\vu{e}_z \vdot \vu{s})$ this Jones matrix is
\begin{align}
\bm{J}(\nu, \vu{s}) & = B(\nu, \vu{s}) \mqty*(\vu{e}_x \vdot \vu{e}_\delta & \vu{e}_x \vdot \vu{e}_\alpha \\
\vu{e}_y \vdot \vu{e}_\delta & \vu{e}_y \vdot \vu{e}_\alpha), \label{eqn:AiryDipoleJones} \\
B(\nu, \vu{s}) & = \frac{2 J_1\qty( \frac{2 \pi a \nu }{c} \sin(\theta) )}{\frac{2 \pi a \nu }{c} \sin(\theta) }
\end{align}

%
where $J_1(x)$ is the Bessel function of the first kind of order one and $2a = 14.6$m is the diameter of a HERA dish. The purpose of this simpler and less realistic model is to illustrate how the details of the instrumental response affect the simulations. 
The HERA antenna simulation used here is that of an early model which is still in the process of development. We expect that the final model of the antenna far-field response will differ slightly from the one available to us now, but not significantly so. The difference will be much smaller than the difference that can be seen in Figure \ref{fig:fullHERAmueller} between the current HERA model and this simple Airy dipole construction. Further, although the polarization properties of the antenna beam have not been measured, other more accessible properties have been measured and their agreement with the CST simulated model suggests it is quite realistic. Therefore, despite the similarity with the HERA model, the Airy dipole should probably be thought of as a large change to the instrument model, rather than a small one.

\subsection{Sky Model}

\begin{figure*}[th]
\centering
\includegraphics[scale=1.0]{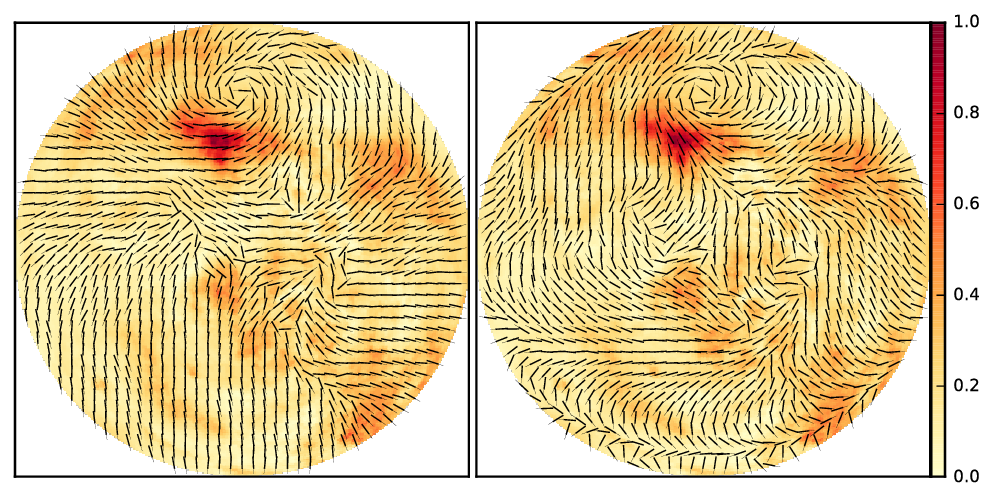}
\caption{An image of the polarized power $L = \sqrt{Q^2 + U^2}$ in the sky model over half of the sky as would be observed by an antenna instantaneously, meaning the edge of the image is the local horizon 90 degrees from zenith. The linear color scale is normalized to the peak of the image and both panels show the same map. Overlaid is a unit tensor field that shows the orientation of the linear polarization state $(Q,U)$.The tensor field in the left panel shows the initial polarization orientation field while the right panel shows the polarization orientation after ionospheric Faraday rotation at $150$MHz (i.e the polarization orientation field of the polarization state $(Q_n, U_n)$ in Equation \ref{eqn:MuellerIntegral}). Since the RM field $\varphi(\vu{s})$ is spatially smooth, the orientation field after Faraday rotation maintains it's initial spatial correlation.}
\label{fig:CoraExample}
\end{figure*}

While a good model of the actual diffuse Stokes I emission is available in the form the Global Sky Model, there is currently no equivalent full sky model of the polarization state of the diffuse galactic synchrotron emission. Therefore the best method available is to use a random realization generated from a statistical model with constrained parameters. We use the {\tt CORA} \footnote{\url{github.com/radiocosmology/cora}} software to generate a set of Stokes $I$,$Q$, and $U$ diffuse maps. The CORA package was developed for use in \citet{shaw15} and the details of its physical motivation and implementation are discussed there. Briefly, CORA generates random realizations of the polarization state of the diffuse emission by rotation measure synthesis \citep{brentjens2005, jelic2010}
\begin{equation}
Q(\nu, \vu{s}) + iU(\nu, \vu{s}) \propto \int_{-\infty}^{\infty}\dd{\phi} F(\phi, \vu{s}) e^{2i\phi \frac{c^2}{\nu^2}}
\end{equation}
where $F$ describes the distribution and polarization angle ($F$ is a complex-valued function) of polarized emission as a function of the Faraday depth $\phi$. The function $F$ is in turn decomposed into spherical harmonic components as
\begin{equation}
F(\phi, \vu{s}) = w(\phi, \vu{s}) \sum_{l,m} f_{lm}(\phi) Y_{lm}(\vu{s}).
\end{equation}
The components $f_{lm}(\phi)$ are Gaussian-random complex-valued functions, while $w(\phi, \vu{s})$ is a fixed function of $\phi$ and $\vu{s}$. Realizations of diffuse linear polarization components $Q,U$ are thus generated by drawing realizations of the components $f_{lm}(\phi)$. Figure \ref{fig:CoraExample} shows an example of the diffuse polarized power $L(\vu{s}) = \sqrt{Q^2(\vu{s}) + U^2(\vu{s})}$ and polarization orientation tensor field generated by this model.

Stokes-I is generated by \texttt{CORA} as an extrapolation of the Haslam map at 408 MHz, but we subtract the Stokes-I term from the Vokes parameters when analyzing the simulated visibilities in Section \ref{sec:results}. The one exception is in Figure \ref{fig:Waterfalls} where, for context, we show the simulated $\mathcal{V}_I(t,\nu)$ function including the Stokes-I term.

While the constraints on \texttt{CORA}'s model parameters are, in the author's words, "crude", the model is sufficiently realistic to capture the important characteristic features of diffuse polarized emission, namely the unsmooth frequency structure and spatial correlation. In particular our results are by construction independent of the absolute level of polarized power present in the sky model and somewhat insensitive to the particular frequency spectrum of the polarization. We purposefully avoid speculation about the absolute level of polarization leakage that may be observed with HERA.

\subsection{Simulation Parameters}

\begin{itemize}
	\item We compute the visibility matrix for a single 30 meter East-West
	
	\item The two dipole-feed orientations $a$ and $b$ are East-West and North-South as is the case for the HERA antenna elements.
	
	\item The visibilities are computed for each of 201 equally spaced frequency points $\nu = \nu_j$ (i.e a $0.5$MHz channel width) in the band $100-200$MHz; the smallest frequency is $100$MHz, the largest is $200$MHz. From this band five $20$MHz sub-bands are used: \begin{multline}\mathcal{B} \in \{(100,120),(120,140),\\(140,160),(160,180),(180,200)\}.\end{multline}
	
	\item In the delay transform we use a Blackman-Harris window function.
	
	\item We compute visibilities using the historical ionospheric data for the 100-day sequence starting on September 10 in each year of interest; this is the sequence for which RM data is shown in Section \ref{sec:radionopy}. The LST-hour range 1-4 was chosen so that all local times in this range are between sunset and sunrise for each of the 100 days at the geographic location of the HERA array.

	\item \textit{Fiducial simulations}: We picked a single realization of the sky model to use for a set of fiducial simulations. In these simulations $N_t=96$ equally spaced time samples $t=t_l$ in the LST-hour range 1-4 were computed. Since each of the functions modeled in our simulation is, by construction, smooth on the scale of our sky pixelization, this is sufficient to completely sample the time dependence of the visibilities. Visibilities were computed using the historical ionosphere data from the years 2009,2011,2012, and 2014, and for both instrumental response models.
	
	\item \textit{Sky model variance simulations}: We also performed simulations using many realizations of the statistical sky model. In order to save computational time in these simulations 6 equally spaced time samples were computed in the same LST range. For each of 12 years from 2003 to 2014, visibilities for 100 different realizations of the sky model were computed i.e. the 100 realization are different for each year. The reduced cadence of the time sampling has an effect on the results but we found from resamplings of the fiducial simulations that it was negligible compared to the change due to the sky model.
\end{itemize}

\section{Results from Simulations}
\label{sec:results}

\subsection{Attenuation of Linear Polarization in a Vokes-parameter delay spectrum}

Power spectrum estimators based on the delay spectrum will generally average visibility measurements taken over multiple days at fixed $(t, \nu)$ in order to attenuate thermal noise. We follow this procedure by averaging the simulated visibilities over a set $S_k$ of $N$ sidereal days. The simulated visibilities $\bm{\mathcal{V}}(n,t,\nu)$ are computed on an LST grid for each of $N_d$ consecutive days index by $n \in S = \{1,2, \ldots, N_d\}$. We can then choose a subset $S_k \subset S$ and compute the average at fixed $t$ as
\begin{equation}
\bm{\mathcal{\overline{V}}}(S_k, t,\nu) = \frac{1}{N}\sum_{n \in S_k} \bm{\mathcal{V}}(n, t,\nu)
\label{eqn:AvgVisMatrix}
\end{equation}
and then the corresponding averaged Vokes parameters are
\begin{equation}
\mathcal{\overline{V}}_\mathcal{S}(S_k, t, \nu) = \Tr(\bm{\sigma}_\mathcal{S} \bm{\mathcal{\overline{V}}}(S_k, t, \nu)).
\label{eqn:AvgVokes}
\end{equation}
Here we find the averaging process for visibilities that was alluded to in Section \ref{sec:theory}. Now, instead of the polarization state of a single source, the averaged quantity is $\overline{\mathcal{V}}_\mathcal{S}$ which can be considered a functional over the sequence of functions $\{\varphi(n)\}_{n \in S_k}$. For the simple model in Section \ref{sec:theory} of a single polarized point source the notion of attenuation of the polarized power was clear, but extending the idea to the delay spectrum of visibilities warrants some additional consideration. 

For each individual day $n$ the effect of the ionospheric Faraday rotation of the polarization state on the sky is a small change to the frequency spectrum $\bm{\mathcal{V}}(\nu)$. We can make the consequences more apparent by considering Equations \ref{eqn:AvgVisMatrix} and \ref{eqn:AvgVokes} in greater detail. The instrumental response $\bm{J}$ is independent of $n$, so we take the sum in Equation \ref{eqn:AvgVisMatrix} inside the integral defining $\bm{\mathcal{V}}(\nu)$ in Equation \ref{eqn:VisMat}:
\begin{align}
\bm{\overline{\mathcal{V}}}(S_k,t,\nu) & = \int_{\mathbb{S}^2} \bm{J} \qty( \frac{1}{N} \sum_{n \in S_k} \bm{R}_n \bm{\mathcal{C}} \bm{R}_n^\dagger ) \bm{J}^\dagger e^{-2 \pi i \frac{\nu}{c} \va{b} \vdot \vu{s}}
\end{align}
The sum in parenthesis may always be expressed as
\begin{align}
\frac{1}{N} \sum_{n \in S_k} \bm{R}_n \bm{\mathcal{C}} \bm{R}_n^\dagger &  = I \bm{\sigma}_I + \bm{\mathcal{T}} \qty\big(Q \bm{\sigma}_Q + U \bm{\sigma}_U) \bm{\mathcal{T}}^\dagger
\end{align}
since the cumulative effect of summing the $N$ different rotations may be described by a single rotation by an angle $2\mu$ where
\begin{equation}
\mu(S_k, t, \nu, \vu{s}) = \frac{1}{2}\Arg\bigg(\sum_{n \in S_k} e^{-2 i \varphi(n,t,\vu{s}) \frac{c^2}{\nu^2}} \bigg)
\end{equation}
and an amplitude factor 
\begin{equation}
A(S_k, t,\nu, \vu{s}) = \abs{\frac{1}{N} \sum_{n \in S_k} e^{-2 i \varphi(n,t,\vu{s}) \frac{c^2}{\nu^2}}} \label{eqn:AttenuationAmplitudeFactor}
\end{equation}
which define the resultant matrix
\begin{align}
\bm{\mathcal{T}} & = \sqrt{A} \mqty*( \cos(\mu) & -\sin(\mu) \\ \sin(\mu) & \cos(\mu) ).
\end{align}
Note that $\bm{\mathcal{T}}^2$ is the matrix representation of the complex number $Z(N)/N$ (Equation \ref{eqn:ComplexWalk}). Then for each $\mathcal{S} \in \{I,Q,U,V\}$ the averaged Vokes-$\mathcal{S}$ is
\begin{widetext}
\begin{align}
\mathcal{\overline{V}}_\mathcal{S} & = \int_{\mathbb{S}^2} M_{\mathcal{S}I} I e^{-2 \pi i \frac{\nu}{c} \va{b} \vdot \vu{s}} + \mathcal{\overline{V}}_{\mathcal{S}L}\\
& = \int_{\mathbb{S}^2} M_{\mathcal{S}I} I e^{-2 \pi i \frac{\nu}{c} \va{b} \vdot \vu{s}} + \int_{\mathbb{S}^2} A \mkern2mu{\cdot} \qty\Big(\cos(2 \mu) \qty(M_{\mathcal{S}Q} Q + M_{\mathcal{S}U} U) +  \sin(2 \mu) \qty(M_{\mathcal{S}Q} U - M_{\mathcal{S}U} Q) )e^{-2 \pi i \frac{\nu}{c} \va{b} \vdot \vu{s}} \label{eqn:VokesIntegralExpand}
\end{align}
\end{widetext}
This form exposes the fact that the ionospheric Faraday rotation need not reduce the Vokes-I polarization leakage, in fact it can increase it. Suppose that $M_{IQ} Q + M_{IU} U = 0$ for some $\nu$ and some $\vu{s}$ so that the polarization leakage term is - by cosmic accident of alignment - intrinsically zero. Then any rotation by a small angle $2\mu \neq 0$ will make the polarization leakage term non-zero. On the other hand, the rotation by itself can reduce the polarization leakage. Suppose now that $M_{IQ} Q + M_{IU} U \neq 0$. Then there is always a choice of rotation angle $2\mu$ which will null the polarization leakage term given by 
\begin{align}
\tan{2\mu} = \frac{M_{IQ} Q + M_{IU} U}{M_{IU} Q - M_{IQ} U}
\end{align}
Of course, generally the change in the magnitude of the leakage terms at each point $\vu{s}$ will fall between these two extremes and the change in the visibility will be the result of integrating over all such changes. This may be visualized by comparing the polarization orientation of the model sky in Figure \ref{fig:CoraExample} and the instrumental response in Figure \ref{fig:HERAmueller}. The spatial coherence of the fields means that merely changing the polarization angle over the whole sky can have dramatic effects on the polarization leakage terms.

For small $N$ where the ionosphere does not change very much between different days, the amplitude factor $A$ is generally fairly close to unity, but the resultant effective rotation of the polarization state by the angle $2 \mu$ can produce a stronger (or weaker) instrumental coupling to the polarization state if the original state $(Q,U)$ was not maximally (or minimally) aligned with the instrument. While $A$ and $\cos(2 \mu),\sin(2 \mu)$ are fairly smooth functions of frequency, the change in the frequency spectrum due to the realignment term (the term proportional to $\sin(2 \mu)$) including non-smooth $Q$,$U$  will tend to make the power in any given $\tau$ mode of the leakage delay spectrum fluctuate slightly. As $N$ increases and the variation between successive ionospheric Faraday screens becomes significant, the amplitude factor $A$ decreases enough to attenuate the polarization leakage regardless of the relative orientation of the sky's polarization state to the instrumental response. The result is that the precise attenuation may be somewhat variable as a function of $N$ for different $\tau$ modes, but as $N$ increases should tend converge to an overall trend. For this reason taking the ratio of a mean over modes of the delay spectrum of the polarization leakage provides a good summary measure of the attenuation.

With these considerations in mind we define a metric to assess the overall level of attenuation of the polarization leakage. Since we are interested in how polarization will affect power spectrum measurements we define the attenuation in terms of the delay spectra that would be used in such measurements.

The attenuation of polarization leakage due to the ionosphere is defined as the ratio of the total power in the leakage function after averaging visibilities over different ionospheric Faraday rotations, to the total power in the intrinsic leakage that would occur if the observation was made in the absence of ionosphere rotation
\begin{equation}
\xi_I(S_k, \mathcal{B}) = \frac{ \sum\limits_{j=1}^{N_\tau} \mathcal{L}_I(S_k,\tau_j,  \mathcal{B})}{\sum\limits_{j=1}^{N_\tau} \mathcal{L}_{I,int}(\tau_j, \mathcal{B})}
\label{eqn:AttenuationFunction}
\end{equation}
where $\mathcal{B}$ is the band over which the delay transform is computed. Then the leakage delay-power spectrum $\mathcal{L}_I$ is computed as
\begin{align}
\mathcal{L}_I(S_k, \tau_j, \mathcal{B}) & = \frac{1}{N_t}\sum_{l=1}^{N_t} \abs{\widetilde{\overline{\mathcal{V}}}_{IL}(S_k,t_l, \tau_j, \mathcal{B})}^2 
\label{eqn:LeakageFunction}
\end{align}
and $\widetilde{\mathcal{V}}_{IL}(n,t,\tau,  \mathcal{B})$ is the delay transform (Equation \ref{eqn:DelayTransform}) of the leakage terms $\mathcal{V}_{IL}(n, t, \nu)$ in the Vokes-I visibility for the $n$-th day:
\begin{align}
\mathcal{V}_{IL} & = \int_{\mathbb{S}^2} \qty(M_{IQ} Q_n + M_{IU} U_n) e^{-2 \pi i \frac{\nu}{c} \va{b} \vdot \vu{s}}.
\end{align}
The intrinsic leakage function $\mathcal{L}_{I,int}$ is formally obtained from Equation \ref{eqn:LeakageFunction} by setting the ionospheric rotation measure as $\varphi = 0$, so
\begin{equation}
\mathcal{V}_{IL,int} = \int_{\mathbb{S}^2} \qty(M_{IQ} Q + M_{IQ} U) e^{-2 \pi i \frac{\nu}{c} \va{b} \vdot \vu{s}}.
\end{equation}
is the Vokes-I polarization leakage that would be observed without ionospheric interference and is constant as a function of $n$, which we think of as the intrinsic leakage. 

\begin{figure*}[h]
	\centering
	\includegraphics[width=1.0\textwidth]{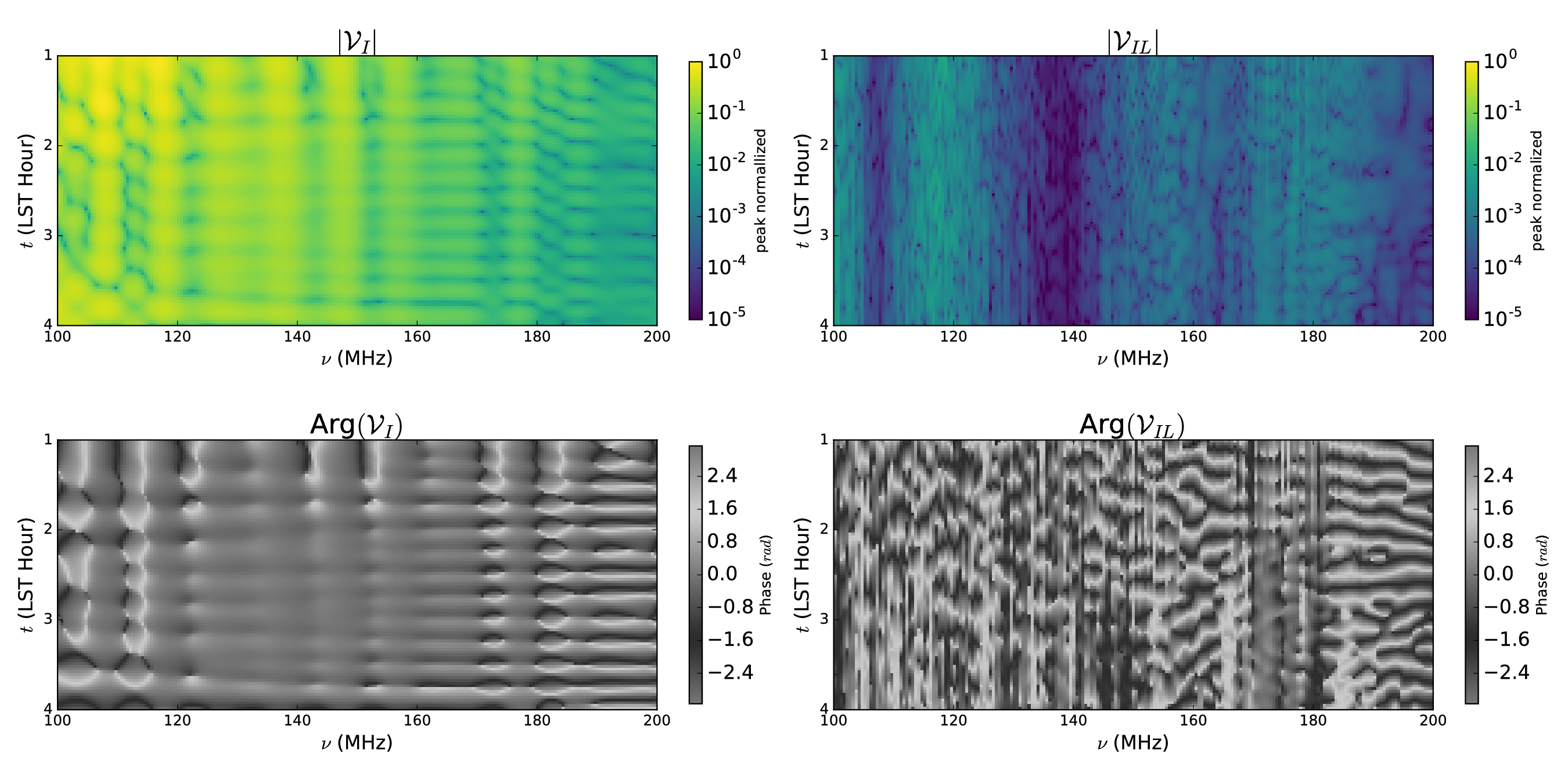}
	\caption{Amplitude and phase of $\mathcal{V}_I(t,\nu)$ (\textit{left}) and $\mathcal{V}_{IL}(t, \nu)$ (\textit{right}) for the intrinsic Vokes-I visibility ($\varphi = 0$) computed in our fiducial visibility simulation. The amplitude plotted in both figures is relative to the maximum of $\abs{\mathcal{V}_I(t,\nu)}$.}
	\label{fig:Waterfalls}
\end{figure*}

\begin{figure*}[h]
	\centering
	\begin{tabular}{cc}
	\includegraphics[width=0.5\textwidth]{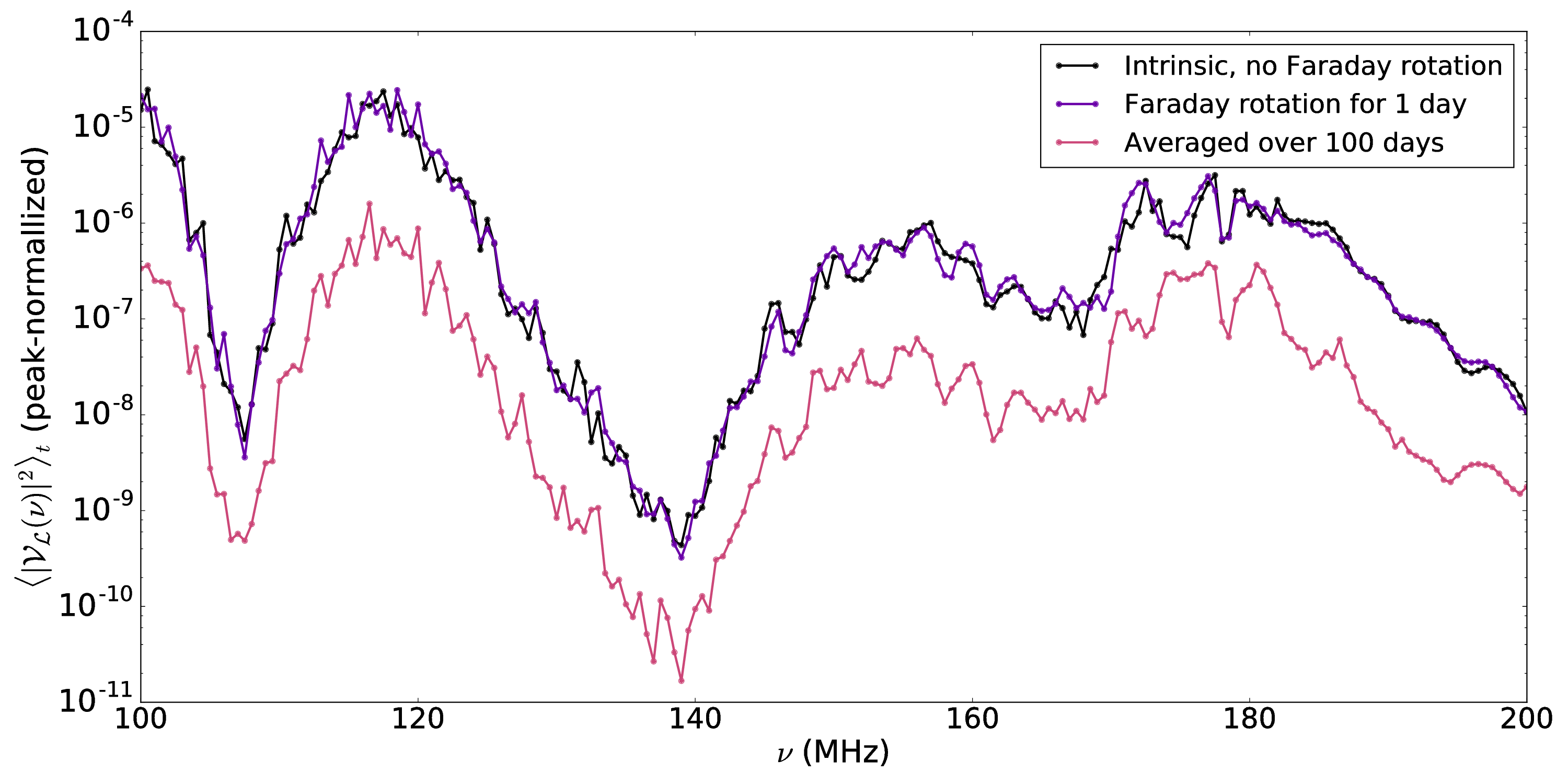} &
	\includegraphics[width=0.5\textwidth]{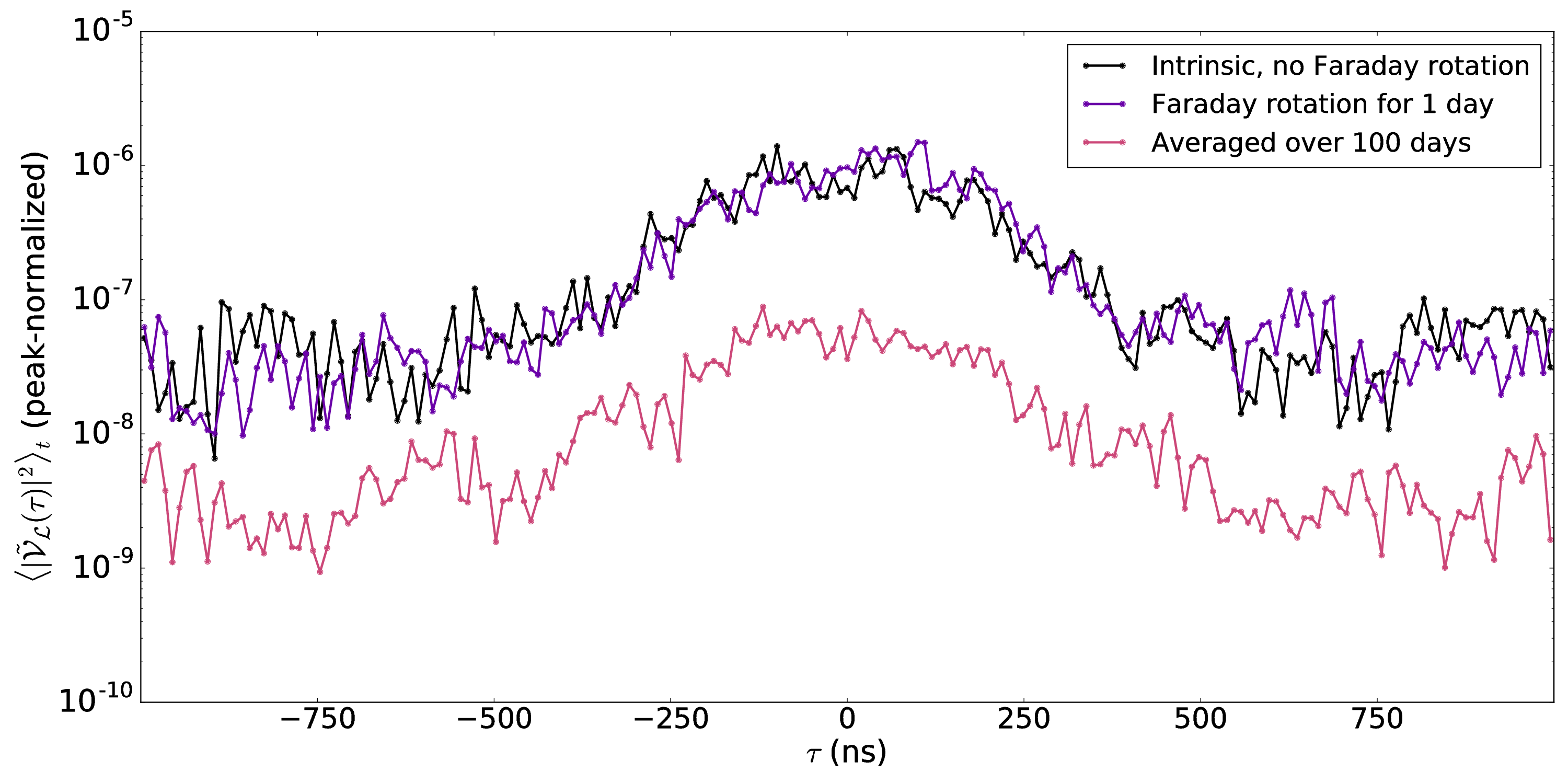}\\
	\end{tabular}
	\caption{\textit{Left:} The LST-averaged square-magnitude of the simulated Vokes-I polarization leakage as a function of frequency $\nu$, e.g. the result of averaging over $t$ in the upper-right panel of Figure \ref{fig:Waterfalls}. Summing the black "Intrinsic" and pink "Averaged" curves over any of the sub-bands $\mathcal{B}$ would produce the numerator and denominator for $S_k = S_{100}$ in Equation \ref{eqn:AttenuationFunction}.  \textit{Right:} The delay-power spectra over the full 100-200 MHz band of the intrinsic Vokes-I polarization leakage and a representative sample of the leakage when Faraday rotation for a single day is included. In both $\nu$ and $\tau$ representations there is a characteristic amplitude and shape, but the ionospheric Faraday rotation for a single day perturbs the spectra. The spectra resulting from averaging the visibilities over 100 days of different Faraday rotations produces an average attenuation, but additionally perturbs the spectra mode-by-mode due to the resultant effective rotation of the polarization state on the sky.}
	\label{fig:AveragedSpectraCuts}
\end{figure*}

Figures \ref{fig:Waterfalls} shows the simulated $\mathcal{V}_I(t,\nu)$ including the intrinsic leakage term, as well as $\mathcal{V}_{IL, int}(t,\nu)$ on it's own, and \ref{fig:AveragedSpectraCuts} shows some examples of the effect of the ionospheric Faraday rotation on the frequency and delay spectra of the visibilities.

It is worth noting explicitly that by Parseval's theorem the sums over the delay $\tau$ in Equation \ref{eqn:AttenuationFunction} are equivalent to simply summing the squared (and windowed) visibility amplitude over the frequency sub-band $\mathcal{B}$. We write the definition in terms of delay spectra in anticipation of modifications to the definition for use with real data or more realistic simulations in which we have more confidence in the detailed frequency-frequency covariance. In particular, with real data we cannot easily subtract the Stokes-I term from our Vokes-I data, but we could instead apply a high-pass delay filter by restricting the sum over $\tau$-modes in Equation \ref{eqn:AttenuationFunction} to $\abs{\tau} > \abs{\tau_{filter}}$ since $\widetilde{I}(\tau)$ is compact in delay compared to $\widetilde{Q}(\tau)$ and $\widetilde{U}(\tau)$. More generally, we reiterate that the notion of "attenuation" is not uniquely defined and we have made a particular choice here, though a well-motivated one, that is only applicable to simulated data.


\begin{figure*}[p]
	\centering
	\begin{tabular}{cc}
		\includegraphics[width=0.45\textwidth]{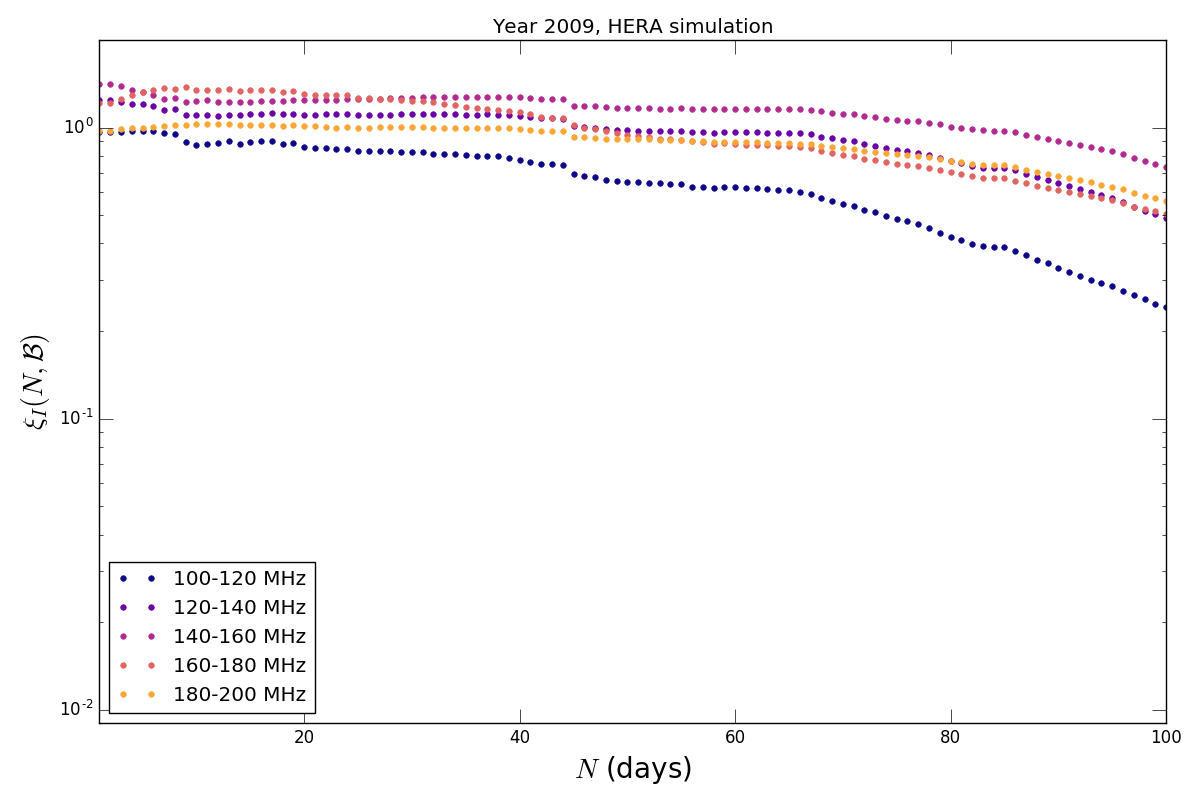}&
		\includegraphics[width=0.45\textwidth]{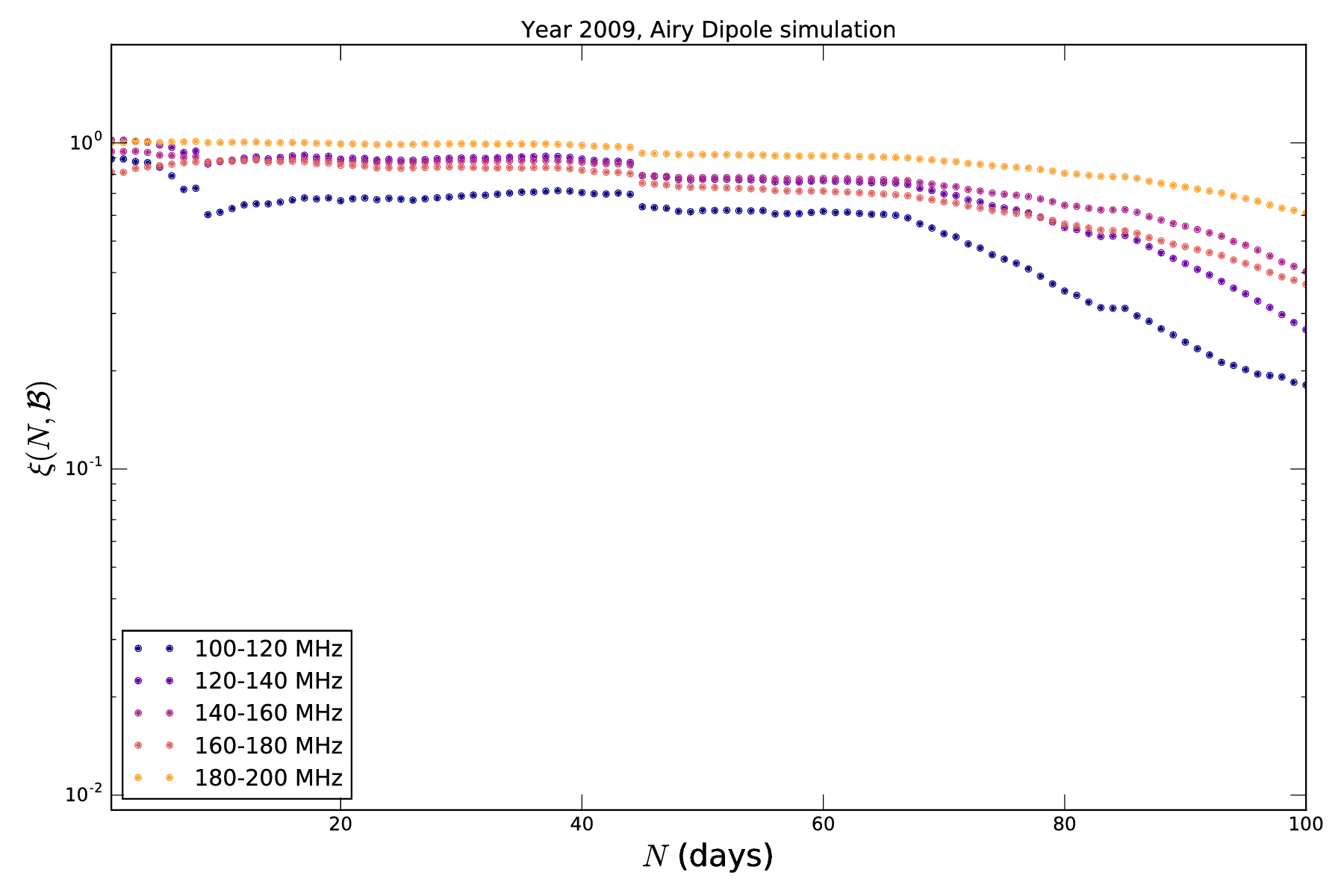}\\
		\includegraphics[width=0.45\textwidth]{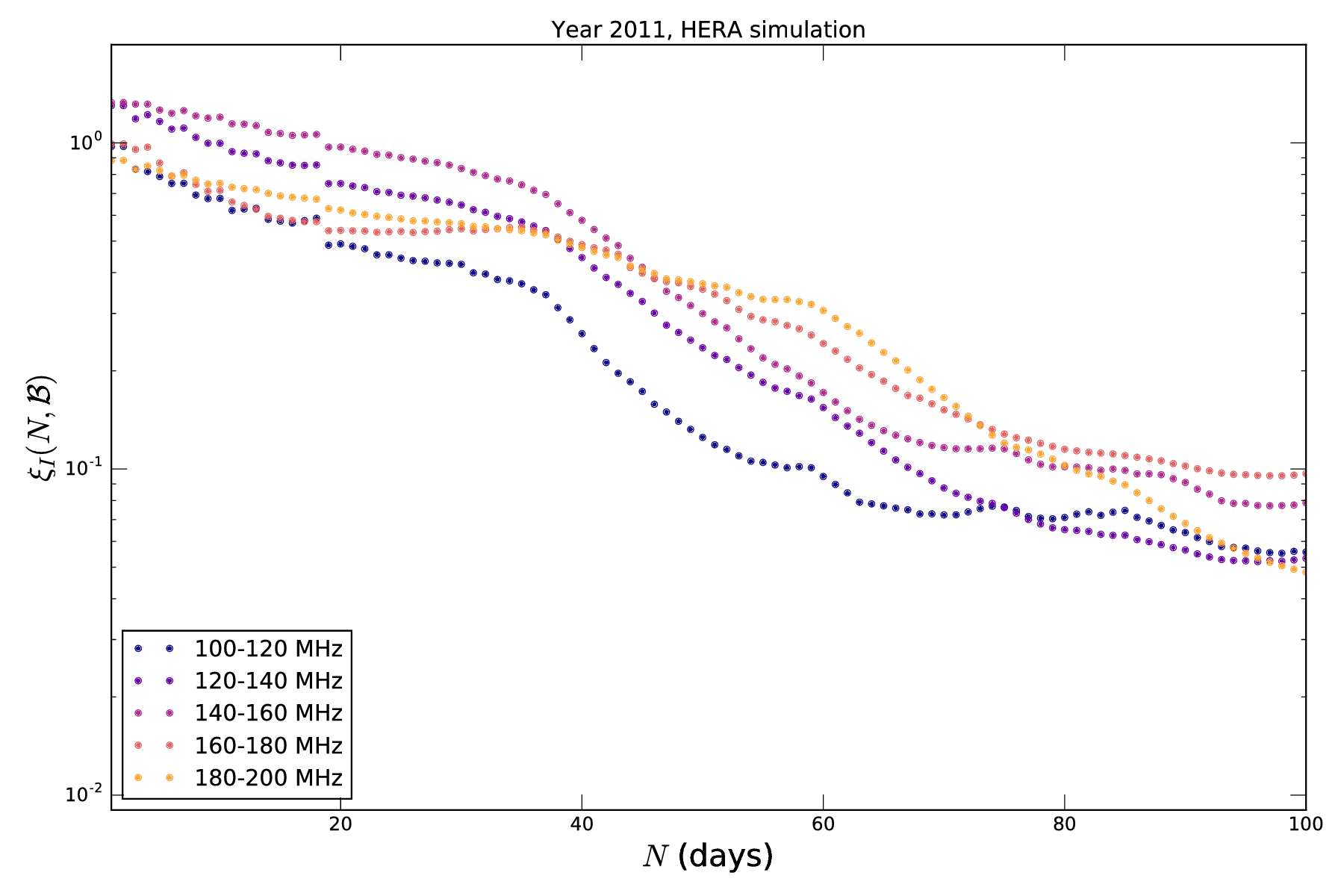}&
		\includegraphics[width=0.45\textwidth]{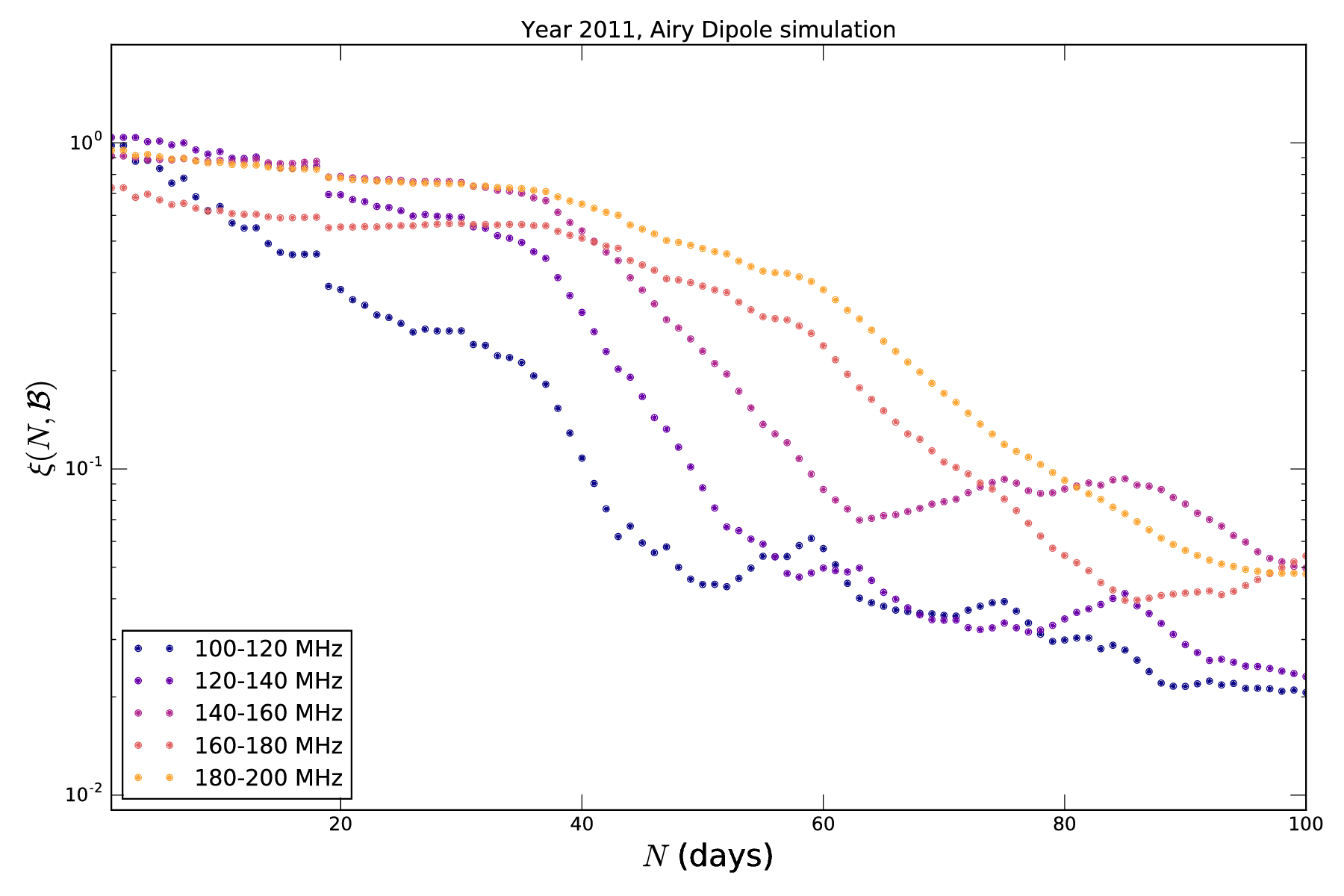}\\
		\includegraphics[width=0.45\textwidth]{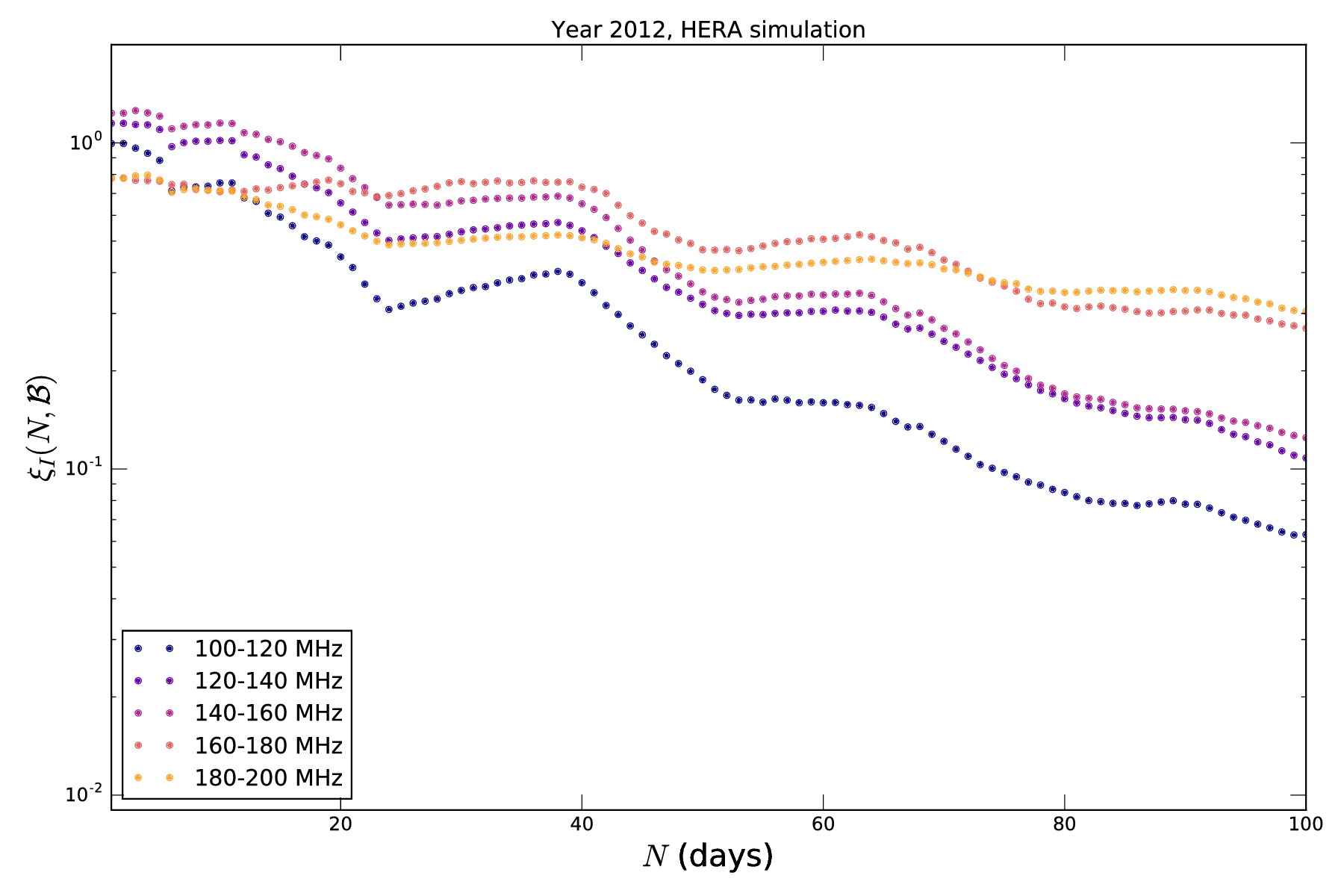}&
		\includegraphics[width=0.45\textwidth]{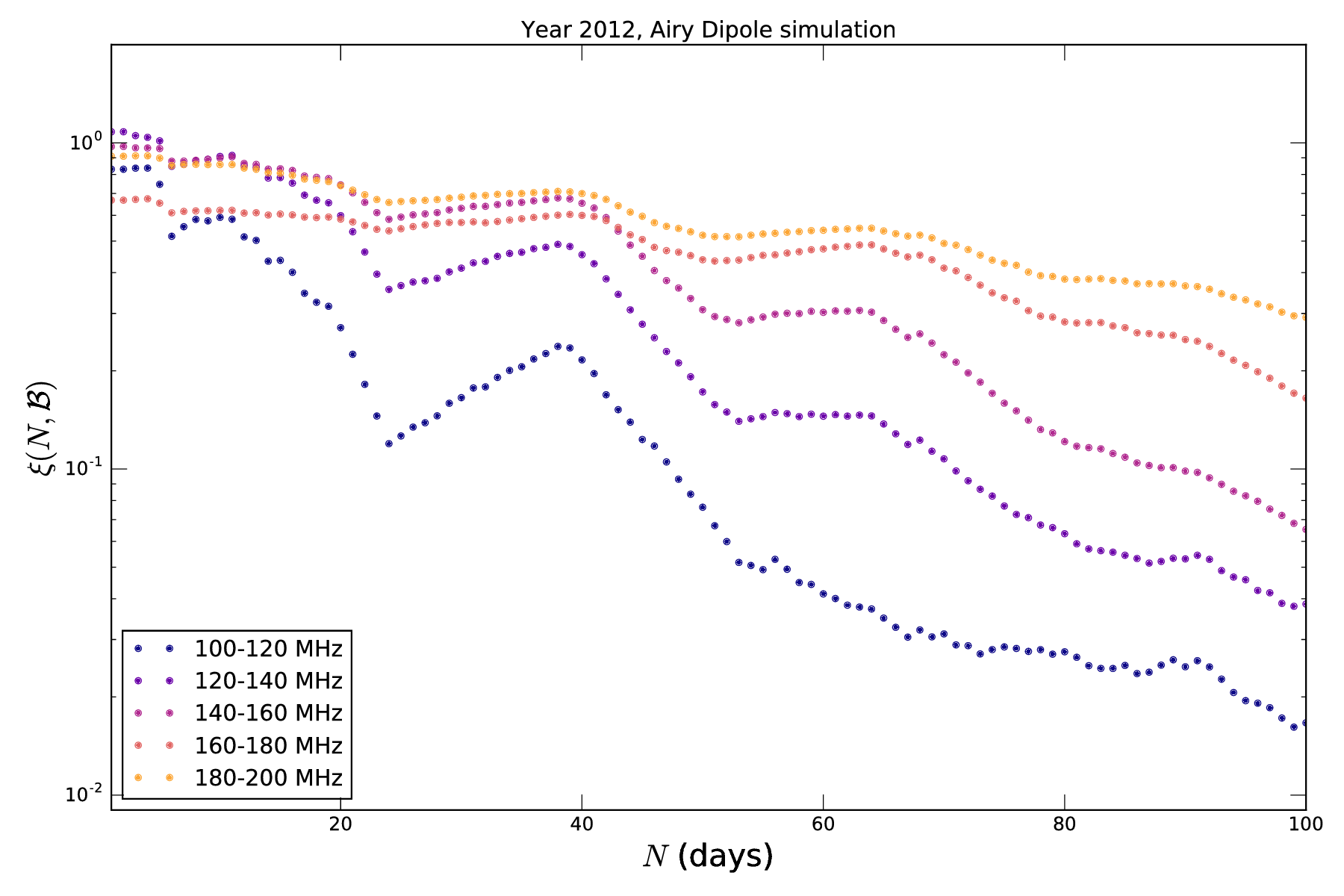}\\ \includegraphics[width=0.45\textwidth]{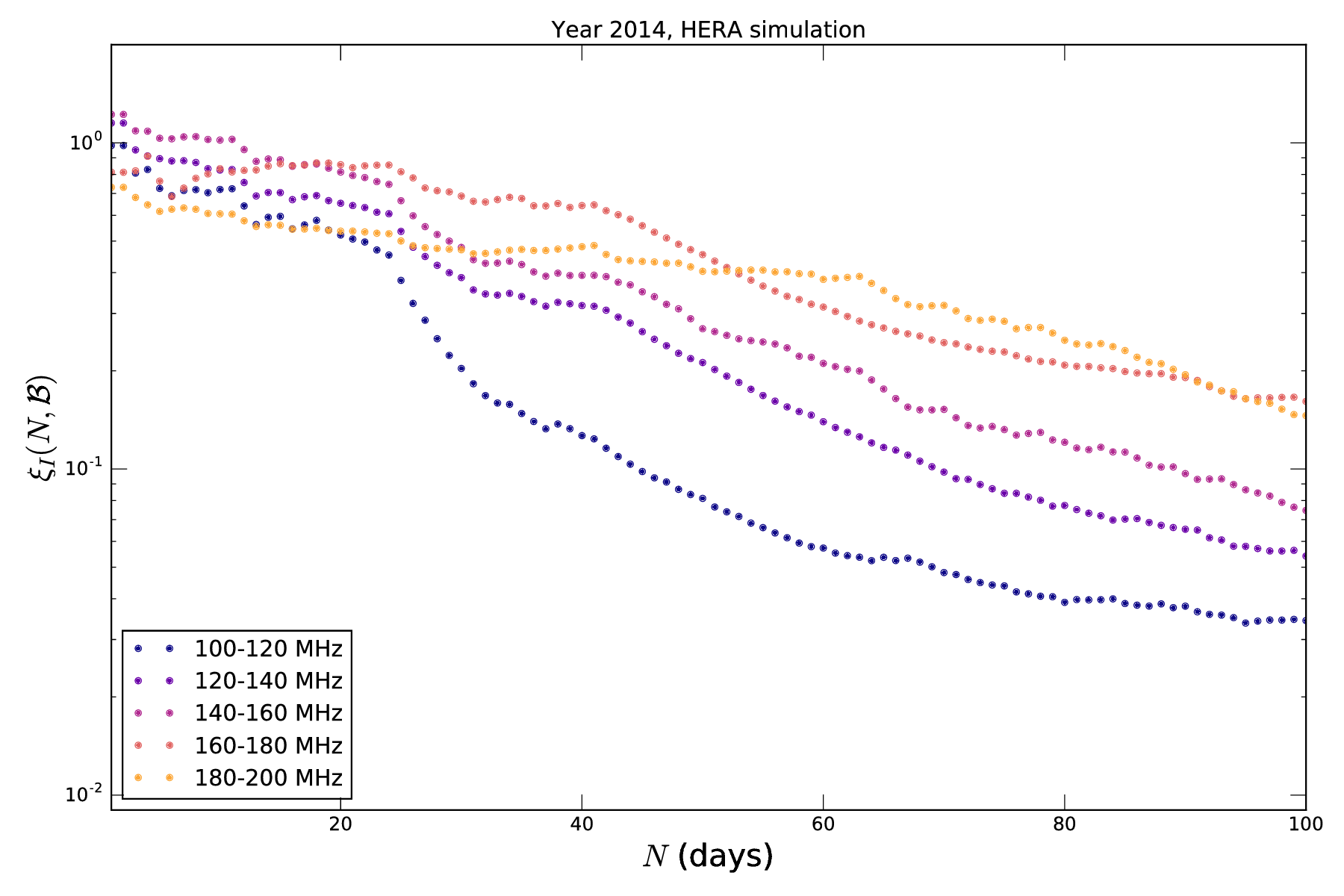}&
		\includegraphics[width=0.45\textwidth]{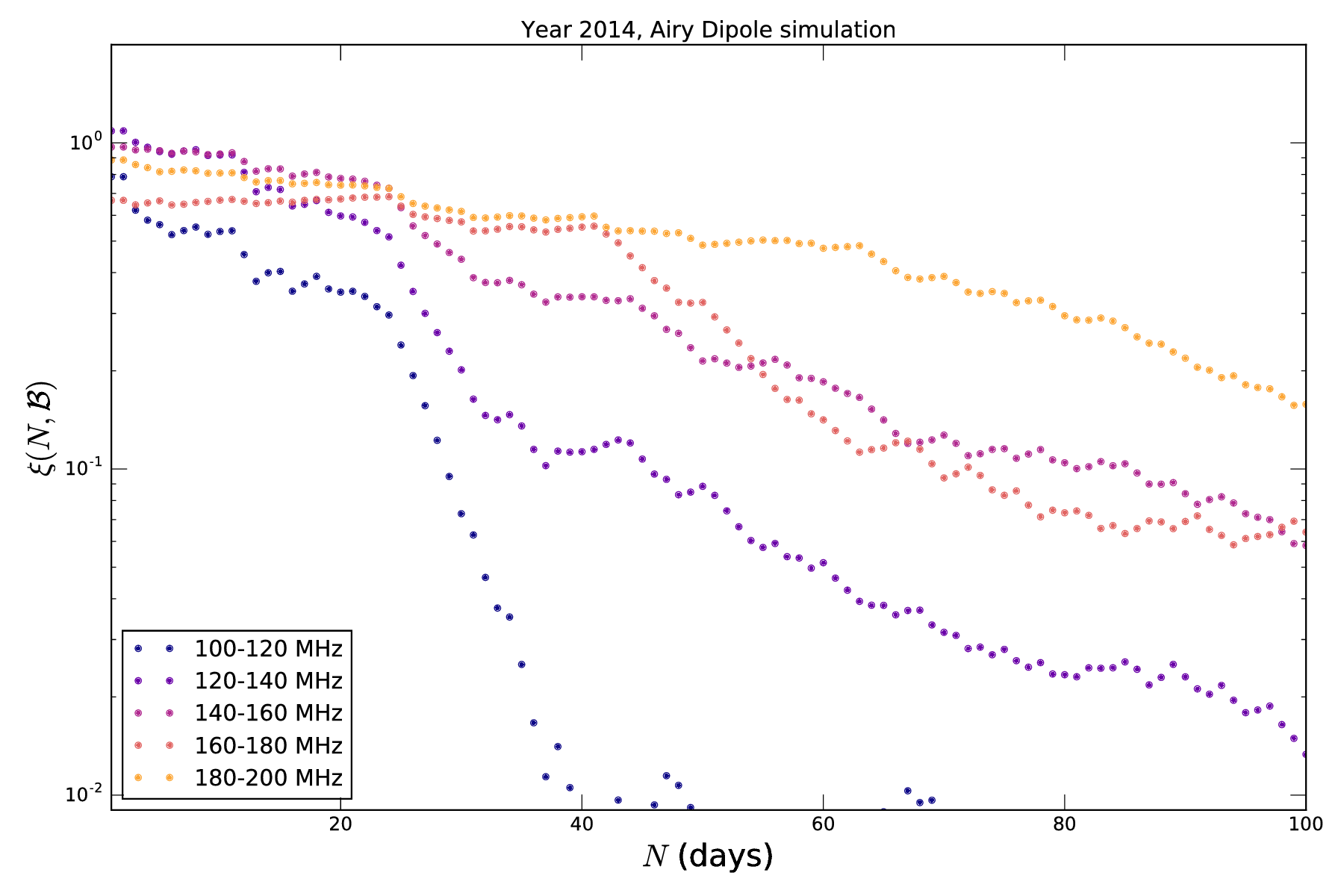}\\
	\end{tabular}
	
	\caption{Natural attenuation of the Vokes-I polarization leakage as defined in Equation \ref{eqn:AttenuationFunction} as a function of the number $N$ of consecutive days averaged over, up to 100 days from Sept. 10 for the four years 2009,2011,2012,and 2014. \textit{Left}: From the fiducial simulations with the HERA antenna response model. \textit{Right}: From the fiducial simulations with the simple Airy-dipole model. It is interesting to see that the large change in the magnitude of the RM during the year 2014 (Figure \ref{fig:RMskewers}) does not produce a correspondingly large change in the attenuation.}
	\label{fig:PspecAttenuation}
\end{figure*}

In Figure \ref{fig:PspecAttenuation} we plot the relative leakage factor $\xi_I$ in each of five frequency bands over the 100 cumulative subsets $S_k$ of $S$ the set of 100 different days for which visibilities were simulated. The cumulative subsets are $S_1 = \{1\}, \ldots S_N = \{1,2, \ldots, N\}, \ldots S_{100} = \{1,2,\ldots, 100\}$ so in Figure \ref{fig:PspecAttenuation} we can write the attenuation $\xi_I$ as a function of $N$, the number of consecutive sidereal days in the sum. This will be refereed to as the "natural attenuation", since it is "natural" to simply average up all available data after observing for $N$ days. Each panel shows the set of five such discrete attenuation curves for each of the four calendar years and two instrumental response models for which we computed visibilities as discussed in Section \ref{sec:numerical}. 

The smooth nature of the attenuation curves is reminiscent of the correlated and trending walks in the plane considered in Section \ref{sec:theory} and reflects the trends of $\varphi(n)$ shown in Figure \ref{fig:RMskewers}. We can also recognize the occasional oscillatory behavior as a natural feature of the polarization state undergoing a trending walk in the tangent plane at each point on the sky. The effect of the changing relative orientation of the polarization state to the instrumental response is evident as $\xi_I > 1$ for some of the curves when $N$ is small. 

The lack of uniform ordering by frequency band is due to the interplay between the intrinsic oscillation in the random walk of the polarization state and the frequency dependence of the instrumental coupling over each band $\mathcal{B}$. Although the instrumental response changes only slowly over each 20MHz sub-band, we see there is significant variation in Figure \ref{fig:HERAmueller} over the full 100-200MHz band. This contributes to the deviations from the ordering of attenuation curves by central frequency that would might naively expect from the simple formulas in Section \ref{sec:theory}. Additionally there is the changing relative importance of the factor $A$ and the $\mu$-rotation as a function of $N$. For small $N$, the amplitude factor $A \sim 1$ but for the highest frequency band the rotation is changing the relative polarization angle between the sky and the instrumental response such that the Faraday rotated leakage is smaller in magnitude than the intrinsic leakage. This is in contrast to the lower frequency bands, some of which have the leakage amplified ($\xi_I > 1$) somewhat by the $\mu$-rotation for small $N$. As $N$ increases, the amplitude factor $A$ becomes more significant. In particular the magnitude of $A$ decreases more quickly at lower frequencies. As a result, eventually the attenuation curves of the lower frequency bands drop into a monotonic, or nearly monotonic, ordering with frequency band. Of course, $\mu$ is still a function of $N$, so its variation should be expected to contribute to eccentricities in the curves as the alignment of the effective polarization state to the instrumental response continues to gradually change. 

The attenuation curves for the HERA model show that in 100 days the polarization leakage in our simulations is attenuated by a factor of $10$ or more in the three bands in the $100-160$MHz range, and the two higher frequency bands are not far behind. Comparison with the Hertzian dipole model shows that the details of the coupling between the instrument and the polarization state of the sky do affect the resulting visibility enough to noticeably change the attenuation factor as a function of $N$. This makes clear that accurate prediction of the attenuation factor is dependent on an accurate beam (and sky) model.

\subsection{Sky Model Variance}

There is significant variation in the natural attenuation curves with the changing ionospheric Faraday rotations for different years which reflects the underlying variation of the visibilities. The fact that different rotations of the polarization state of the sky can produce such a change in the power spectrum suggests that if the Faraday rotation field and instrument model is held fixed, different polarized skies could also produce significantly different results.

The attenuation curves we have considered so far are the results of simulations using a single realization of a statistical model of the diffuse polarization. To understand how much our simulations could vary with the choice of sky model we compute visibilities for 100 realizations of the diffuse polarization generated with CORA.

\begin{figure*}[p]
	\centering
	\includegraphics[scale=1.0]{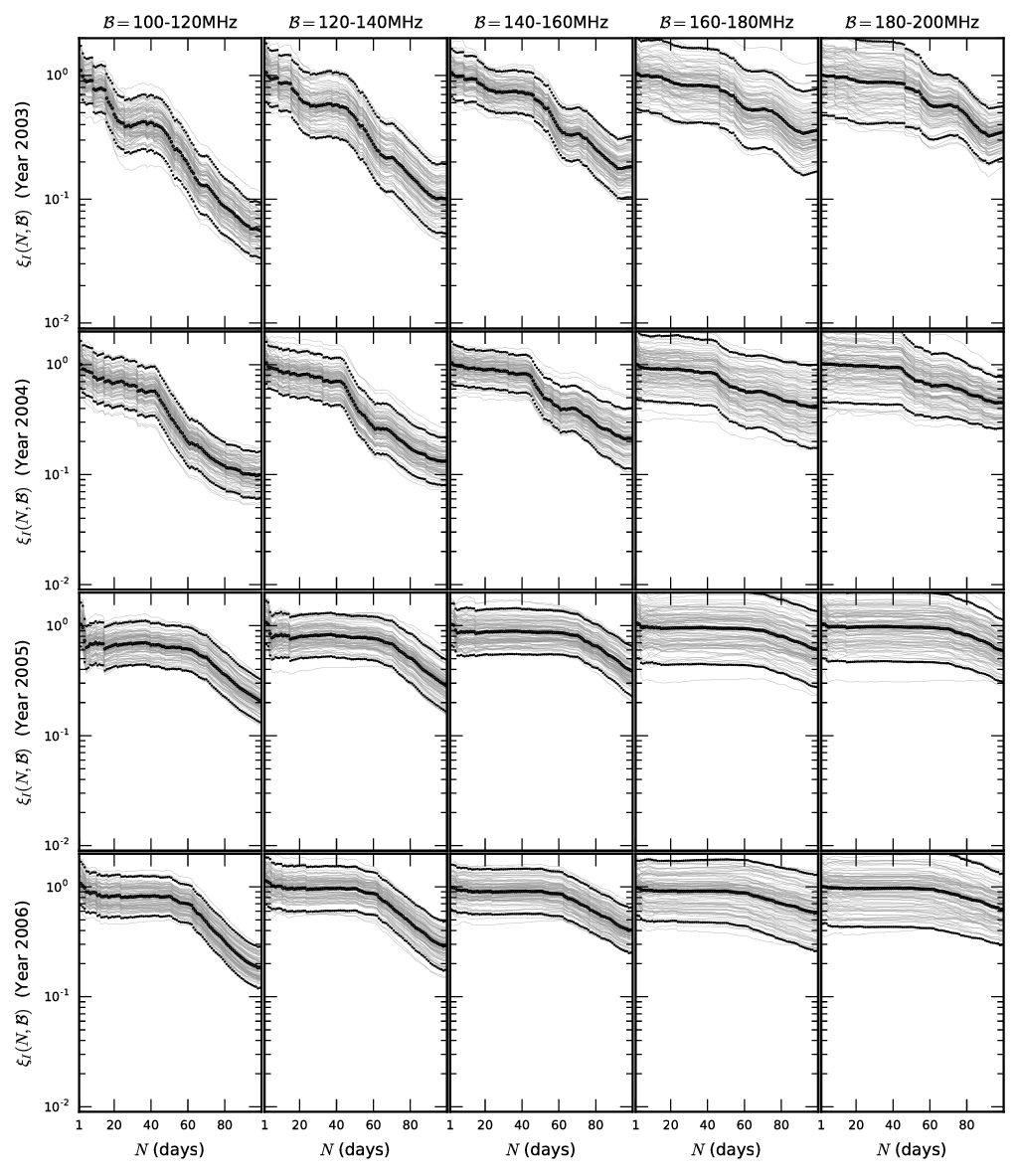}
	\caption{Distributions of the Vokes-I polarization leakage attenuation factor obtained in simulations with different realizations of the sky model. For each year (row) 100 different sky realizations are generated and the attenuation factor computed in each sub-band (column). The thick black points show the geometric mean and geometric 2-$\sigma$ deviation of the distribution of $\xi(N)$ for each $N$.}
	\label{fig:MCAttenuationGrid1}
\end{figure*}

\begin{figure*}[p]
	\centering
	\includegraphics[scale=1.0]{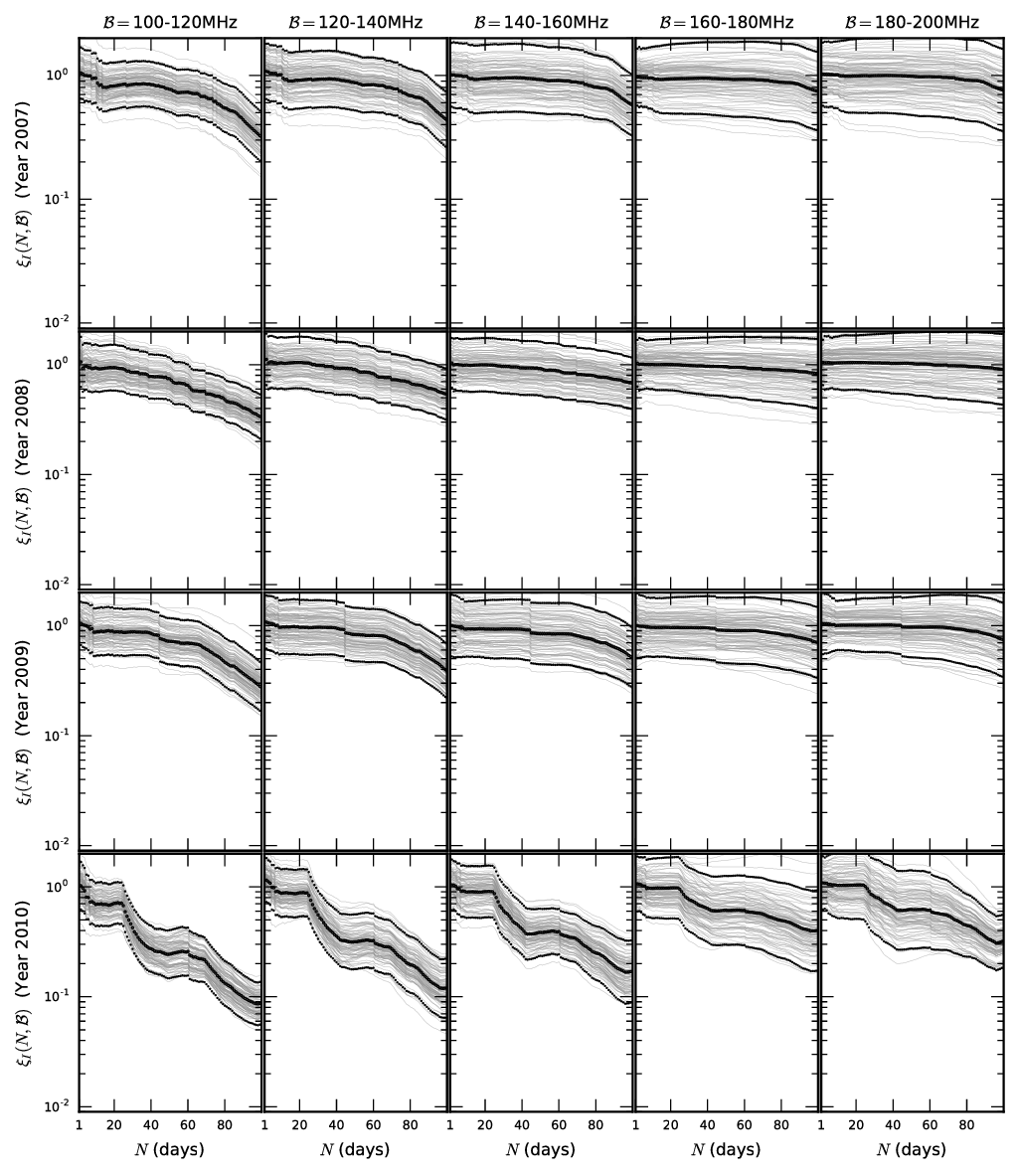}
	\caption{Plots as in Figure \ref{fig:MCAttenuationGrid1} for the years 2007-2010.}
	\label{fig:MCAttenuationGrid2}
\end{figure*}

\begin{figure*}[p]
	\centering
	\includegraphics[scale=1.0]{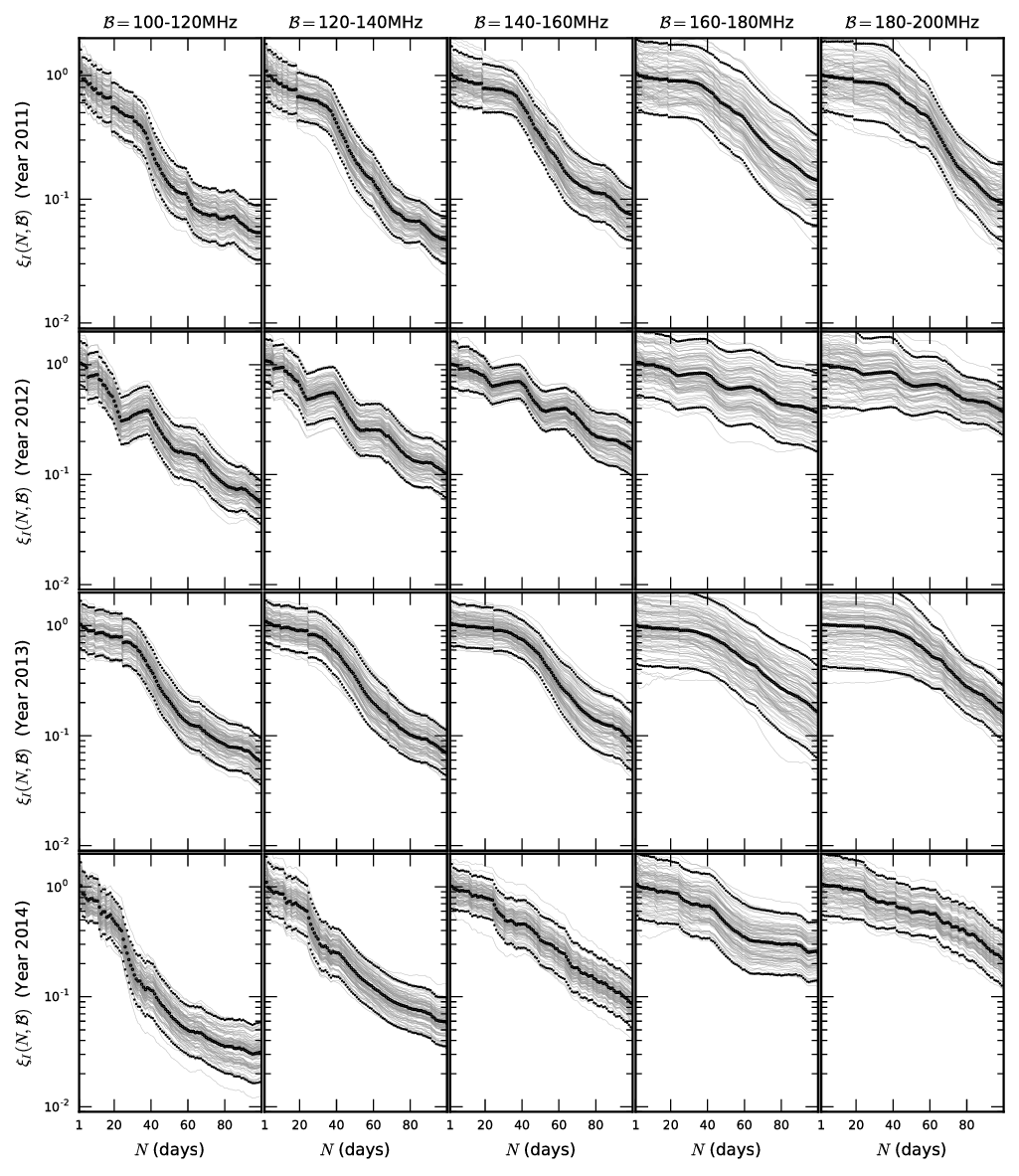}
	\caption{Plots as in Figure \ref{fig:MCAttenuationGrid1} for the years 2011-2014.}
	\label{fig:MCAttenuationGrid3}
\end{figure*}

Figures \ref{fig:MCAttenuationGrid1} - \ref{fig:MCAttenuationGrid3} show the resulting natural attenuation curves of 100 different sky model realizations using the HERA instrument model over the same 100 day sequence in each year from 2003 to 2014. For each year and sub-band the geometric mean and geometric variance of the sample of attenuation curves are estimated and the resulting mean curve and 2-$\sigma$ intervals are also shown.

There is a significant variance over the different realizations of the sky. Thus is it not possible to predict with high accuracy the attenuation that occurs for a specific set of measurements without an accurate sky (and instrument) model.  Nevertheless, for the purpose of forecasting the likely range of attenuation that might be obtained in HERA data, the variation is constrained enough that we still get a good idea of what to expect. By considering the mean attenuation curves for each year we see a large variation as a function of the year. This is due to the solar cycle.

\subsection{The Solar Cycle}

\begin{figure*}[ht]
	\centering
	\includegraphics[width=\textwidth]{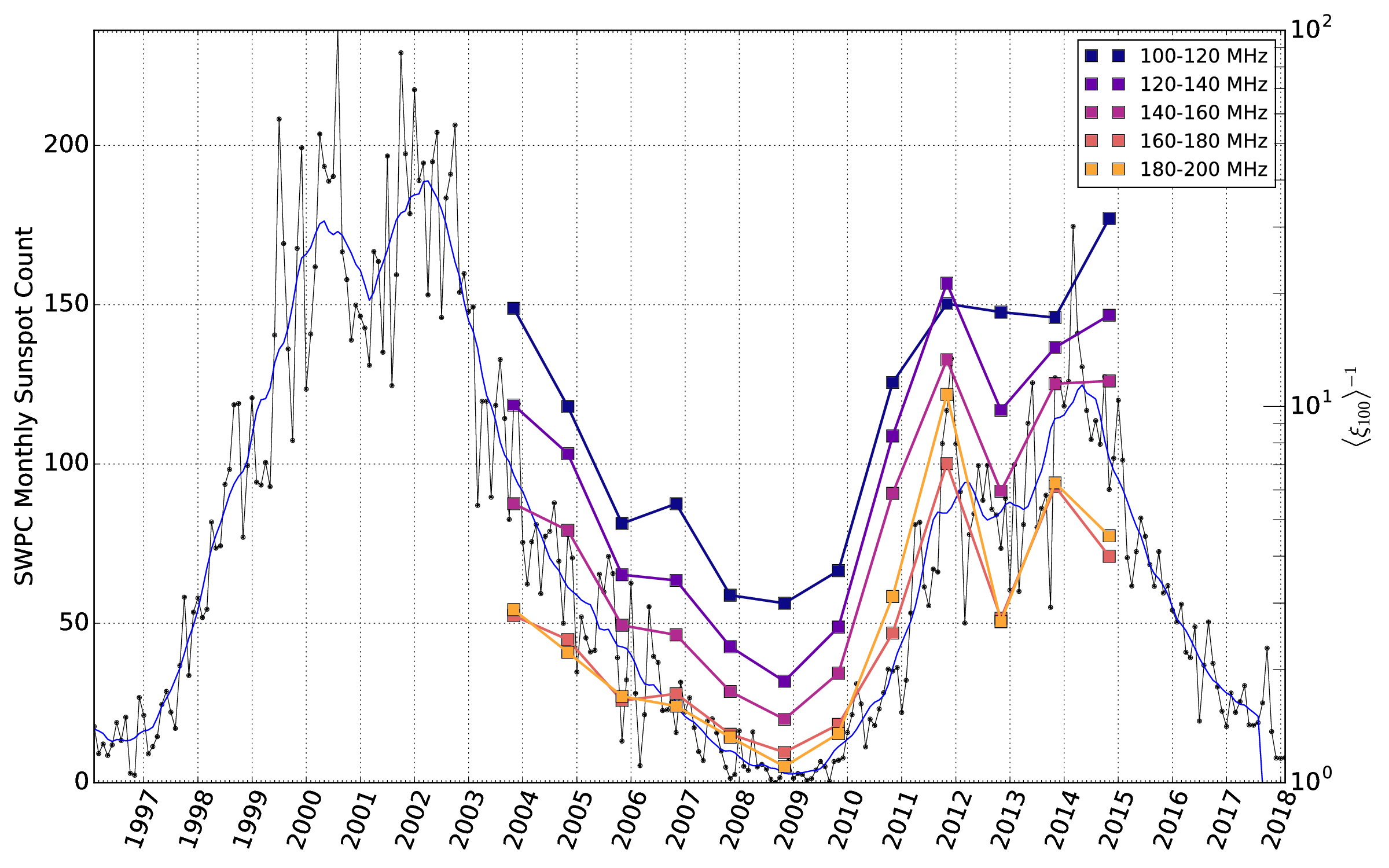}
	\caption{This figure shows how the ionospheric attenuation in our simulations follows the solar cycle by overlaying sunspot number data with the inverse of the expected attenuation factors obtained in our sky model variance simulations. Note the two different vertical axes on the left and right. The thin blue and thin black curves show the historical monthly sunspot numbers from the NOAA Space Weather Prediction Center\textsuperscript{a} over solar cycles 23 and 24. The black points show the count for each month, while the blue curve is a smoothed version of the data obtained by averaging the counts over 13 months. Overlaid are the inverse of the mean over sky realizations of the power spectrum attenuation for each band obtained in the sky model variance simulations at $N=100$ i.e. the last point on the curves in Figures \ref{fig:MCAttenuationGrid1} and \ref{fig:MCAttenuationGrid2}. }
	\small\textsuperscript{a} \url{https://www.swpc.noaa.gov/products/solar-cycle-progression}
	\label{fig:SolarCycle}
\end{figure*}

Solar activity -- the rate of ionizing flux and charged particle emission from the sun -- waxes and wanes with the $\sim$11 year solar cycle, and is highly correlated with the rate of sunspot occurrence.  Thus there is a correlation between the average TEC and solar activity \citep[e.g.,][]{sotomayor-beltran13} and we expect a similar correlation with the attenuation factor. Historical sunspot data as well as future projections are available from the NOAA Space Weather Prediction Center. In Figure \ref{fig:SolarCycle} we compare the mean attenuation over the variable sky model simulations at 100 days with the solar cycle as tracked by the number of sunspots observed in each month.

It turns out that the years 2011-2014 approximately span the peak of the current solar cycle 24. In contrast the year 2008 - the year of the previous solar minimum - exhibits little or no attenuation even after 100 days. The HERA array should reach its full complement of antennas and observe for 3 years from 2021 - 2023.
Thus we expect the next few seasons will pass through solar minimum, and be near solar maximum by the end of data taking.

\subsection{Fluctuating polarized power in the Vokes parameters}

The fluctuation of the ionospheric Faraday rotation turns the otherwise constant (for each $t$) polarization leakage term into something resembling a stochastic noise term which is suppressed similarly to the thermal noise by averaging over multiple days of observation. However, unlike the thermal noise we should not expect the Faraday rotated polarization to be optimally suppressed by averaging over all the available visibility samples. 

For example, this is obvious in the attenuation curve for the year 2012 in Figure \ref{fig:PspecAttenuation} where there is a significant lack of monotonicity in the attenuation curves. In this case, if one had visibilities for only the first 40 days, the optimal attenuation is not obtained by averaging over all 40 days; more attenuation could be obtained by only including the first $\sim$20 days in the average, i.e. by using only half the available data. The reason for this can be observed in Figure \ref{fig:RMskewers} where the value of the RM for the year 2012 between days 20 and 40 can be seen to have a corresponding trend reversal, resulting in more coherent averaging of the polarization leakage over this range of days and thus less attenuation. On the other hand because of the linear trend over 10's of days it is reasonable to expect that an average over a set of days with more separation between them will produce a greater attenuation, since the difference between each day's RM tends to increase with the number of intervening days. 

Given the set of indices of consecutive sidereal days
\begin{equation}
S = \{1,2,\ldots,N_d\}, 
\end{equation}
it is then likely that there are subsets $S_k \subset S$ of non-consecutive days that will produce significantly more attenuation of the polarization terms in the Vokes parameters than the natural attenuation produced by summing over all consecutive days. There will equivalently be subsets that produce significantly less attenuation or even amplification of the polarization terms. 

\begin{figure*}[th]
\begin{center}
	\begin{tabular}{ccc}
		\includegraphics[width=0.5\textwidth]{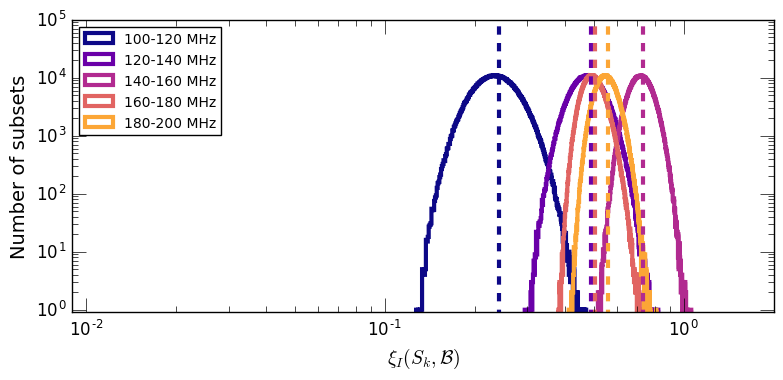}& \vline &
		\includegraphics[width=0.5\textwidth]{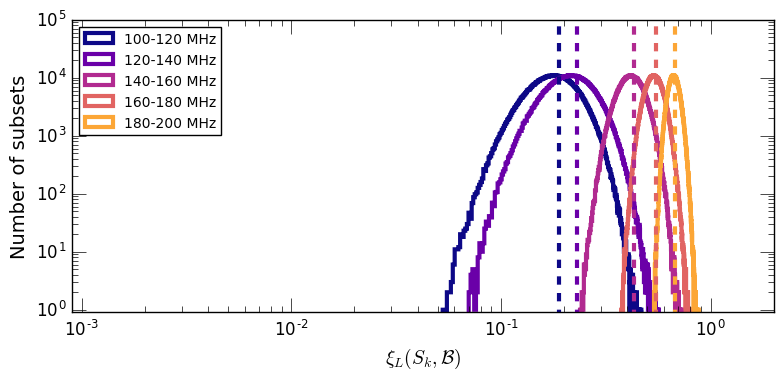}\\
		\includegraphics[width=0.5\textwidth]{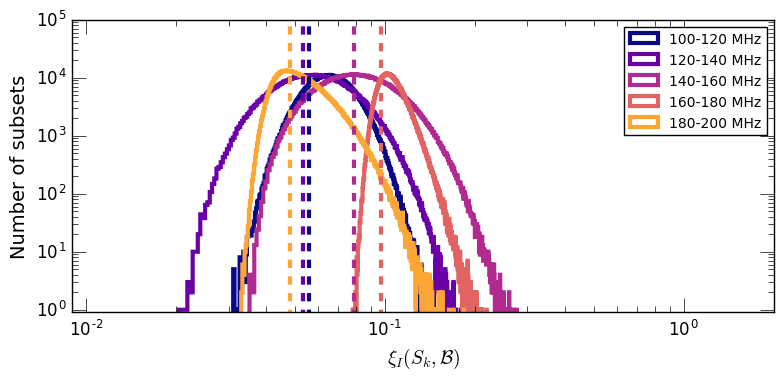} & \vline & 
		\includegraphics[width=0.5\textwidth]{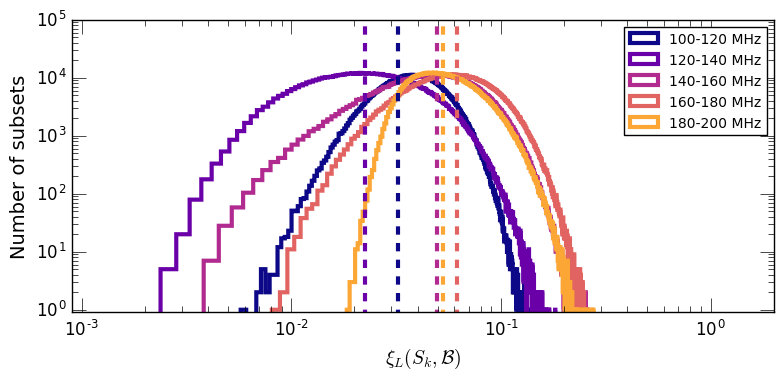}\\						
		\includegraphics[width=0.5\textwidth]{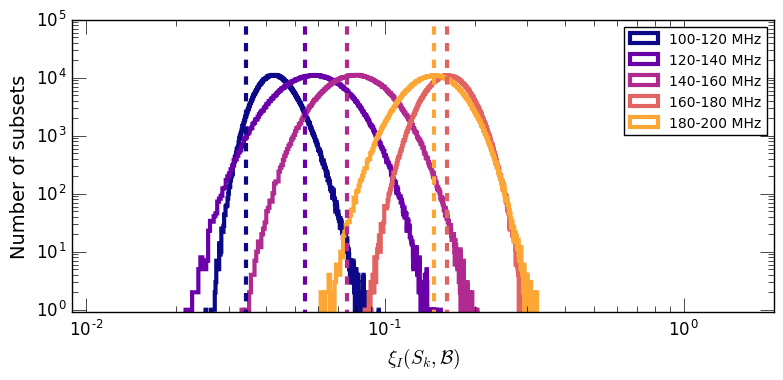}& \vline &
		\includegraphics[width=0.5\textwidth]{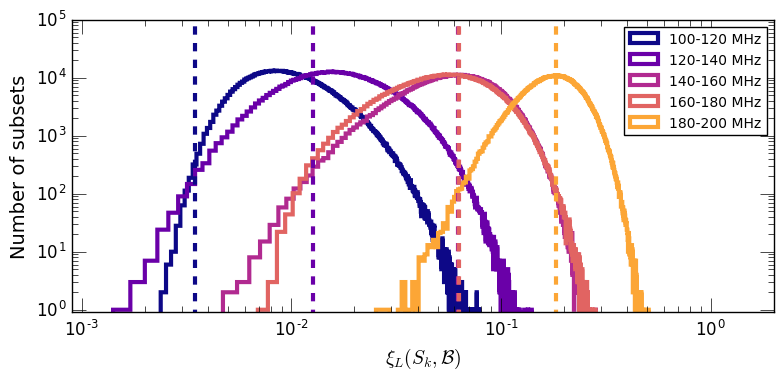}\\
	\end{tabular}
\end{center}
	\caption{\textit{Left:} Distributions of  $\xi_I(S_k, \mathcal{B})$. \textit{Right:} Distributions of $\xi_L(S_k, \mathcal{B})$. The distributions are obtained by taking $10^6$ random subsets $S_k$ of length $N = 50$ days from the fiducial HERA simulations in the years 2009 (\textit{top}), 2011 (\textit{middle}), and 2014 (\textit{bottom}). The dashed lines indicate the 100-day natural attenuation value (i.e. $S_k = S$, $N=100$) value of $\xi_I$ or $\xi_L$, not the mean of the distribution. Note the difference in range on the horizontal axes between the left and right panels.}
	\label{fig:SubsetDistributions}
\end{figure*}

Figure \ref{fig:SubsetDistributions} shows some distributions of $\xi_I(S_k, \mathcal{B})$ obtained from the fiducial visibility simulations over a collection $\mathfrak{C} = \{S_k\}_{k=1}^{N_s}$ of $N_s = 10^6$ subsets of $S$ for which the natural attenuation is shown in Figure \ref{fig:PspecAttenuation}. The number of elements $N$ in each subset $S_k$ is held fixed at $N=50$ - that is, the elements of each $S_k$ are drawn from $S$ without replacement.

The existence of these wide distributions of attenuation factors suggests a null test that is particularly sensitive to polarization leakage in the delay spectrum estimator of the EoR power spectrum. If the thermal noise is sufficiently suppressed for a given subset size $N$, then the fluctuation of any problematic polarization leakage should dominate the variation of the different power spectrum estimates. Since variation in the ionospheric Faraday rotation of polarization leakage implies there will be a distribution of delay-power spectra similar to those in Figure \ref{fig:SubsetDistributions}, the absence of such a distribution rules out polarization leakage as the source of a detection above the expected thermal noise. The converse is not necessarily true - there may be other sources of contamination that could also produce such a distribution, so the presence of such a distribution does not imply that the excess power is due to polarization on the sky. 

The best-case scenario is that the magnitude of any polarization leakage is below the cosmological signal level. In this case the effect of ionospheric fluctuations would never be observed directly in the Vokes-I spectrum. Therefore as a consistency check we will want to simultaneously observe the ionospheric fluctuation of the Vokes-Q and U parameters. Observing fluctuating polarized power in the Vokes-Q/U parameters will then show that there is a fluctuation that would have been observed in Vokes-I if the magnitude of polarization leakage had been large enough.

While considering the delay spectra of $\mathcal{V}_Q$ and $\mathcal{V}_U$ individually can be useful for assessing instrumental systematics \citep{kohn2016}, we can continue to proceed by analogy to the Stokes parameters to define a quantity analogous to $L^2 = Q^2 + U^2$ that is maximally sensitive to the magnitude of linear polarization on the sky:
\begin{align}
P_L(t, \tau, \mathcal{B}) = \abs{\widetilde{\mathcal{V}}_Q(t, \tau, \mathcal{B})}^2 + \abs{\widetilde{\mathcal{V}}_U(t, \tau, \mathcal{B})}^2.
\end{align}

For real data this will include a bias due to Stokes-I and the instrumental polarization impurity that is unaffected by the changing Faraday rotations (c.f. Equation \ref{eqn:VokesIntegralExpand}) and at low frequencies where the polarization fraction on the sky is small Stokes-I will be at least even with Stokes-Q/U in contribution to Vokes-Q/U. For the purpose of assessing the attenuation of observed polarized power in our simulation we follow the same philosophy here as was applied to Vokes-I leakage and subtract the Stokes-I term from Vokes-Q and Vokes-U to isolate the terms that are sensitive to the ionospheric Faraday rotation .

We then define a Vokes-polarization delay-power spectrum

\begin{widetext}
\begin{equation}
\mathcal{L}_L(S_k, \tau_j, \mathcal{B}) = \frac{1}{N_t}\sum_{l=1}^{N_t} \qty( \abs{\frac{1}{N}\sum_{n \in S_k} \widetilde{\mathcal{V}}_{QL}(n, t_l, \tau_j, \mathcal{B})}^2 + \abs{\frac{1}{N}\sum_{n \in S_k} \widetilde{\mathcal{V}}_{UL}(n, t_l, \tau_j, \mathcal{B})}^2 ),
\end{equation}
\end{widetext}
and thus the relative attenuation factor for the Vokes-polarization power
\begin{align}
\xi_L(S_k, \mathcal{B}) = \frac{\sum\limits_{j=1}^{N_\tau}\mathcal{L}_L(S_k, \tau_j, \mathcal{B})}{\sum\limits_{j=1}^{N_\tau} \mathcal{L}_{L,int}(\tau_j, \mathcal{B})}.
\end{align}

Examples of $\xi_L$ as computed from our simulations are also shown in Figure \ref{fig:SubsetDistributions} - the $\xi_L$ distributions are generated from the exact same collection of subsets $S_k$ that was used for the $\xi_I$ distributions. We can see that the mean over $\mathfrak{C}$ , and the $N=100$ values of $\xi_L$ (dashed vertical lines in the plots) are comparable to the same quantities for $\xi_I$, but they are not perfectly correlated. It is notable that the variance of the $\xi_L$ distributions is significantly larger than those of $\xi_I$, because the definitions of $\xi_I$ and $\xi_L$ already divide out an absolute magnitude. Note that this increased variance is not symmetric relative to the peaks of the $\xi_I$ distributions - the distributions of $\xi_L$ are relatively skewed toward smaller values. It appears that it is easier to find a combination of Faraday rotations that can significantly attenuate the Vokes-polarization power spectrum than it is for the Vokes-I polarization leakage.

Another potential issue with this null test is the computational cost in sampling the collection of possible subsets. The number of possible subsets $S_k$ of any length is $2^{N_d}$, which is $\sim 10^{30}$ for $N_d = 100$, while the number of subsets of size $N=50$ is $\frac{100!}{50!50!} \sim 10^{29}$. It is thus not possible to construct the full distributions over all possible subsets even in simulation, much less for actual measurements which generally require more processing to produce power spectrum estimates. In reality the number of samples $N_s$ that can be used may only be in the hundreds or thousands. 

The distributions in Figure \ref{fig:SubsetDistributions} were generated by uniformly sampling the full collection of subsets of length 50. But we are not ignorant of the ionosphere's behavior, and we would like to be able to use this knowledge to bias the sampling toward the tails of these distributions. 

In principle we might like to use simulations such as the ones done in this paper to inform a selection of subsets with which to estimate the power spectrum on. But we have already seen how sensitive the attenuation is to the detailed coupling of the polarization states of the sky to the instrumental response which casts doubt on our ability to make accurate predictions from simulations given our current levels of knowledge about the relevant functions. Fortunately, for the purpose of this null test we do not need perfect accuracy, only to do a little better than completely random guessing. Additionally, we need not precisely predict the actual magnitude of the attenuation in a particular subset, only its relative place in the distribution.

We attempt to approximate the distribution of $\xi_I(S_k, \mathcal{B})$ in our fiducial HERA simulations by a simpler functional of the ionospheric RM that is independent of the observed sky, and includes only an approximate and generic model of the instrumental response. Define 
\begin{widetext}
\begin{align}
\mathcal{A}^2(S_k, \nu_*) & = \frac{1}{t_b - t_a} \int_{t_a}^{t_b} \dd{t} \frac{1}{4\pi}\int_{\mathcal{S}^2} \Big( M_{IQ}^2(\nu_*, \vu{s}) + M_{IU}^2(\nu_*, \vu{s}) \Big) A^2(S_k,t,\nu_*, \vu{s})\\
& \approx \frac{1}{N_t} \sum_{l = 1}^{N_t} \frac{1}{N_p} \sum_{p=1}^{N_p} \Big( M_{IQ}^2(\nu_*, \vu{s}_p) + M_{IU}^2(\nu_*, \vu{s}_p) \Big) A^2(S_k,t_l,\nu_*, \vu{s}_p)
\end{align}
\end{widetext}
where $\nu_*$ is the central frequency of each subband $\mathcal{B}$, the function $A^2$ is given by Equation \ref{eqn:AttenuationAmplitudeFactor}, and the sums are computed over an \texttt{nside} $=8$ \textsc{HEALPix} map as the integrand does not vary as much on small scales as the functions in our visibility calculation. The Mueller matrix elements used are those of the analytically defined Airy-dipole model, computed from the definition of this Jones matrix (Equation \ref{eqn:AiryDipoleJones}) and the formula for the Mueller matrix elements (Equation \ref{eqn:MuellerComponents}).

This quantity need not predict the value of the attenuation precisely. We are only interested here in finding subsets that correspond to attenuation factors in the tails of the distributions of $\xi_I$ in Figure \ref{fig:SubsetDistributions}. Thus to compare the distributions for $\xi_I$, $\xi_L$, and $\mathcal{A}^2$ we compute the z-scores for each variable from the distribution over the chosen collection $\mathfrak{C}$ of subsets $S_k$. The z-score for the variable $X \in \{\xi_I(S_k,\mathcal{B})$,$\xi_L(S_k,\mathcal{B})$,$\mathcal{A}^2(S_k,\mathcal{B})\}$ is
\begin{align}
\mathcal{Z} = \frac{X - Mean(X)}{Std(X)}
\end{align}
where $Mean()$ and $Std()$ are the mean and standard deviation of $X$ over $\mathfrak{C}$. We denote the z-scores for each of these variables by $\mathcal{Z}_I$, $\mathcal{Z}_L$,$\mathcal{Z}_A$, respectively. 

\begin{figure*}[ht]
	\centering
	\begin{tabular}{c}
		\includegraphics[width=\textwidth]{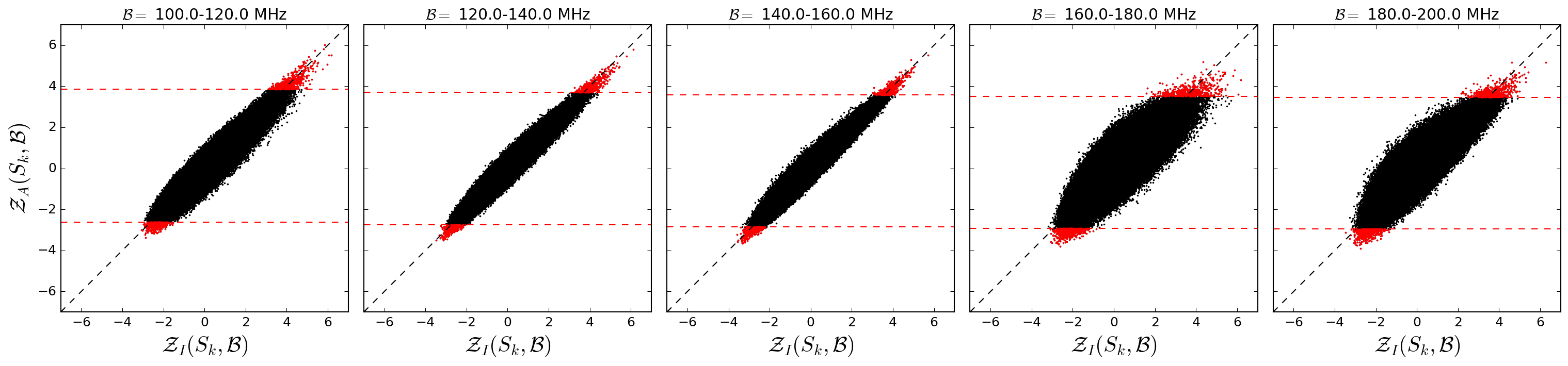}\\
		\includegraphics[width=\textwidth]{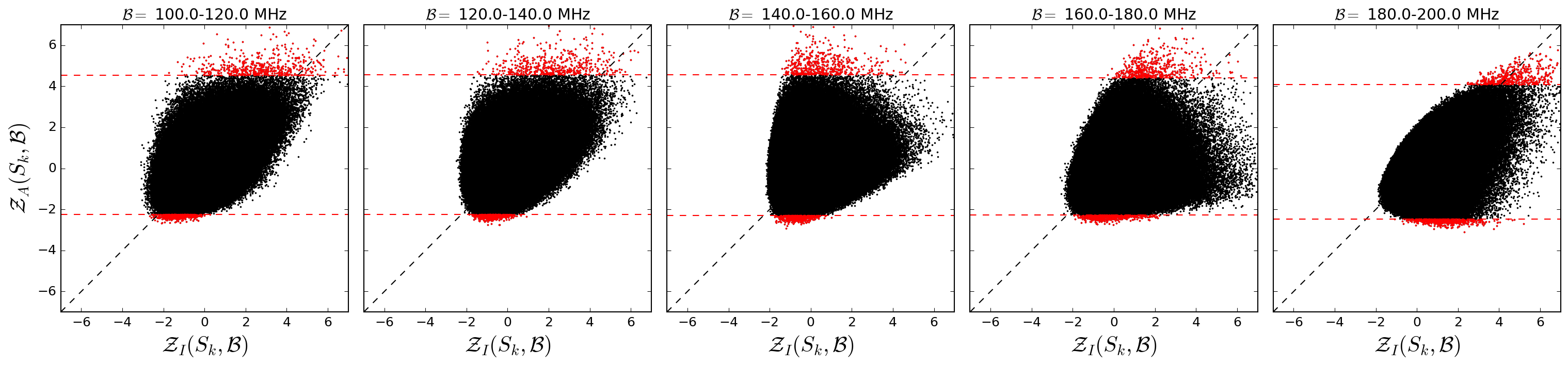}\\
		\includegraphics[width=\textwidth]{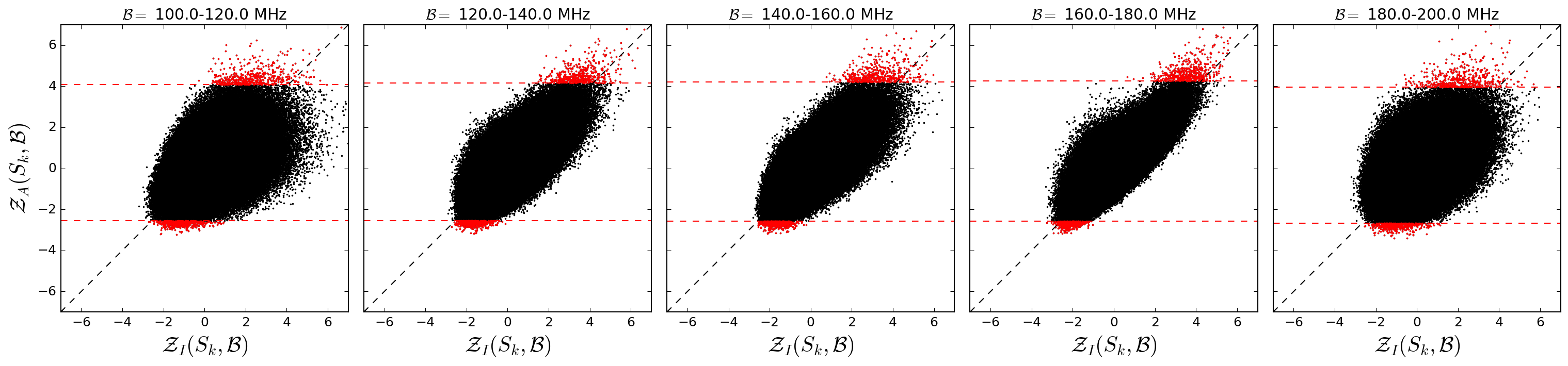}\\
	\end{tabular}
	\caption{Correlation of the ionospheric fluctuation tracer $\mathcal{A}^2$  with the Vokes-I polarization leakage in the fiducial visibility simulation for the years 2009 (\textit{top}), 2011 (\textit{middle}), and 2014 (\textit{bottom}). The collection of $10^6$ subsets used here is the same as the one used to make Figure \ref{fig:SubsetDistributions}. The red points show a cut on the 500 largest and smallest values of $\mathcal{Z}_A$ for each $\mathcal{B}$. Such a cut would select the subsets to be used to estimate the power spectrum in our null test.}
	\label{fig:VokesILvsA2}
\end{figure*}

Figure \ref{fig:VokesILvsA2} shows the correlation of $\mathcal{Z}_A(S_k, \mathcal{B})$ with $\mathcal{Z}_I(S_k, \mathcal{B})$ in the same years and for the same collection of subsets as used in Figure \ref{fig:SubsetDistributions}. We can see that the subsets that produce values of $\mathcal{A}^2$ in the tails of the distribution tend to also find values of $\xi_I$ in the tails of the distribution. The correlation is far perfect, but as noted the point is merely to improve the statistical power of the null test - any correlation helps compared to completely uniform sampling. Additionally, the proxy function $\mathcal{A}^2$ is simply an inspired guess based on Equation \ref{eqn:VokesIntegralExpand}. It seems likely that an improved method of sampling these distribution based on the ionospheric RM data could be found; in particular we have not used the fact that the RM has a significant trend as a function of $n$.

\begin{figure*}[ht]
	\centering
	\begin{tabular}{c}
		\includegraphics[width=\textwidth]{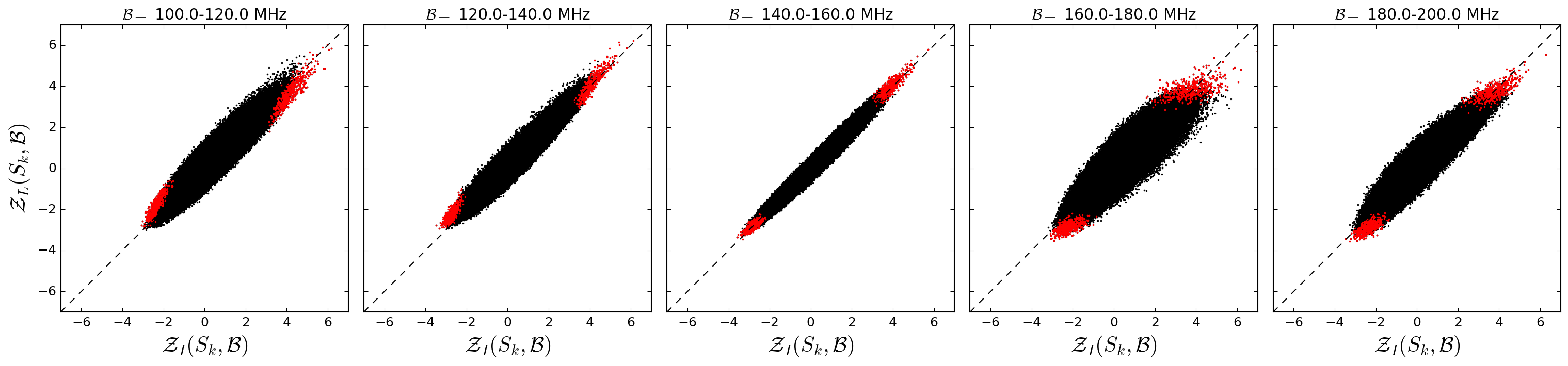}\\
		\includegraphics[width=\textwidth]{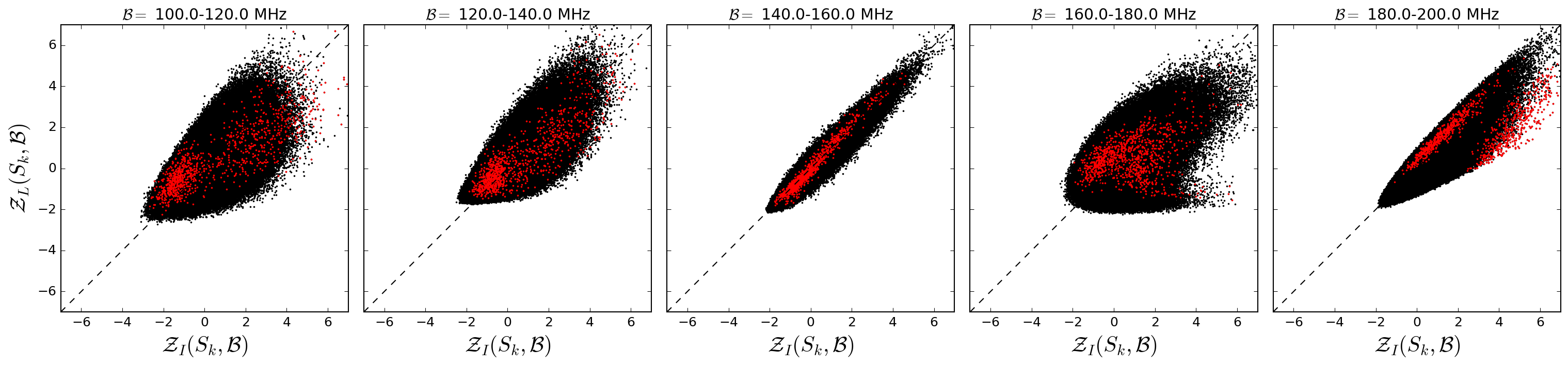}\\
		\includegraphics[width=\textwidth]{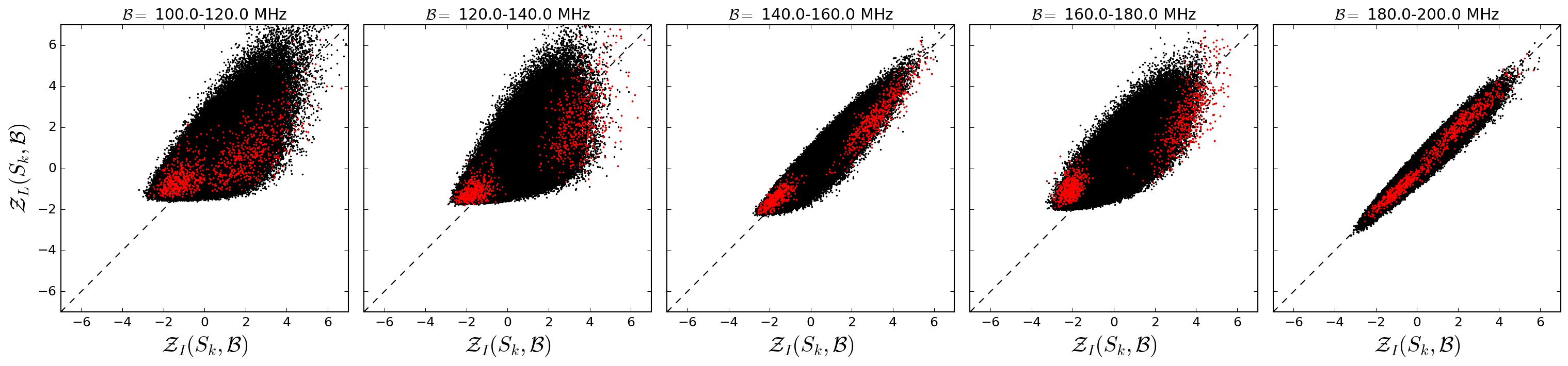}\\
	\end{tabular}
	\caption{Correlation of the ionospheric fluctuations in the Vokes-polarization band power with the Vokes-I polarization leakage band power from the fiducial HERA visibility simulations for the years 2009 (\textit{top}), 2011 (\textit{middle}), and 2014 (\textit{bottom}). The red points here correspond to the subsets from the cut on $Z_A$, i.e the red points in Figure \ref{fig:VokesILvsA2}. This shows that if we see a distribution in the Vokes-polarization we can infer that there exists a distribution in the Vokes-I polarization, though we should not necessarily expect to see the \textit{same} distribution.}
	\label{fig:VokesILvsVokesQUL}
\end{figure*}

\section{Discussion and Conclusions}
\label{sec:conc}

\begin{enumerate}
	\item  Ionospheric attenuation cannot be counted on to suppress polarization leakage in the power spectrum. Given what little is known about the level of polarized power on the sky in the 100-200 MHz frequency band, even at solar maximum it seems as likely as not that this attenuation might suppress polarization leakage to a negligible level. This increases the importance of precise modeling of this systematic, either to show that it will indeed be small relative to the EoR signal, or for the purpose of subtraction.
	
	\item Our simulations suggest a definitive test for polarization leakage in the power spectrum. This test comprises the following:
	\begin{enumerate}
		\item \label{SubsetSelection} From the set $S$ of $N_d$ available sidereal days select a collection $\mathfrak{C}$ of subsets $S_k \subset S$ with the number $N < N_d$ of elements in each $S_k$ held fixed. The number $N_d$ must be large enough to allow significant ionospheric variation over $S$. Additionally, the fraction $N/N_d$ must be chosen to strike a balance between allowing the ionospheric attenuation to vary significantly between subsets, while also ensuring that each subset represents sufficient integration time on the thermal noise.
		
		\item Compute the power spectra $P_I(S_k)$ and $P_L(S_k)$ for each of the subsets. This produces a distribution of power spectra over $\mathfrak{C}$.
		
		\item \label{ISame} If the Vokes-I power spectrum estimator is dominated by Stokes-I on the sky, then the changing ionospheric Faraday rotations between different subsets will have no effect and each subset will produce the same spectrum up to an expected distribution due to the thermal noise.
		
		\item \label{LChange} The distribution of $P_L$ should be significantly and obviously inconsistent with the expected thermal noise distribution.
		
		\item The null test is passed when both \ref{ISame} and \ref{LChange} are satisfied, as \ref{LChange} demonstrates that the effective polarized power on the sky has an observable variation over $\mathfrak{C}$, while \ref{ISame} shows that there is no corresponding variation of what is supposed to be Stokes-I.
	\end{enumerate}
	
	The method by which the elements of $\mathfrak{C}$ should be chosen remains open to further investigation. We have shown that a simple proxy function for ionospheric attenuation can reliably bias the sampling toward subsets with relatively high or low attenuation factors. Additional consideration could produce an improved method.
	
	The sensitivity of this test as a function of the thermal noise level is explored in a schematic way in Appendix \ref{sec:NoiseAppendix}, but detailed consideration should be the subject of further simulations and analysis that can explore in detail the parameter space of cosmological signal level, thermal noise level, and polarized foreground power level.
	
	Additionally, the method of quantifying the consistency of these distributions with an expected thermal noise distribution need not be limited to simply computing the variance. For example, we showed that using our simple proxy function to select subsets can often produce distinctly bimodal distributions. The difference in the mean of the high-attenuation collection to the low-attenuation collection could be a useful discriminating statistic. More generally, an advanced subset selection method may go hand-in-hand with a more robust way of distinguishing the resulting distributions from the expected thermal noise.

	\item The simulations we have used of the polarized sky are intended to be reasonably accurate representations of the expected sky, but their fidelity could certainly be improved. This is necessary for accurate prediction, since we have shown that the level of leakage is sensitively dependent on the correlated structure in the sky model and its alignment with the polarized antenna response, and this does produce large variations in the potential level of leakage.  Given this uncertainty, we have purposely avoided considerations of the details of the absolute level of polarization leakage by considering ratios, and demonstrate that these do show systematic trends independent of the details of the sky model. 
	
	\item Averaging over sidereal days at fixed LST may still be a useful method for suppressing polarized foregrounds even in the situation in which one tries to model and subtract them directly from the visibilities, as the residual (unmodeled) polarization leakage will be attenuated by averaging over many days. This may ease the requirements on the completeness of the polarized model.  On the other hand, an increasing level of ionospheric attenuation goes hand-in-hand with increasing complexity of the ionosphere, and thus increasing complexity of the model that must be constructed in order to perform the subtraction. It remains to be seen whether the global model of the ionospheric Faraday rotation which we have presented here would be adequate for such a task.

	\item The variance in the visibility and resulting power spectrum can be quite large when the polarization angle on the sky is not constrained. While preliminary, the results of our simulations suggest that a statistical foreground model which does not constrain the orientation of the polarization on the sky may be inadequate for predicting polarization leakage levels to the accuracy required for HERA, and possibly other EoR experiments. Determining the extent to which this is true or not through more careful consideration of the parameterization the sky model and the mapping into the visibility will require further research.  Obviously, it is necessary to determine the polarization angle accurately to be able to subtract a model from the visibilities.
	
\end{enumerate}

\acknowledgements
This material is based upon work supported by the National Science Foundation under Grants \#1440343 and \#1636646, the Gordon and Betty Moore Foundation, and institutional support from the HERA collaboration partners. SAK is supported by a University of Pennsylvania SAS Dissertation Completion Fellowship. JEA acknowledges support from NSF CAREER award \#1455151.

\appendix

\section{Comparing ionospheric RM outputs}
\label{sec:ComparisonAppendix}

\begin{figure*}[h]
\centering
\begin{tabular}{cc}
\includegraphics[scale=0.2]{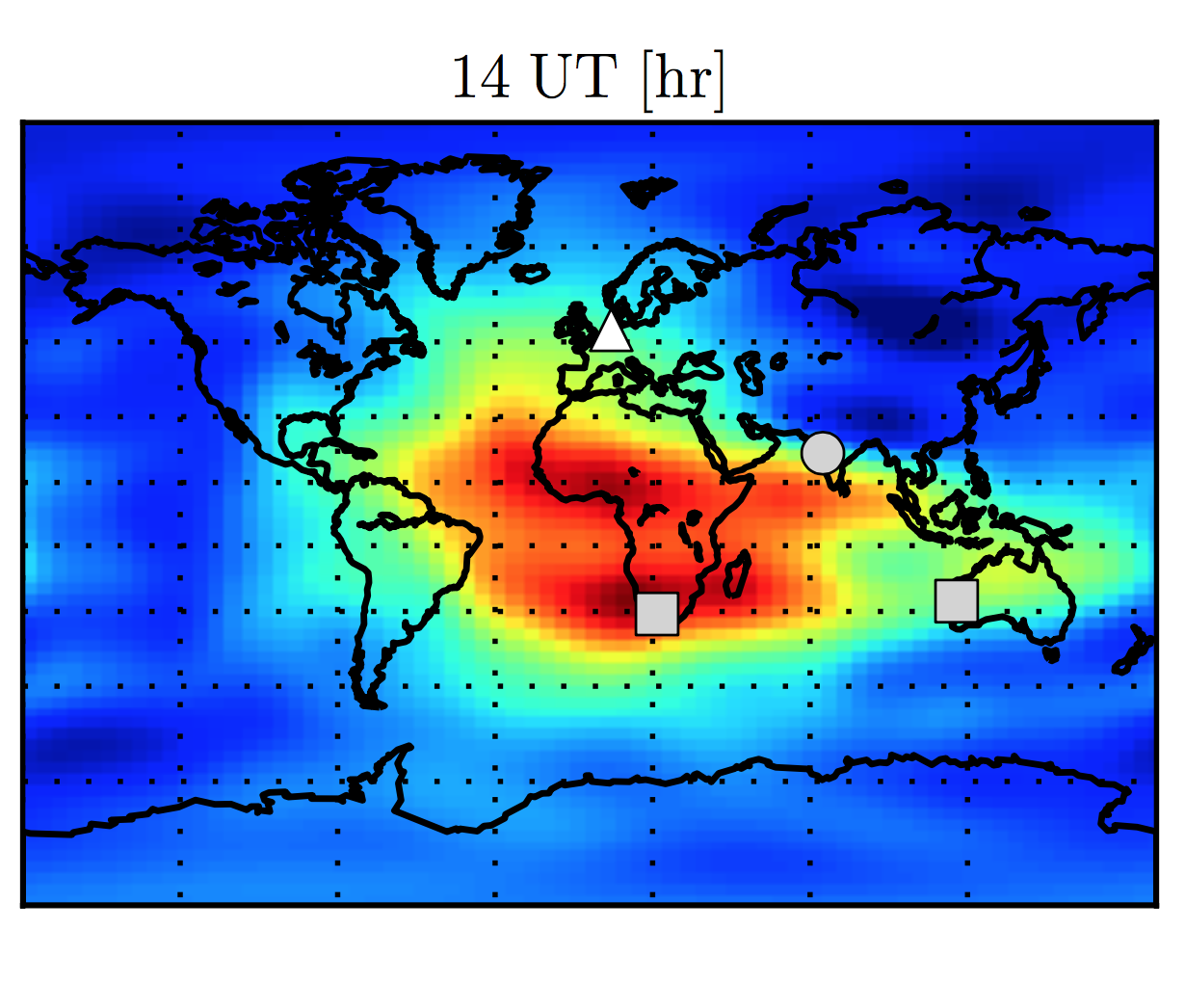}&
\includegraphics[scale=0.2]{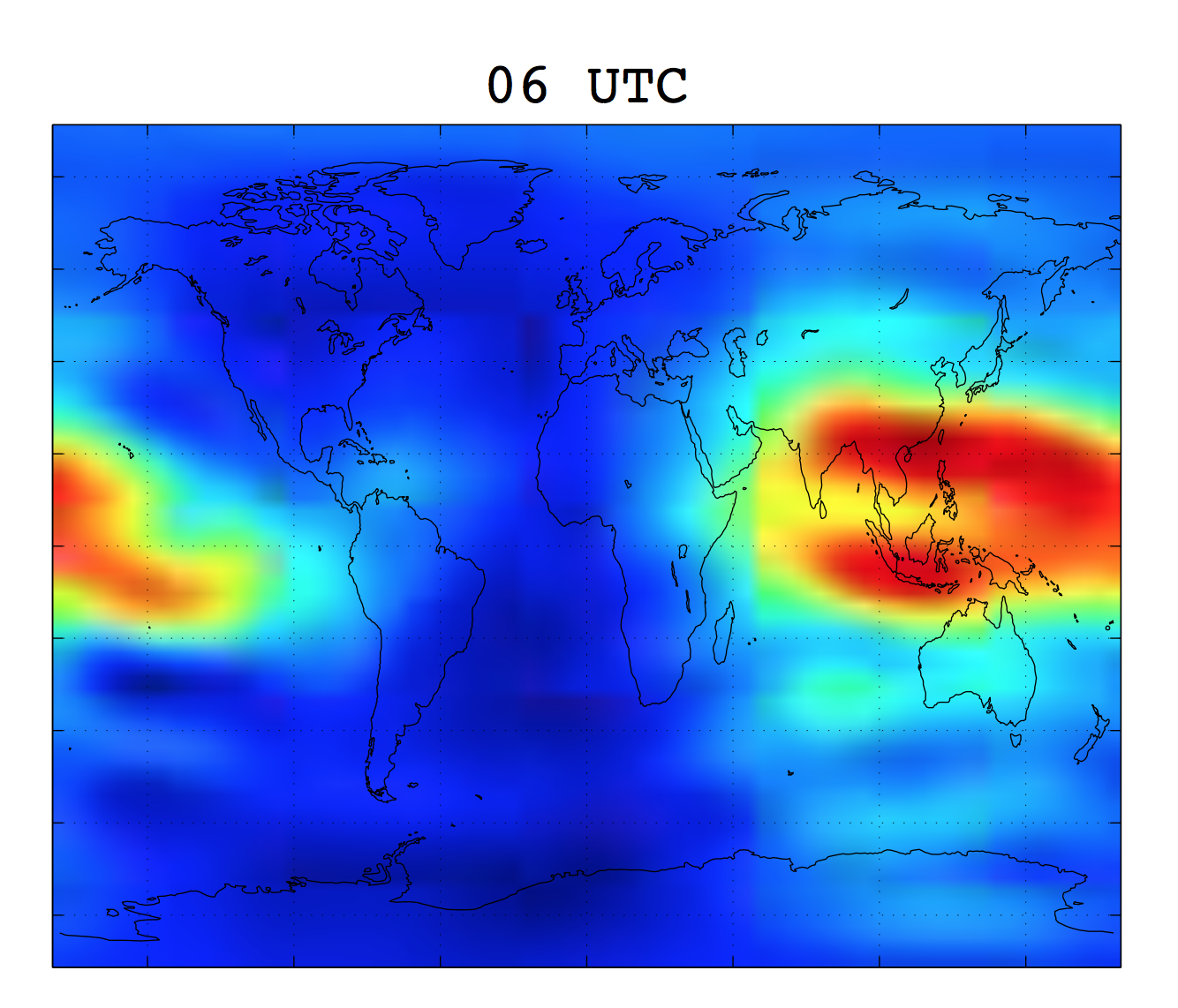}\\
\includegraphics[scale=0.3]{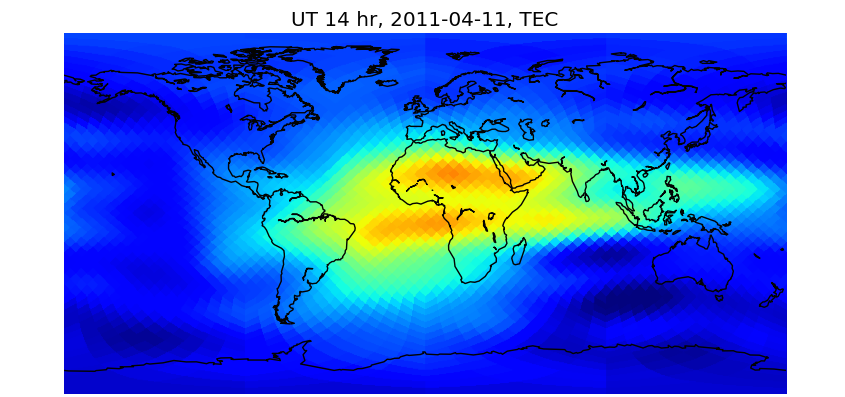}&
\includegraphics[scale=0.3]{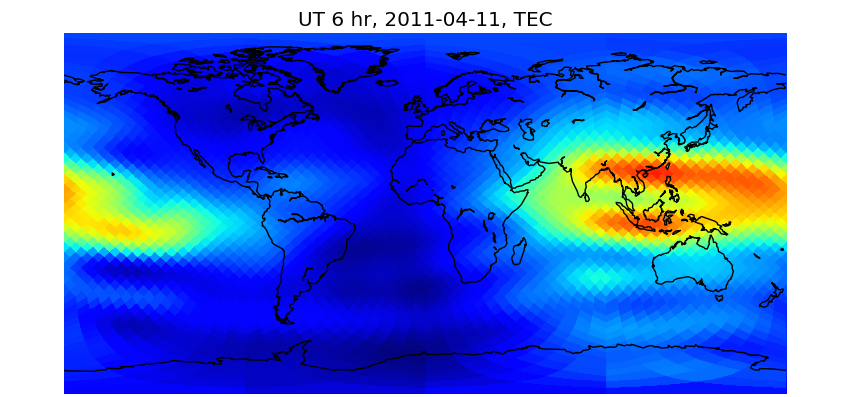}\\
\end{tabular}
\caption{\textit{Top}: The vertical TEC from the CODE {\sc ionex} file for April 11th, 2011, over-plotted on the globe in a Cartesian projection, as measured in \citet{sotomayor-beltran13} and \citet{Arora15} (left and right, respectively). \textit{Bottom}: The {\tt radionopy} output for the same times and day. There is qualitative agreement, save for a error resulting in upside-down maps in \citet{sotomayor-beltran13}, as pointed-out by \citet{Arora15}.}
\label{fig:compareTEC}
\end{figure*}

\begin{figure}[h]
\centering
\includegraphics[width=0.4\textwidth]{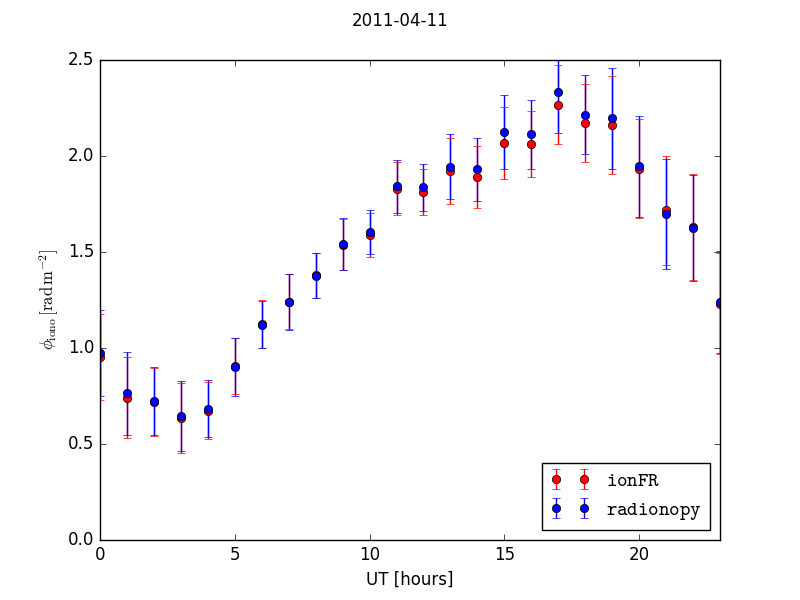}
\caption{The RM of Cas A as viewed from the LOFAR Core site in the Netherlands on April 11th, 2011, according to {\tt ionFR} and {\tt radionopy}. The two codes show quantitative agreement; this demonstrates that {\tt radionopy} can be used for single-pointing as well as full-sky RM measurements.}
\label{fig:compareRM}
\end{figure}

There are now several software packages that interpret CODE {\sc ionex} files specifically for the use of low frequency radio interferometers. Two of these are {\tt ionFR} \citep{sotomayor-beltran13} and the results shown in \citep{Arora15}. In Figure~\ref{fig:compareTEC} we show qualitative agreement with both of these works by comparing maps of vertical TEC values over the globe. In Figure~\ref{fig:compareRM} we show {\tt radionopy} and {\tt ionFR} RM output for a single pointing towards Cassiopeia A (Cas A; RA = $\rm 23^h23^m27.9^s$, Dec = +$58\arcdeg 48\arcmin 42.4\arcsec$) from the LOFAR Core site in the Netherlands, which exhibit quantitative agreement. Slight offsets at the highest RM values that day can be attributed to differences in our interpolation schemes.

\begin{figure}[h]
	\centering
	\includegraphics[scale=0.25]{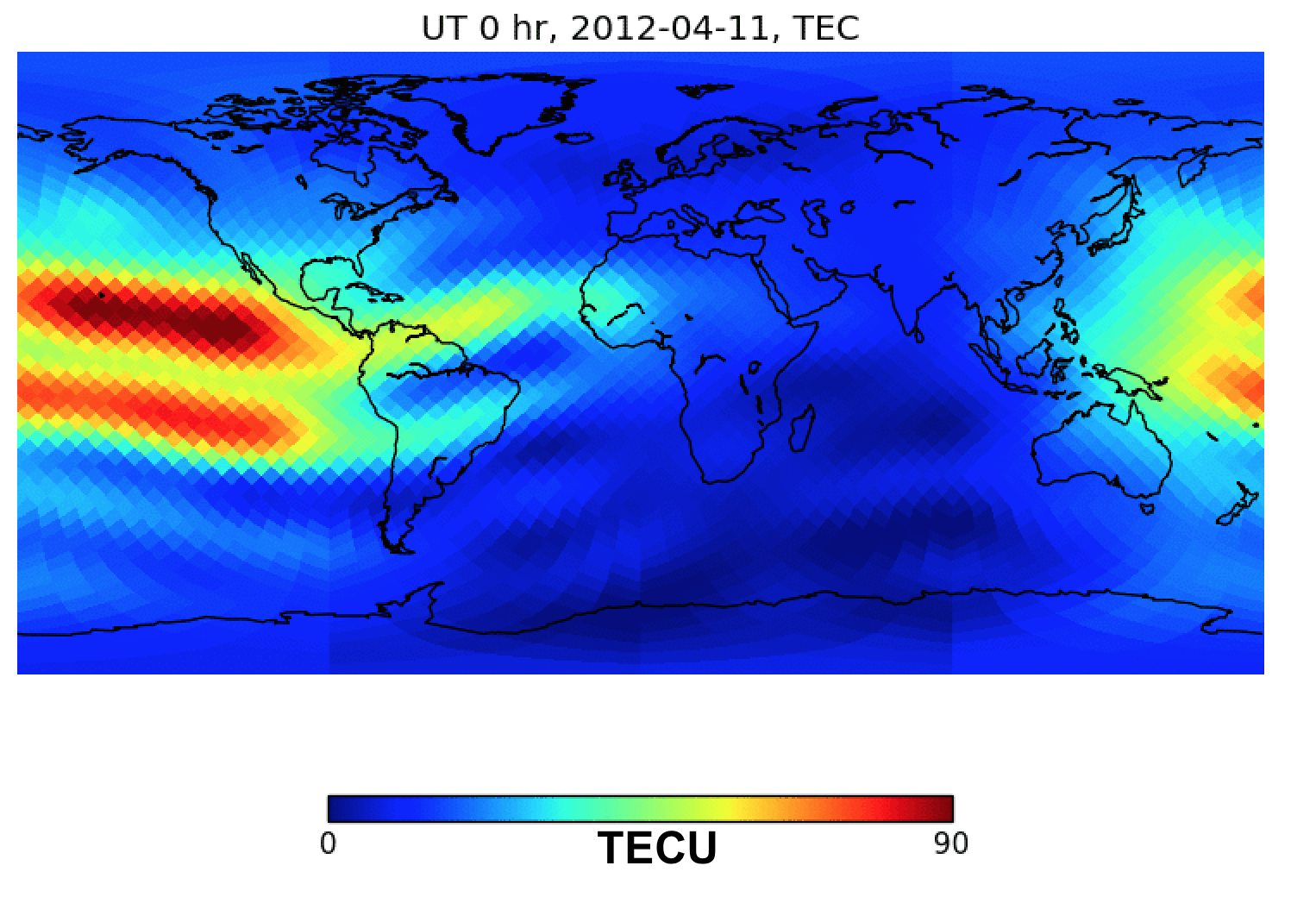}
	\caption{Example output from {\tt radionopy}: the full-sky TEC content of the ionosphere (in TECU; factors of $10^{16}$ electrons per m$^2$) on a {\sc Healpix} grid, in this case projected onto a {\sc python} {\tt Basemap}. This particular snapshot shows the ionosphere at UT 0 hours on 11th April 2012.}
	\label{fig:radiono_example}
\end{figure}

\section{The instrumental Jones matrix and basis transformation}
\label{sec:InstrumentAppendix}

While the instrumental Jones matrix $\bm{J}$ happens to be a 2x2 matrix in the case of the antenna with two different feed polarizations, it is better thought of as a list of rank-1 tensor fields $\va{\mathcal{F}}_k(\nu, \vu{s})$ for the $k$'th feed of $N$ feeds
\begin{equation}
\bm{J}(\nu, \vu{s}) = \mqty*(\va{\mathcal{F}}_1^T(\nu, \vu{s}) \\ \vdots \\ \va{\mathcal{F}}_N^T(\nu, \vu{s}) ) = \mqty*(  
\mqty*( J_1^{\delta} & J_1^{\alpha}) \\
\vdots \\
\mqty*(J_N^{\delta} & J_N^{\alpha } ) \\
)
\end{equation}
that correspond to the far-field electric-vector fields generated by each feed operated in transmission. Each row of the matrix corresponds to the normalized electric vector field pattern of a single feed of the antenna.

In order to compute Equation \ref{eqn:VisMat} the instrumental Jones matrix $\bm{J}$ and the coherency matrix $\bm{\mathcal{C}}$ must be specified in the same basis. Because of the cylindrical symmetry of the $\vu{e}_\alpha, \vu{e}_\delta$ basis we can specify the instrumental response in this basis, rather than the alternative of performing a basis transformation on the observed coherency matrix. Observe that the integrand in Equation \ref{eqn:VisMat} is invariant under a transformation
\begin{align}
\bm{\mathcal{J}} & \rightarrow \bm{\mathcal{J}} \bm{\mathcal{U}}\\
\bm{\mathcal{C}} & \rightarrow \bm{\mathcal{U}}^\dagger \bm{\mathcal{C}} \bm{\mathcal{U}}
\end{align}
where $\bm{\mathcal{U}}(\vu{s})$ is a 2x2 unitary matrix field. Any basis transformation (a point-by-point 2x2 rotation) is such a unitary matrix. Since the coherency matrix is specified in the $\vu{e}_\alpha, \vu{e}_\delta$ basis we thus require the instrumental response to be specified in this basis. 

However, it is generally practical to specify the instrumental response in a basis of spherical coordinates local to the antenna so that the representation is independent of the telescope's geographic location, and a standard choice of coordinates is the zenith angle $\theta \in (0, \pi)$ and local azimuthal angle $\phi \in [0, 2 \pi)$. Explicitly, the means that the electric field data generated from an EM simulation of the antenna is specified as the complex coefficient functions of the vector field
\begin{align}
\va{E}(\nu, \vu{s}) = E_\theta(\nu, \vu{s}) \vu{e}_\theta + E_\phi(\nu, \vu{s}) \vu{e}_\phi,
\end{align}
which defines the instrumental response as
\begin{align}
\va{\mathcal{F}}(\nu, \vu{s}) & = \mathcal{F}_\theta(\nu, \vu{s}) \vu{e}_\theta + \mathcal{F}_\phi(\nu, \vu{s}) \vu{e}_\phi\\
& = \frac{1}{|\va{E}(\nu, \vu{s}_b)|} \qty( E_\theta^*(\nu, \vu{s}) \vu{e}_\theta + E_\phi^*(\nu, \vu{s}) \vu{e}_\phi )
\end{align}
where $\vu{s}_b$ denotes the direction of the antenna's bore-sight. There is an equivalent representation of this vector field $\va{\mathcal{F}}$ in the equatorial basis 
\begin{equation}
\va{\mathcal{F}} = \mathcal{F}_\delta(\nu, \vu{s}) \vu{e}_\delta + \mathcal{F}_\alpha(\nu, \vu{s}) \vu{e}_\alpha. \\
\end{equation}
The components in the two different bases are related by 
\begin{align}
\mathcal{F}_\delta & = \vu{e}_\delta \vdot \va{\mathcal{F}}\\
& = \qty(\vu{e}_\delta \vdot \vu{e}_\theta) \mathcal{F}_\theta + \qty(\vu{e}_\delta \vdot \vu{e}_\phi ) \mathcal{F}_\phi\\
\mathcal{F}_\alpha & = \vu{e}_\alpha \vdot \va{\mathcal{F}}\\
& = \qty(\vu{e}_\alpha \vdot \vu{e}_\theta) \mathcal{F}_\theta + \qty(\vu{e}_\alpha \vdot \vu{e}_\phi) \mathcal{F}_\phi
\end{align}
which defines a rotation matrix field $\mathcal{U}(\vu{s})$ with elements
\begin{align}
\mathcal{U}(\vu{s}) & = \mqty*(\vu{e}_\delta \vdot \vu{e}_\theta & \vu{e}_\delta \vdot \vu{e}_\phi \\ \vu{e}_\alpha \vdot \vu{e}_\theta & \vu{e}_\alpha \vdot \vu{e}_\phi \\) \\
& = \mqty*(\cos(\chi(\vu{s})) & \sin(\chi(\vu{s})) \\ -\sin(\chi(\vu{s})) & \cos(\chi(\vu{s}))\\ )
\end{align}
For two feeds $a$ and $b$ with the response of each given by the vector fields $\va{\mathcal{F}}_a$ and $\va{\mathcal{F}}_b$ the instrumental Jones matrix is then specified in the equatorial basis as
\begin{equation}
\bm{J} = \mqty*( \mathcal{F}_{a\delta} & \mathcal{F}_{a \alpha} \\ \mathcal{F}_{b \delta} & \mathcal{F}_{b \alpha}).
\end{equation}

\section{Effect of thermal noise in polarization null test}
\label{sec:NoiseAppendix}

Since we have not included the effect of thermal noise or an absolute scale for the polarized power in our analysis, we consider a schematic model of how these variables would affect the statistics of the proposed null test. The point is to argue that if polarization leakage were the limiting systematic in the power spectrum the variance in our null test due to fluctuations in the polarized power will eventually dominate the variance due to thermal noise. 

Let $\mathcal{P}_I$ be the Stokes-I contribution to the power spectrum, $\mathcal{P}_L$ the contribution of Stokes-Q and U, and $\mathcal{N}$ the thermal noise with mean $\expval{\mathcal{N}} = 0$ - for simplicity of exposition we neglect cross-terms between Stokes parameters. The power spectrum
\begin{equation}
\mathcal{P} = \mathcal{P}_I + \mathcal{P}_L + \mathcal{N}
\end{equation}
can then be considered a random variable over the collection $\mathfrak{C}$ of subsets of sidereal days, as each subset produces a different realization of the noise, and changing ionospheric attenuation produces a fluctuation in $\mathcal{P}_L$. The $\mathcal{P}_I$ term which represents the cosmological signal is taken to be constant over the subsets. If $\widehat{\mathcal{P}}_L$ is the intrinsic polarized power then the attenuation factor is
\begin{align}
\xi & = \frac{\mathcal{P}_L}{\widehat{\mathcal{P}}_L}\\
& = \overline{\xi} + \delta \xi
\end{align}
where $\overline{\xi}$ is defined by the mean of $\mathcal{P}_L$ over $\mathfrak{C}$,
\begin{equation}
\overline{\mathcal{P}}_L = \expval{\mathcal{P}_L} = \widehat{\mathcal{P}}_L \expval{\xi} = \widehat{\mathcal{P}}_L \overline{\xi}
\end{equation}
The polarized power can also be written as
\begin{align}
\mathcal{P}_L & = \overline{\mathcal{P}}_L + \delta \mathcal{P}_L\\
& = \overline{\mathcal{P}}_L \qty(1 + \frac{\delta \xi}{\overline{\xi}})
\end{align}
so we can see that
\begin{equation}
\delta \mathcal{P}_L = \overline{\mathcal{P}}_L \frac{\delta \xi}{\overline{\xi}}
\end{equation}
The mean and variance of $\mathcal{P}$ are then
\begin{equation}
\expval{\mathcal{P}} = \mathcal{P}_I + \overline{\mathcal{P}}_L
\end{equation}
\begin{align}
\expval{\mathcal{P}^2} - \expval{\mathcal{P}}^2 & = \expval{\delta \mathcal{P}_L^2} + \expval{\mathcal{N}^2}\\
& = \overline{\mathcal{P}}_L^2 \expval{\frac{\delta \xi^2}{\overline{\xi}^2} } + \expval{\mathcal{N}^2}\\
& = \overline{\mathcal{P}}_L^2 \qty(\frac{\expval{\delta \xi^2}}{\overline{\xi}^2}  + \frac{\expval{\mathcal{N}^2}}{\overline{\mathcal{P}}_L^2})
\end{align}
If $\mathcal{P}_I >> \overline{\mathcal{P}}_L$, then we detect the cosmological signal with an uncertainty dominated by the thermal noise, and any other small systematics. If $\mathcal{P}_I << \overline{\mathcal{P}}_L$, then we can see that the ionospheric fluctuation of $\mathcal{P}_L$ in our null test will dominate the variance due to thermal noise - we will have been thwarted from observing cosmological reionization, but we will not be fooled into thinking otherwise.

In a regime where polarization leakage is comparable to the cosmological signal we would have $\overline{\mathcal{P}}_L \approx \mathcal{P}_I$ and thus the second term is approximately the thermal-noise-to-signal on a detection in the absence of $\mathcal{P}_L$. The HERA experiment is designed to detect the EoR power spectrum at high signal-to-thermal-noise so even in a regime where $\overline{\mathcal{P}}_L$ is slightly smaller than $\mathcal{P}_I$ the excess variance in the null test should still be detectable.



\end{document}